\documentclass[aps,prd,tightenlines,preprintnumbers,showpacs,superscriptaddress,nofootinbib,eqsecnum,floatfix,amsmath,longbibliography,twocolumn]{revtex4-2}

\usepackage{bm}
\usepackage{latexsym}
\usepackage{graphicx}
\usepackage{multirow,soul}
\usepackage[dvipsnames]{xcolor}

\usepackage{hyperref}

\def\b{\beta}
\def\c{\chi}

\def\e{\epsilon}                
\def\g{\gamma}

\def\j{\psi}
\def\k{\kappa}

\def\m{\mu}
\def\n{\nu}

\def\p{\pi}                     

\def\D{\Delta}

\def\cbo{{\,\raise-.15ex\Sc [\,}}                       

\def\svev#1{\left\langle #1\right\rangle}       

\def\ddt#1{{\buildrel {\hbox{\LARGE .\kern-2pt.}} \over {#1}}}

\def\half{{1\over 2}}

\long \def \blockcomment #1\endcomment{}

\definecolor{red}{rgb}{1., 0., 0.}
\definecolor{myred}{rgb}{0.8,0.0,0.0}
\definecolor{green}{rgb}{0.0,0.6,0.0}
\definecolor{darkblue}{rgb}{0.0,0.1,0.7}
\definecolor{brown}{rgb}{0.6,0.1,0.0}
\definecolor{gray}{rgb}{0.6,0.6,0.6}
\definecolor{darkgreen}{rgb}{0.0, 0.545098, 0.0}
\definecolor{verydarkgreen}{rgb}{0.0, 0.4, 0.0}
\definecolor{veryverydarkgreen}{rgb}{0.0, 0.3, 0.0}
\definecolor{purple}{rgb}{0.5,0.0,0.5}
\definecolor{applegreen}{rgb}{0.55, 0.71, 0.0}
\definecolor{babypink} {rgb}{0.64, 0.44, 0.44}
\definecolor{orange}{rgb}{0.9,0.4,0.0}

\def\BS#1{{#1}}

\def\ggf{g^2}
\def\svev#1{\left\langle #1\right\rangle}

\def\vs{{\it vs.}}

\def\LHC{\Lambda_{HC}}
\def\LEHC{\Lambda_{EHC}}
\def\Eq#1{Eq.~(\ref{#1})}
\def\Tab#1{Table~\ref{#1}}
\def\Fig#1{Fig.~\ref{#1}}
\def\Sec#1{Sec.~\ref{#1}}

\def\rcite#1{Ref.~\cite{#1}}
\def\ph{\phantom}

\def\betaS{\beta_\text{S}}
\def\betaW{\beta_\text{W}}
\def\betaC{\beta_\text{C}}
\def\sigmaS{\sigma_\text{S}}

\begin{document}

\title{Infrared fixed point and anomalous dimensions\\
  in a composite Higgs model}

\author{Anna Hasenfratz}
\affiliation{Department of Physics, University of Colorado, Boulder, CO 80309, USA}

\author{Ethan T. Neil}
\affiliation{Department of Physics, University of Colorado, Boulder, CO 80309, USA}

\author{Yigal Shamir}
\affiliation{Raymond and Beverly Sackler School of Physics and Astronomy,
Tel~Aviv University, 69978 Tel~Aviv, Israel}

\author{Benjamin Svetitsky}
\affiliation{Raymond and Beverly Sackler School of Physics and Astronomy,
Tel~Aviv University, 69978 Tel~Aviv, Israel}

\author{Oliver Witzel}
\affiliation{Center for Particle Physics Siegen, Theoretische Physik 1,
  Naturwissenschaftlich-Technische Fakult\"at, Universit\"at Siegen,
  57068 Siegen, Germany}


\begin{abstract}
We use lattice simulations and
the continuous renormalization-group method, based on the gradient flow,
to study a candidate theory of composite Higgs and a partially composite top.
The model is an SU(4) gauge theory with four Dirac fermions in each of
the fundamental and two-index antisymmetric representations.
We find that the theory has an infrared fixed point at $g^2 \simeq 15.5$ in the gradient flow scheme.
The mass anomalous dimension of each representation is large at the fixed point.
On the other hand, the anomalous dimensions of top-partner operators
do not exceed 0.5 at the fixed point.
This may not be large enough
for a phenomenologically successful model of partial compositeness.
\end{abstract}

\preprint{SI-HEP-2023-08}

\maketitle

\newpage
\section{\label{sec:intro} Introduction}
\subsection{\label{Intro1} Background}

Compared to the Planck scale, the mass of the Higgs particle
in the Standard Model is unnaturally small.
A popular solution is to suppose that the Higgs is a
pseudo Nambu-Goldstone boson arising from
the spontaneous breaking of a chiral symmetry
\cite{Georgi:1984af,Dugan:1984hq}, which in turn
is induced by a novel strong interaction sometimes known as hypercolor (HC).
Since the top quark has a mass on the same scale, one might similarly suppose
that the top quark is partially composite \cite{Kaplan:1991dc},
receiving its mass from the direct coupling to a hypercolor baryon
called the top partner.

Composite Higgs models have been extensively studied using
effective field theory techniques  (for reviews,
see Refs.~\cite{Contino:2010rs,Bellazzini:2014yua,Panico:2015jxa}).
It is important, however, to seek out realizations of this paradigm
as a concrete, asymptotically free theory such as hypercolor.
A list of candidate theories
satisfying a number of desirable properties was compiled
by Ferretti and Karateev \cite{Ferretti:2013kya,Ferretti:2014qta,Ferretti:2016upr,Belyaev:2016ftv} (see also \rcite{Vecchi:2015fma}).
As a stand-alone theory---before coupling
to Standard Model fields---each model in the Ferretti--Karateev list is
a vector-like gauge theory with fermions in two different representations
of the gauge group.  The top partner is a {\em chimera}:
a three-fermion bound state made out of fermions of both representations.

Without additional interaction terms, the hypercolor gauge interaction
cannot generate Standard Model fermion masses,
nor can it induce electro-weak symmetry breaking.
The coupling of the top quark to its chimera partner must come from
four-fermion interactions, whose origin lies in a sector,
known as extended hypercolor (EHC), with a still higher energy scale.
Such interactions at the HC scale are naively suppressed
by a factor $\LHC^2/\LEHC^2$, where $\LHC$ is the scale of
the hypercolor theory, while $\LEHC\gg\LHC$ is the scale of the EHC theory.

In the absence of a concrete realization of extended hypercolor
in the literature,%
\footnote{See, however, \rcite{Cacciapaglia:2020jvj}}
it must be assumed that, generically,
the four-fermion interactions induced by the EHC theory can give rise to
unwanted flavor violations.  In order to respect experimental bounds
on flavor violation, $\LHC^2/\LEHC^2$ must be small.
In order to obtain a realistic top mass, however, the effect of
the suppression factor on the top-partner mixing term must be reduced.
To this end, two conditions must be satisfied.
First, some of the four-fermion operators responsible for partial compositeness
must have a large anomalous dimension.  Schematically, these operators
have the form $\bar{q}B$, where $q$ is a Standard Model fermion field,
and $B$ is a hypercolor-singlet chimera operator.  Therefore, one
requires the existence of chimera operators $B$ with
large anomalous dimensions within the hypercolor theory.
The second requirement is that the hypercolor theory itself must be nearly conformal,
allowing the large chimera anomalous dimensions to persist over many scales---ideally, all the way from the EHC scale down to the hypercolor scale.%
\footnote{Since the days of walking technicolor \cite{Bando:1987br,Hill:2002ap}, it has been expected that large anomalous dimensions would appear near the sill of the conformal window. See also \rcite{Kaplan:2009kr}.}

The hope, then, is that successful composite-Higgs models will be found
near the sill of the conformal window.  Theories just below the sill
are obvious candidates, nearly conformal but ultimately confining
and chirally broken.  Theories slightly above the sill, which feature
an infrared fixed point when all the fermions are massless,
are also eligible, as one can induce confinement by giving large masses
to a small subset of the fermions \cite{Brower:2015owo,Hasenfratz:2016gut,Witzel:2019oej,Appelquist:2020xua}.

\begin{figure}[t]
\includegraphics[width=.95\columnwidth]{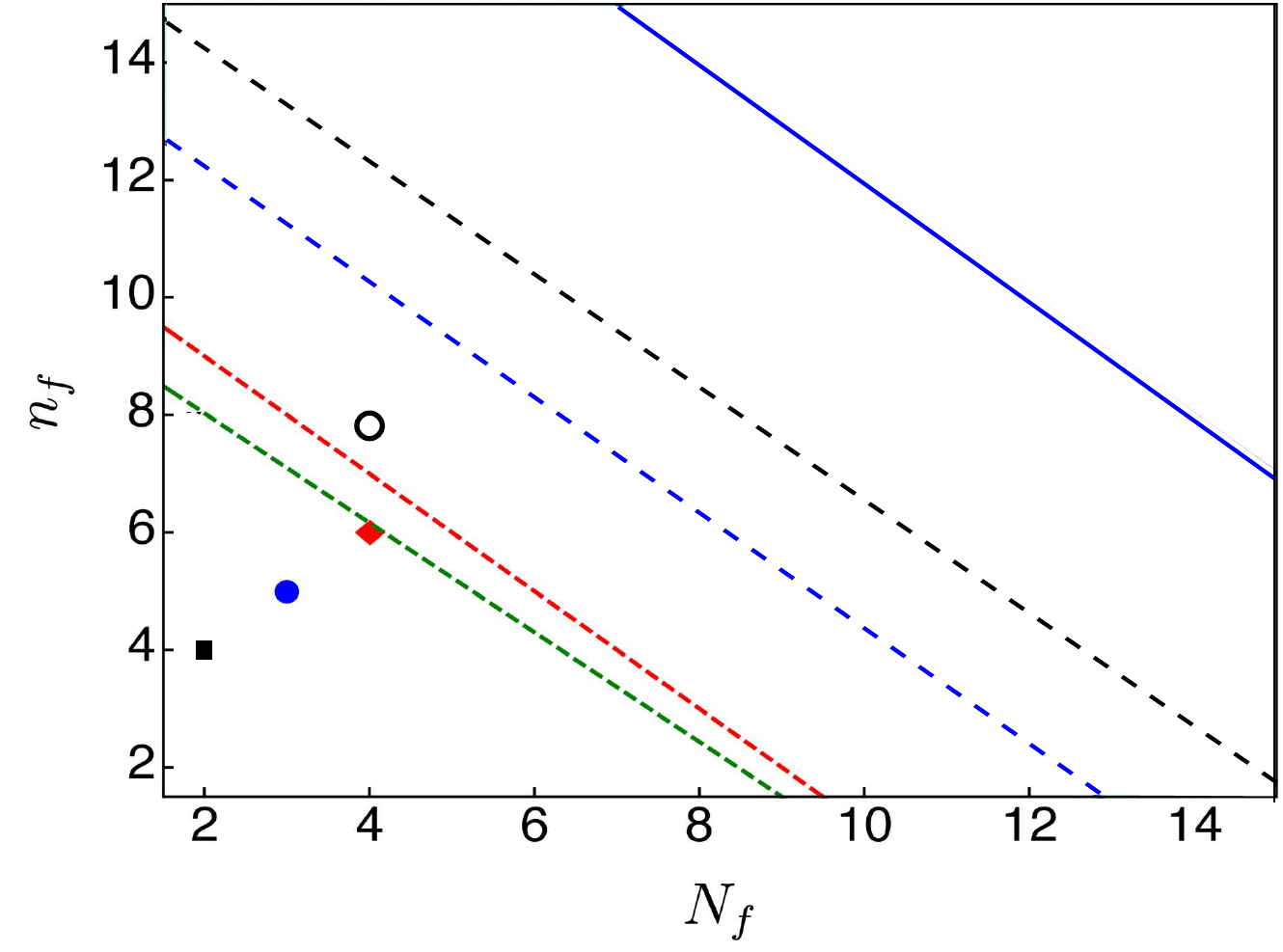}
\caption{\label{fig:theories}
\BS{Estimates for the location of the conformal window
of the SU(4) gauge theory with $N_f$ Dirac fermions in the fundamental
representation and $n_f$ Majorana fermions in the sextet representation.
The uppermost line is the limit of asymptotic freedom, where the 1-loop beta function changes sign.
The other lines are various analytical estimates of the sill
of the conformal window (see text).
Blue circle: M6 model; red diamond: M11 model; black square: 2+2 model;
open circle: 4+4 model, the subject of this paper.
Only the lowest two lines are consistent with the IR fixed point that we find for the 4+4 model.
}}
\end{figure}

Figure~\ref{fig:theories} shows analytic estimates for the location of the
conformal sill for SU(4) gauge theory in the plane spanned by
the number of Dirac fermions in the fundamental {\bf 4} representation
and the number of Majorana fermions in the two-index antisymmetric {\bf 6}
representation, which is real.%
\footnote{
The dashed green line is the sill of the conformal window according to the
2-loop beta function, where it first develops an IR fixed point;
the dashed red line is from the ``all-orders'' beta function; 
the dashed blue line reflects the ``critical condition'' of \rcite{Kim:2020yvr};
the dashed black line stems from long-standing analysis of Schwinger--Dyson equations---for all of these, see \rcite{Kim:2020yvr}, from which this figure was adapted, and references therein.}
Also shown in the figure are two of the Ferretti--Karateev models
that belong to this plane: the so-called M6 and M11 models
\cite{Ferretti:2013kya,Belyaev:2016ftv}.
The ``2+2 model,'' which contains two Dirac fermions in each of the {\bf 4}
and {\bf 6} representations,\footnote{
  The two Dirac fermions in the real {\bf 6} representation are equivalent to four Majorana fermions.
}
has been studied extensively using lattice techniques.  While QCD-like, and
hence far from the conformal sill, the 2+2 model served
as a useful laboratory, providing the first example studied of
an asymptotically free gauge theory with two representations,
in particular one that produces chimera baryons
\cite{Ayyar:2017qdf,Ayyar:2018zuk,Ayyar:2018ppa,Ayyar:2018glg,Ayyar:2019exp,Golterman:2020pyx,Cossu:2019hse,DelDebbio:2022qgu}.\footnote{
For lattice studies of a composite Higgs model based on the Sp(4) gauge group,
see Refs.~\cite{Bennett:2017kga,Bennett:2019jzz,Bennett:2022yfa,Hsiao:2022kxf,Bennett:2023wjw}.
}

In this paper we move on to the ``4+4 model,'' where we increase
the number of Dirac fermions in each representation from two to four.
The 4+4 model has a number of desirable features.  First,
the M6 and M11 models can both be embedded into the 4+4 model, and
can be reached by giving a subset of the fermions a (Dirac or Majorana) mass.
In addition, as can be seen in the figure, the 4+4 model is much more likely
to be close to, or even inside, the conformal window.

\subsection{\label{Intro2} Method and summary of results}

We extract the beta function and anomalous dimensions
using a continuous renormalization group (RG) method
\cite{Hasenfratz:2019hpg,Hasenfratz:2019puu}.%
\footnote{for a slightly different approach
  see Refs.~\cite{Peterson:2021lvb,Hasenfratz:2023bok,Fodor:2017die}.}
The length scale for this RG is $\sqrt{t}$, where $t$ is the parameter of
a gradient flow (GF) generated by integrating a diffusion equation for the
gauge field \cite{Luscher:2010iy}.
The GF running coupling is defined as \cite{Fodor:2012td}
\begin{equation}
\label{ggf}
\ggf = \frac{\mathcal{N}}{C(t;L,T)} t^2 \svev{E(t)} \ .
\end{equation}
Here the energy density at scale $\sqrt{t}$ is
$E=\frac{1}{4}G_{\m\n}^a G_{\m\n}^a$,
where $G_{\m\n}^a$ is the flowed gauge field strength.
\BS{$\mathcal{N}$ is a numerical factor which depends on the gauge group, and $C(t;L,T)$ is a correction for the finite dimensions $L,T$ of the volume being simulated (see below).}
Viewing the gradient flow as an RG transformation, the beta function is
\begin{equation}
\label{betafn}
\b(\ggf) = -t\,\frac{\partial \ggf}{\partial t} \ .
\end{equation}
The extension of the GF technique to fermions was developed in \rcite{Luscher:2013cpa}, while the use of the continuous RG for obtaining anomalous dimensions of fermion operators was introduced in \rcite{Carosso:2018bmz}.

\begin{figure}[t]
\begin{center}
\includegraphics*[width=\columnwidth]{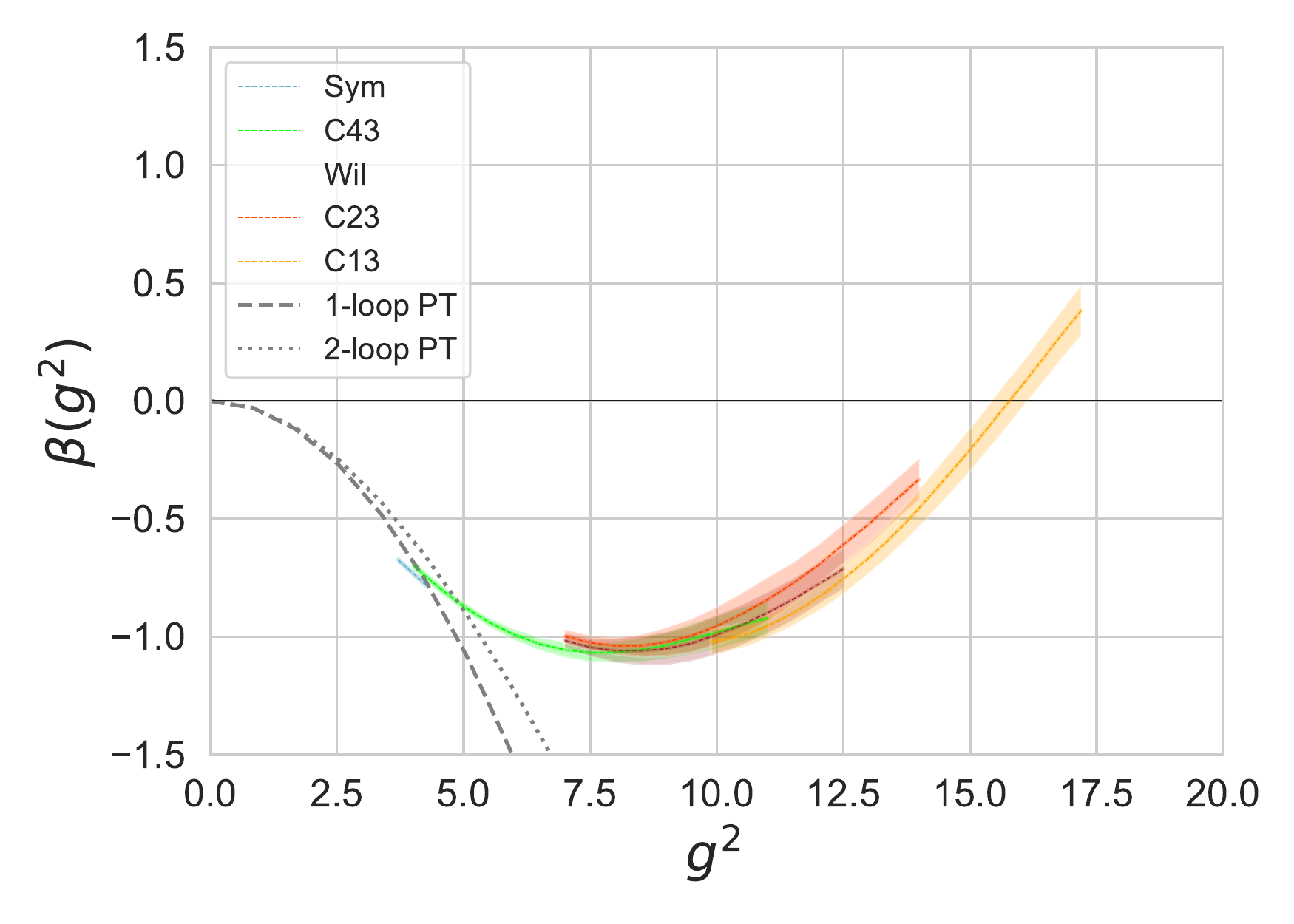}
\end{center}
\begin{quotation}
\caption{\label{betafn-matched}
The $\beta$ function obtained with five different lattice gradient flows.
All flows are the same in the continuum limit;
their regions of validity are different, though overlapping.
For details see Sec.~\ref{sec:GF}.
}
\end{quotation}
\vspace*{-3ex}
\end{figure}

\begin{figure}[t]
\begin{center}
\includegraphics*[width=\columnwidth]{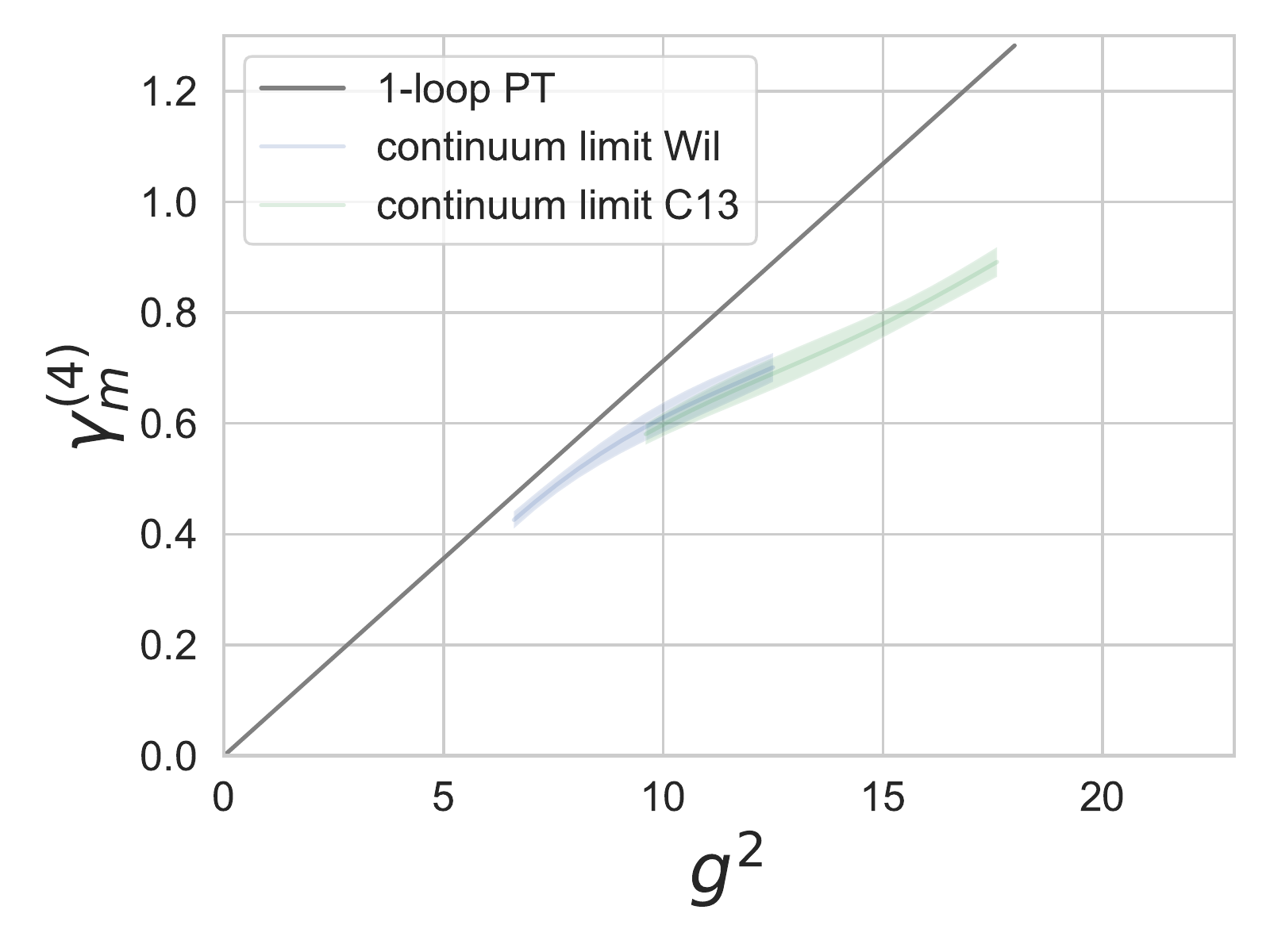}\\
\includegraphics*[width=\columnwidth]{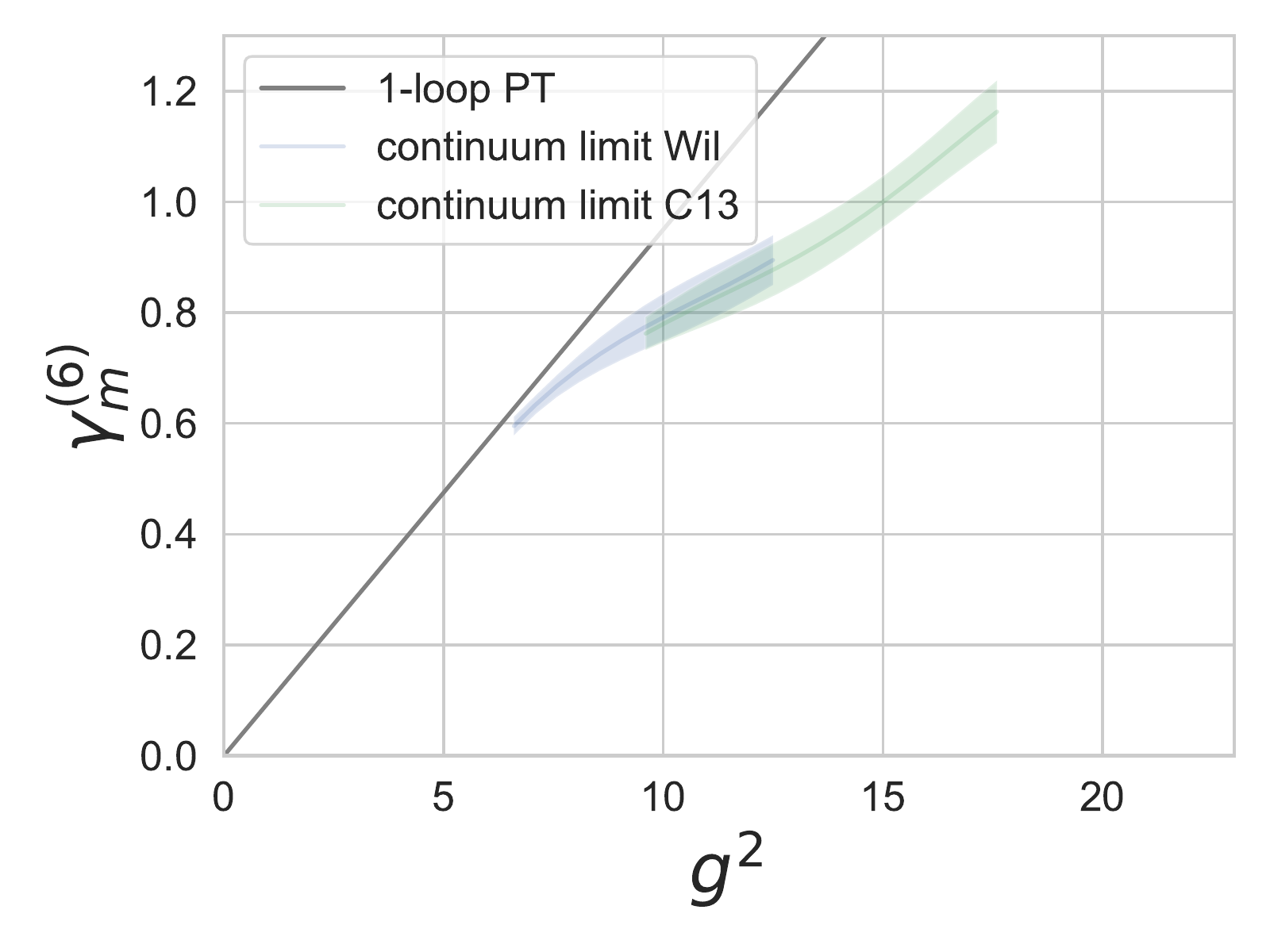}
\end{center}
\caption{\label{gamma-cont} Anomalous dimensions of the two mass operators.
Top: fundamental representation.  Bottom: sextet representation.
See Sec.~\ref{sec:andim} for details.
}
\end{figure}

\begin{figure}[t]
\begin{center}
\includegraphics*[width=\columnwidth]{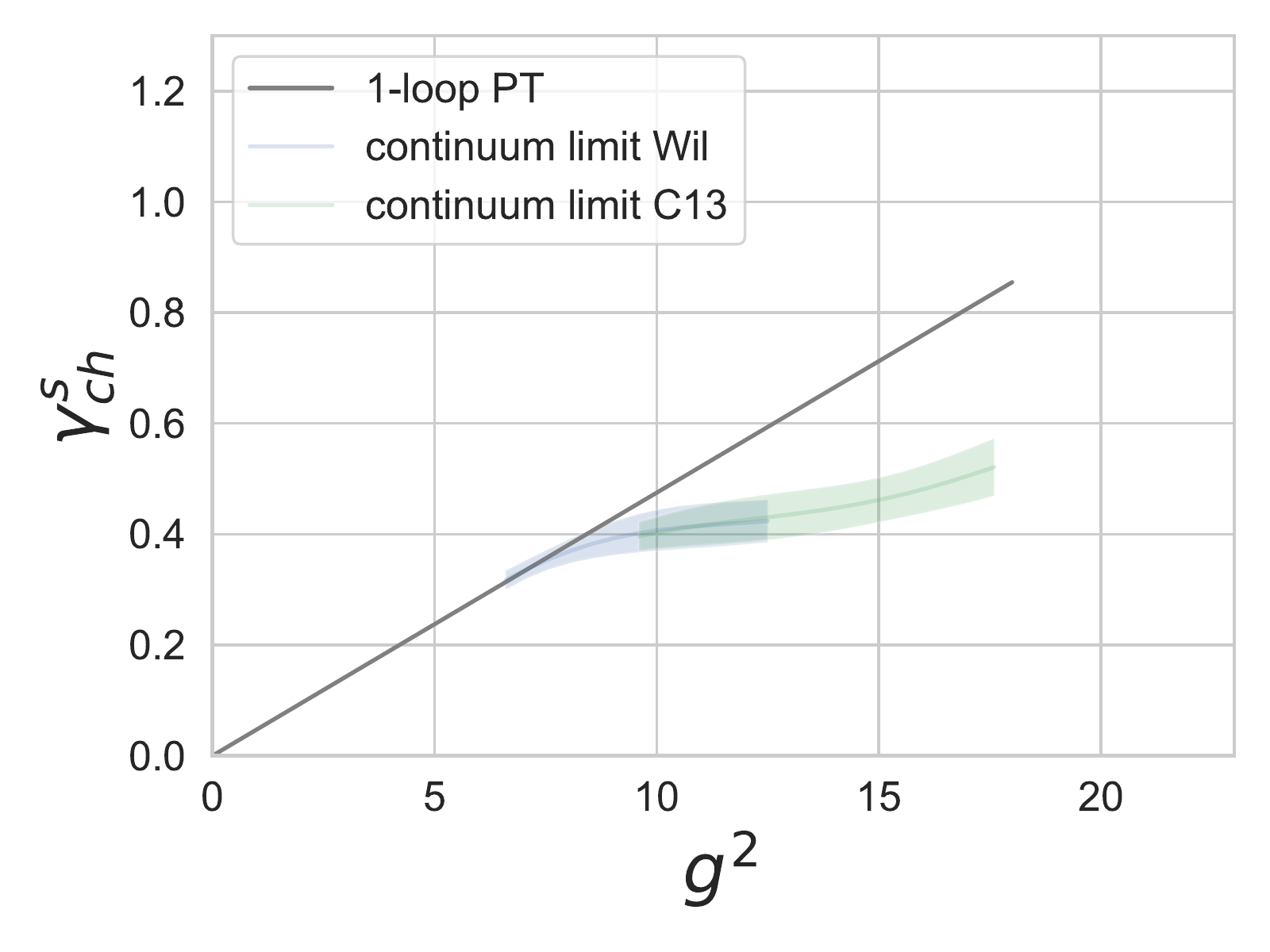}
\end{center}
\caption{\label{andim-chimera}
The largest chimera anomalous dimension, $\g^s_{ch}$.
The other two chimera anomalous dimensions are smaller (see \Fig{andim-chimera2}).}
\end{figure}

Our main findings are the beta function (\Fig{betafn-matched}),
which shows an IR fixed point;%
\footnote{\BS{In comparing to the 2-loop curve, recall that it also crosses the abscissa, giving an IR-stable fixed point, but at a strong coupling well outside the range of the figure.}}
 the mass anomalous dimensions (\Fig{gamma-cont});
 and the anomalous dimensions of the chimera operators with the lowest
mass dimension, namely, three-fermion operators with no derivatives.
The largest chimera anomalous dimension is shown in \Fig{andim-chimera},
while the other two are shown in \Fig{andim-chimera2}.
The IR fixed point places the model inside the conformal window.
Moreover, the anomalous dimensions of the mass operators are large,
for both representations, at the fixed point.
Unfortunately, the anomalous dimensions of all chimera operators
are fairly small at the fixed point, making it unlikely that the model
can successfully account for a partially composite top quark.

We obtain the flowed expectation values in \Eq{ggf} in the presence of a lattice cutoff, generating ensembles of the gauge field with numerical simulations.
We present our lattice methods, including the calculation of the beta function and its extrapolation to the continuum limit, in \Sec{sec:GF}.
The calculation of anomalous dimensions is the subject of \Sec{sec:andim}.
We offer our conclusions in \Sec{sec:conc}.
Further technical details of the lattice calculation are given in the appendix.

\section{\label{sec:GF} Gauge flow and the beta function}

\subsection{\label{sec:lattice} Lattice strategy}
The lattice action, our simulation algorithm, and the ensembles we generated are described in the appendix.
We use Wilson-clover fermions and tune the bare masses such
that the fermions of both representations are essentially massless.

A new ingredient in the lattice action is a set of Pauli--Villars fields
\cite{Hasenfratz:2021zsl}. Without these, the presence of many fermion flavors,
especially with smearing-improved gauge connections, generates a large
screening effect in the effective action for the gauge field. In order to obtain a strong
renormalized coupling, one would be pushed towards large bare coupling
$g_0^2 = N_c/\beta$.  This, in turn, would cause large ultraviolet fluctuations.
As a consequence, these systems often encounter phase transitions
or other discontinuities, lattice artifacts that prevent the approach to
the desired renormalized coupling, especially when the fermions are light.
Our 4+4 system
exhibited such a discontinuity when the original lattice action was used.
The addition of Pauli--Villars fields weakens the induced term and allows us
to reach much further into strong renormalized coupling.

As mentioned in the introduction, we use the continuous RG technique
to determine the beta function.  The 4+4 model that we simulate
is a massless, asymptotically free lattice theory.
The gradient flow acts as a transformation in coupling space:
as the flow time $t$ is increased, irrelevant operators die out
and the flow converges towards the renormalized trajectory (RT)
emerging from the gaussian fixed point in the ultraviolet.
Our task will be to ensure that, given a specific flow
at a specific renormalized coupling, the flow has reached close enough
to the RT that any remaining discretization effects can be removed
by a simple extrapolation.

We begin by applying a gradient flow
to the gauge field of every lattice configuration.
In principle, extracting continuum results requires first taking
the infinite-volume limit and then the continuum limit of zero lattice spacing $a$;
the latter limit involves taking the dimensionless lattice flow time $t/a^2 \to \infty$.
Eschewing a formal infinite-volume limit, we focus on
the continuum limit, which we extract from
eight ensembles on a lattice volume of $24^3\times48$ sites.
The bare couplings of these ensembles were selected
to range from weak to strong coupling.
We have also generated ensembles on volume $28^3\times56$ for two of the bare couplings;
we use these to estimate finite-volume effects.
By using lattice flow time $t/a^2 \le 3.4$ we have limited the
finite volume effects to no more than a few percent.

\subsection{\label{sec:flows} Gauge flows}
The continuum GF equation takes the form \cite{Luscher:2010iy}
\begin{equation}
\label{contGF}
\frac{\partial B_\m}{\partial t} = D_\n G_{\n\m} \ ,
\end{equation}
where $B_\m$ is the flowed gauge field, and $G_{\n\m}$ the associated
field strength.  The initial condition of the flow is the dynamical
gauge field, viz.,
\begin{equation}
\label{bcBA}
B_\m \big|_{t=0} = A_\m \ .
\end{equation}
The right-hand side of \Eq{contGF} is just $-\partial S_g/\partial B_\m(x)$,
where $S_g = \frac14 G_{\m\n}^a G_{\m\n}^a$ is the standard continuum gauge action
in terms of the flowed field.

\begin{table}[t]
\begin{ruledtabular}
\begin{tabular}{c|ccccc}
Flow & Sym & C43 & Wil & C23 & C13 \\
\hline
$c_p$ & 5/3 & 4/3 & 1 & 2/3 & 1/3 \\
$c_r$ & $-1/12$ & $-1/24$ & 0 & $1/24$ & $1/12$ \\
\end{tabular}
\end{ruledtabular}
\caption{\label{tab:flows} Lattice gradient flows.
The first line gives the flow's name used in this paper.
``Sym'' and ``Wil'' are the well-known Symanzik and Wilson flows.
$c_p$ and $c_r$ are the coefficients of the plaquette and rectangle terms
in the lattice action that generates each flow.
}
\end{table}

On the lattice, the continuum gauge field is replaced by
the link variables $U_\m(x)$.  Each saved lattice configuration provides
initial values for the gradient flow.
We also need a discrete version of $\partial S_g/\partial  B_\m(x)$
for the right-hand side of the flow equation (see \rcite{Luscher:2010iy}).
In this paper we apply five different gradient flows, each derived
from a different discretization of the continuum action $S_g$.
The lattice gauge action which generates each flow is a linear combination
of the plaquette ($1\times1$ Wilson loop) and
rectangle ($1\times2$ Wilson loop) terms,
with coefficients shown in \Tab{tab:flows}.
For proper normalization in the weak-coupling limit,
the plaquette and rectangle coefficients $c_p$ and $c_r$ are constrained by
\begin{equation}
\label{cpcr}
c_p + 8c_r = 1 \ .
\end{equation}
As discussed in detail below, we have found that increasing $c_p$ from 1
gives rise to flows with relatively small discretization effects
at weak bare coupling, while $c_p<1$ works well at strong bare coupling.
\BS{In particular, we found the range of validity of the Symanzik flow to be limited to the weak coupling regime. We did not pursue the fully Symanzik-improved Zeuthen flow \cite{Ramos:2015baa},
as it is unlikely that further perturbative improvement would significantly change this conclusion.}

Returning to \Eq{ggf} which defines the gradient flow coupling,
the numerical constant here is
\begin{equation}
\label{C}
\mathcal{N} = \frac{128\pi^2}{3(N_c^2-1)} \simeq 28.4 \ ,
\end{equation}
for $N_c=4$.
\BS{The finite-volume correction term $C(t;L,T)=1+\delta$ corrects for the zero modes of the gauge fields \cite{Fodor:2012td}.
For volume $L^3\times T$ we have%
\footnote{\BS{This expression is not given in \rcite{Fodor:2012td} but was shared privately by D.~Nogradi. We thank him for his assistance.}}
\begin{equation} \delta(t;L,T) = -  \frac{\pi^2 (8t)^2}{L^3 T}  + \vartheta \left(  t/L^2 \right)^3 \vartheta\left(t/T^2 \right) \ .
\end{equation}
}
\BS{The correction $|\delta|$ is smaller than 0.005 for all our volumes in the useful ranges of $t$.}

In addition, one has to select a discretization of the ``energy''
$E=S_g$, which is used to define the coupling in \Eq{ggf}.
We use three different discretizations, two of which correspond
to the Symanzik (S) and Wilson (W) actions of \Tab{tab:flows}.
A third discretization is provided by the ``clover'' (C) operator.
As explained below, we have found that the S operator
gives the smoothest approach to the continuum limit,
and so we will focus on results obtained using this operator.

\begin{figure}[t]
\begin{center}
\includegraphics*[width=\columnwidth]{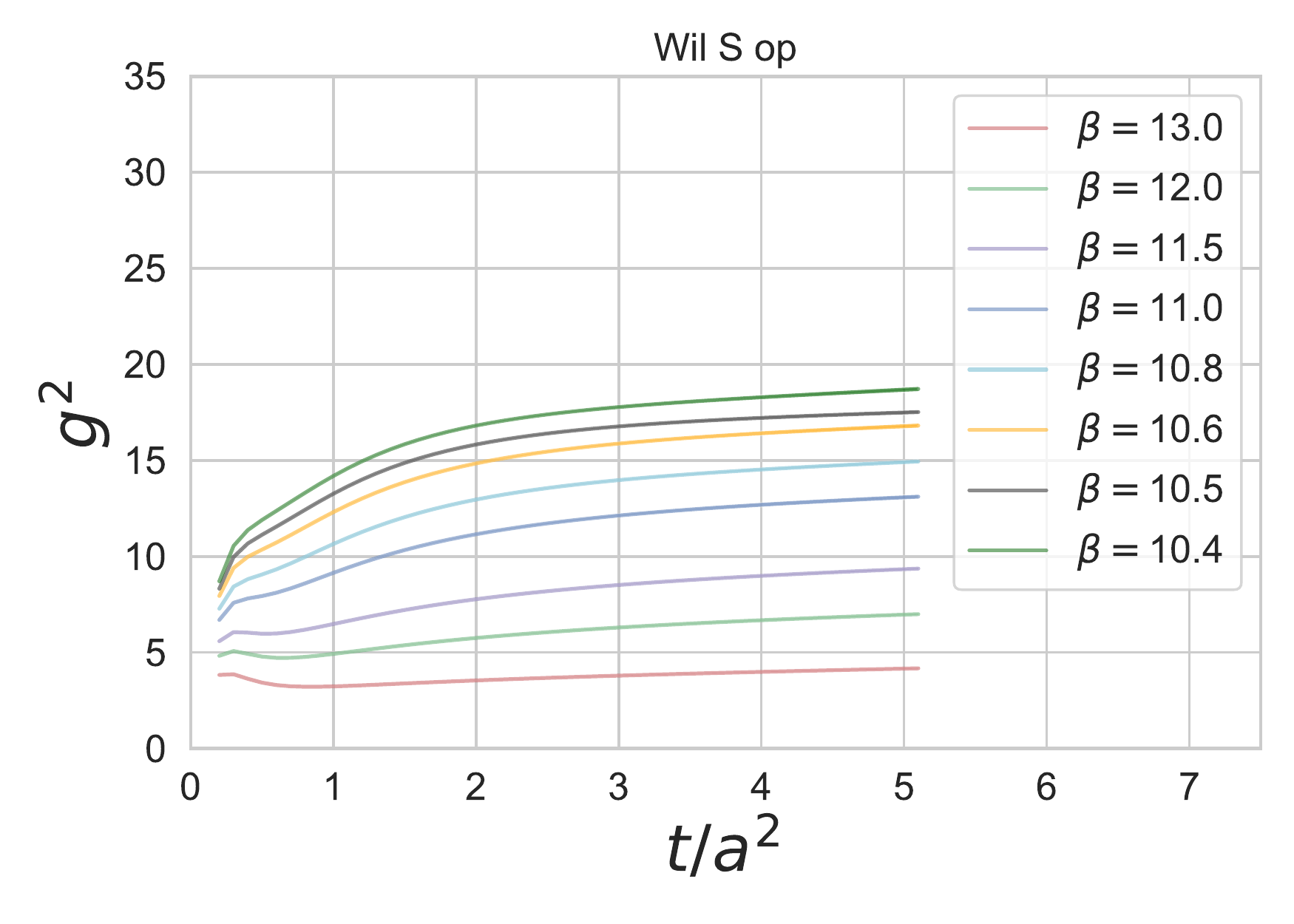}\\
\includegraphics*[width=\columnwidth]{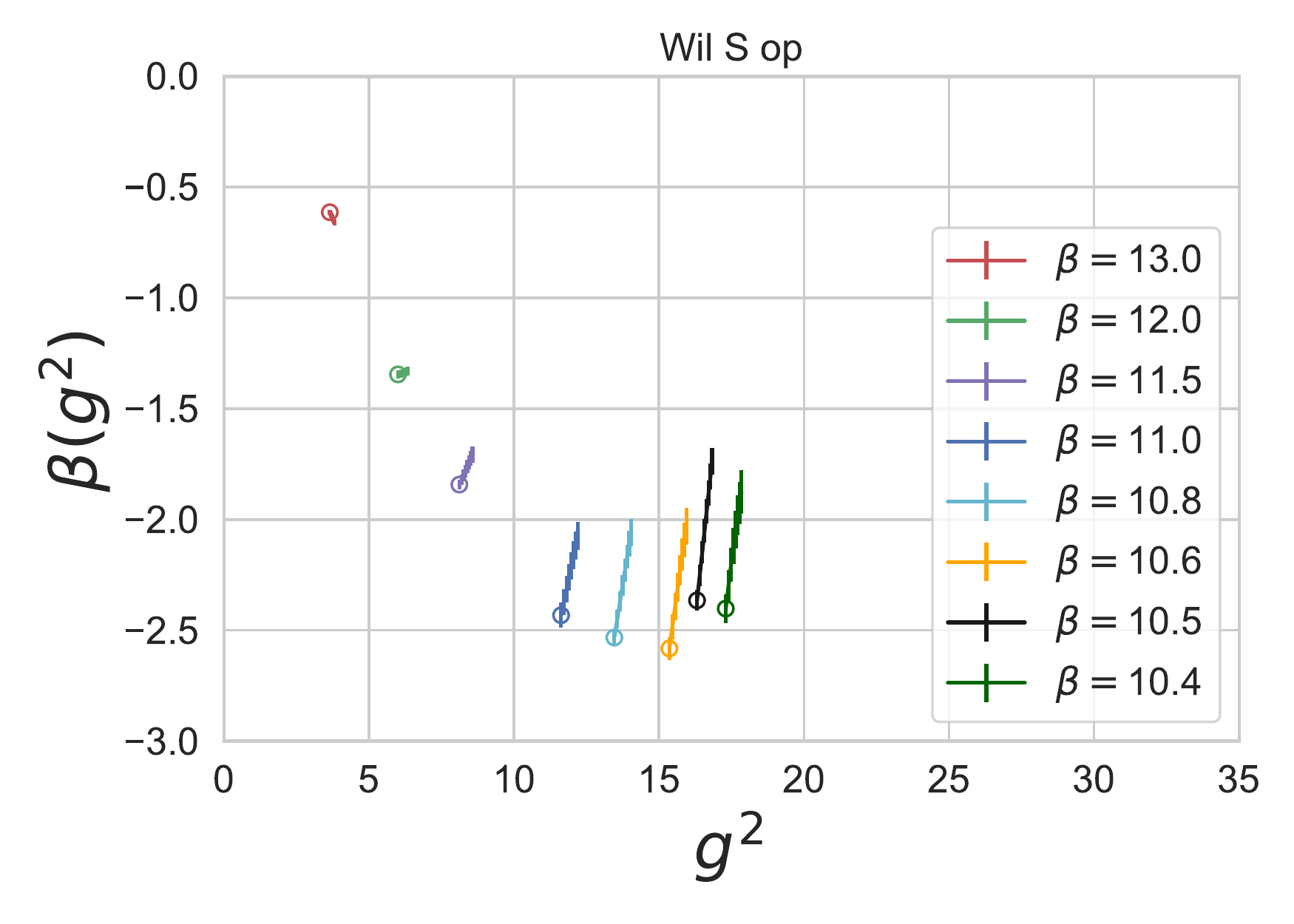}
\end{center}
\begin{quotation}
\caption{\label{fig:Wil-raw}
Results for the S operator measured in the Wilson flow.
Top: Gradient flow coupling $\ggf$ as a function of lattice flow time $t/a^2$.
Bottom: $\beta(g^2)$ \vs~$\ggf$ at flow times $2.4\leq t/a^2\leq3.2$.
Each open circle indicates the smallest flow time shown, $t/a^2=2.4$.
Error bars in the top panel of this figure,
as well as in Figs.~\ref{fig:C13-raw} and~\ref{fig:vol},
are too small to be visible.
}
\end{quotation}
\vspace*{-4ex}
\end{figure}

\subsection{\label{FV} Example flows and infinite volume limit}
We generated each flow by numerically integrating a lattice version
of the flow equation~(\Ref{contGF}) in steps of $dt=0.01$.
The energy $E$ of the flowed gauge field was recorded for the three operators
S, W, and C, at intervals of $\D t=0.1$.  The derivative in the bare
beta function (\ref{betafn}) was estimated with a five-point difference formula.

\begin{figure}[t]
\begin{center}
\includegraphics*[width=\columnwidth]{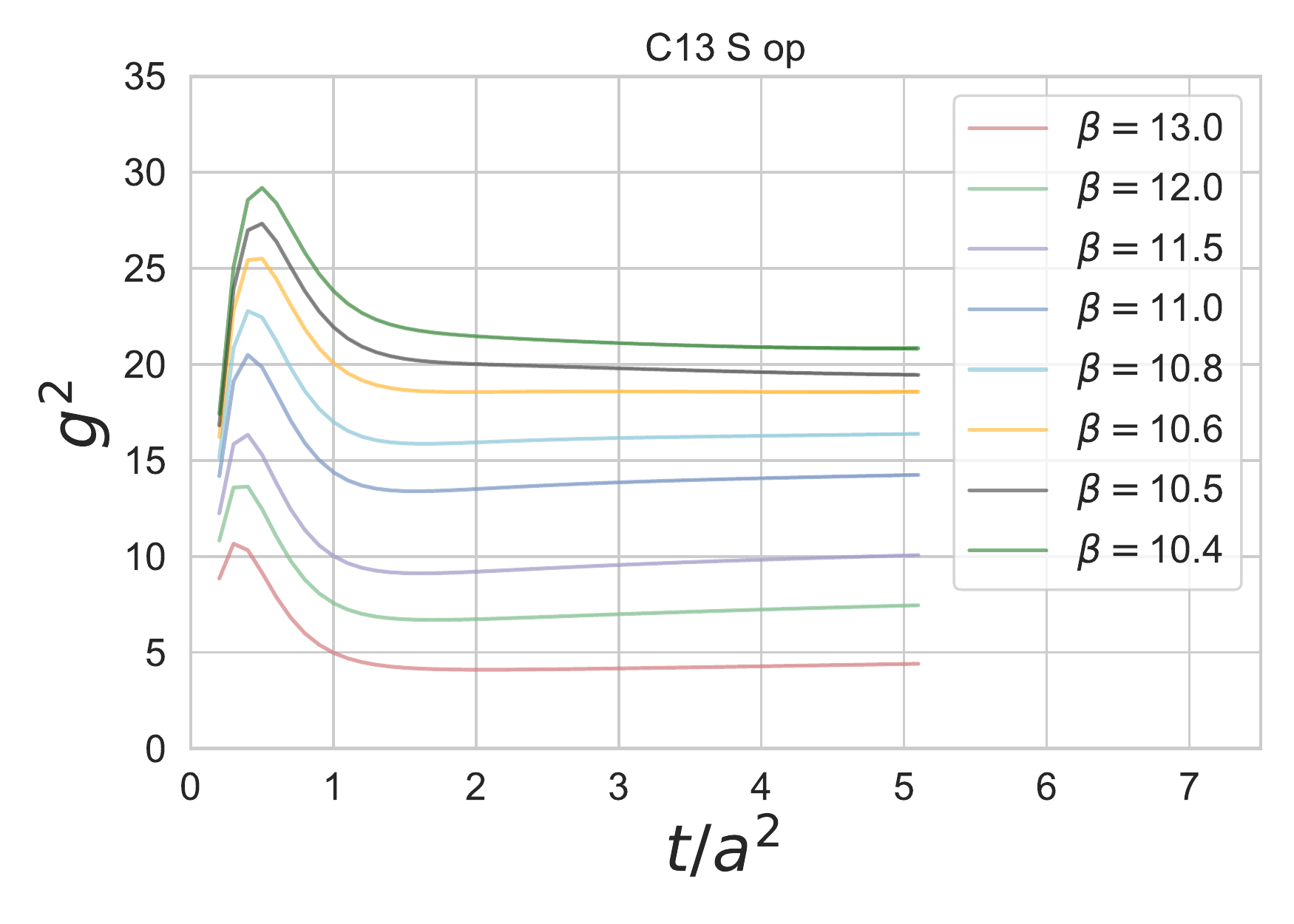}\\
\includegraphics*[width=\columnwidth]{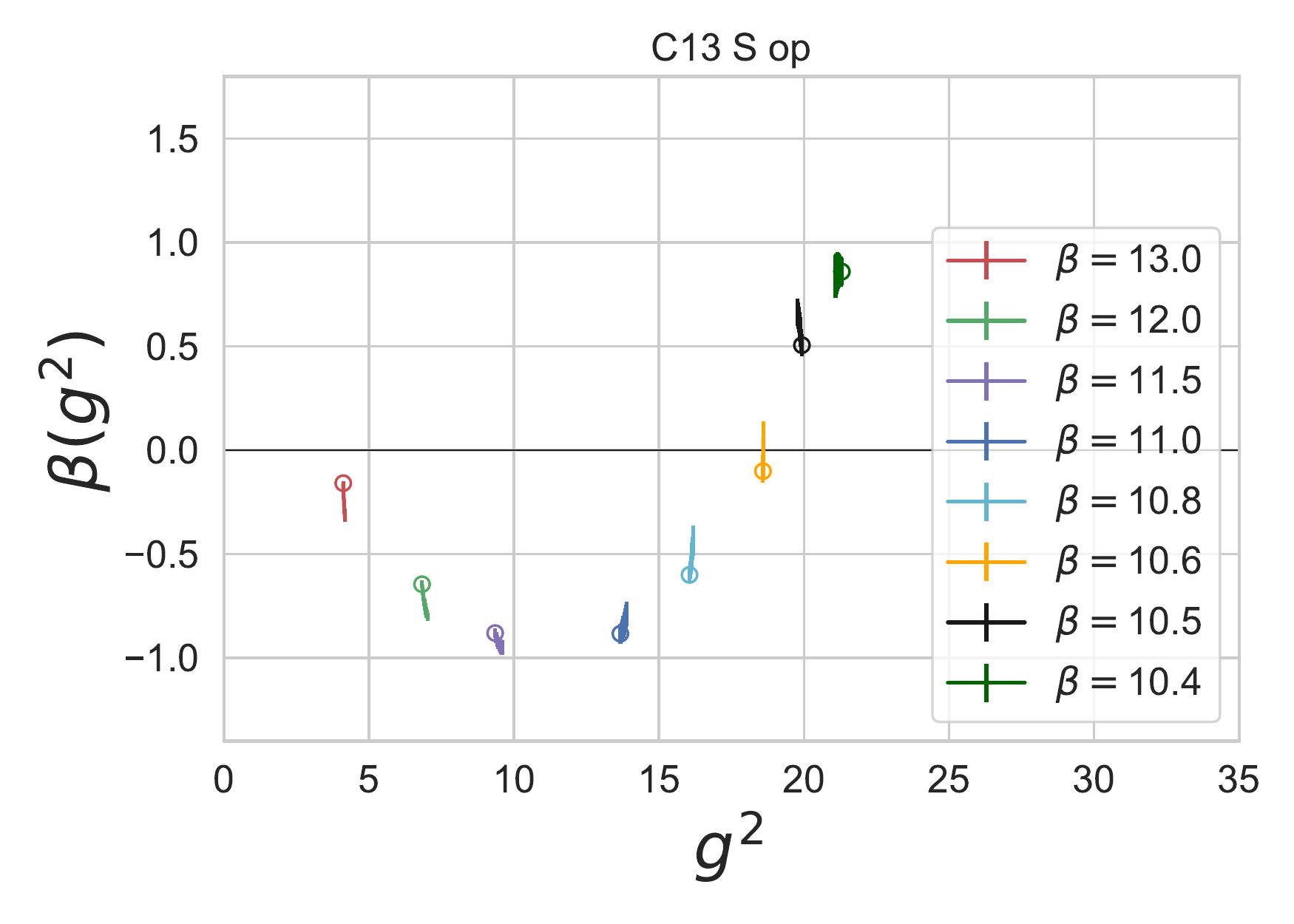}
\end{center}
\begin{quotation}
\caption{\label{fig:C13-raw}
Similar to \Fig{fig:Wil-raw}, for the S operator measured in C13 flow.}
\end{quotation}
\vspace*{-4ex}
\end{figure}

We show the raw flows for all eight of our $24^3\times48$ ensembles
in \Fig{fig:Wil-raw}\ for Wilson flow
and in \Fig{fig:C13-raw}\ for C13 flow.\footnote{
  The C13 flow is equivalent to the AFLOW introduced in
  \rcite{Hasenfratz:2020vta}.
  The motivation  there is quite different from ours.
}
Both present the S operator.
In each figure, the top panel shows the gradient flow coupling $g^2$
as a function of the lattice flow time $t/a^2$.  The bottom panel shows the resulting
raw beta function as a function of $g^2$ for a fixed interval of
flow time $2.4\leq t/a^2\leq3.2$.

A comparison of the two figures reveals that
the C13 flow reaches further into strong coupling than the Wilson flow:
with the S operator, the former
reaches $g^2\simeq 22$ while the latter reaches only $g^2\simeq 18$\@.
There are similar distinctions among the three operators.
For C13 flow, and within the above $t/a^2$ interval,
the C operator reaches $g^2 \simeq 18$ at our strongest bare coupling
and the W operator reaches $g^2 \simeq 20$, whereas the S operator reaches
$g^2 \simeq 22$ as just noted.  In the lower panel of \Fig{fig:C13-raw},
we see that the beta function becomes positive for the largest couplings.
Reaching larger values of the gradient flow coupling
has direct bearing on the ability to confirm the existence of an IR fixed point.

Figures~\ref{fig:Wil-raw} and~\ref{fig:C13-raw} show
raw data---before extrapolation to the continuum.
In \Sec{interpolation} and \Sec{contlim} we will show that
cutoff effects at strong coupling remain small for the S operator,
and almost as small for the W operator.
Moreover, the strongest attainable renormalized coupling of the C13 flow
remains much larger than that of the Wilson flow in the continuum limit.

\begin{figure}[t]
\begin{center}
\includegraphics*[width=\columnwidth]{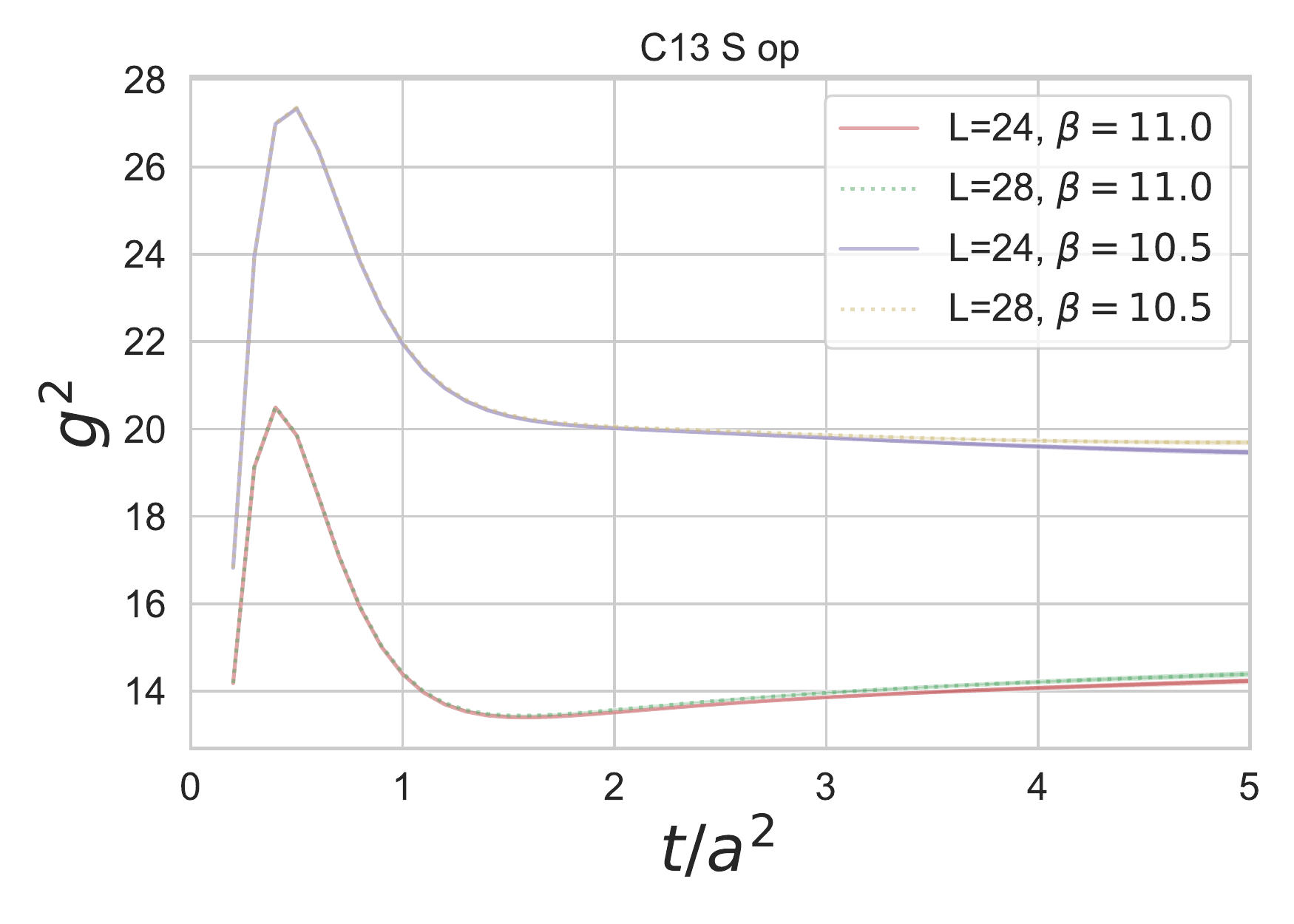}\\
\includegraphics*[width=\columnwidth]{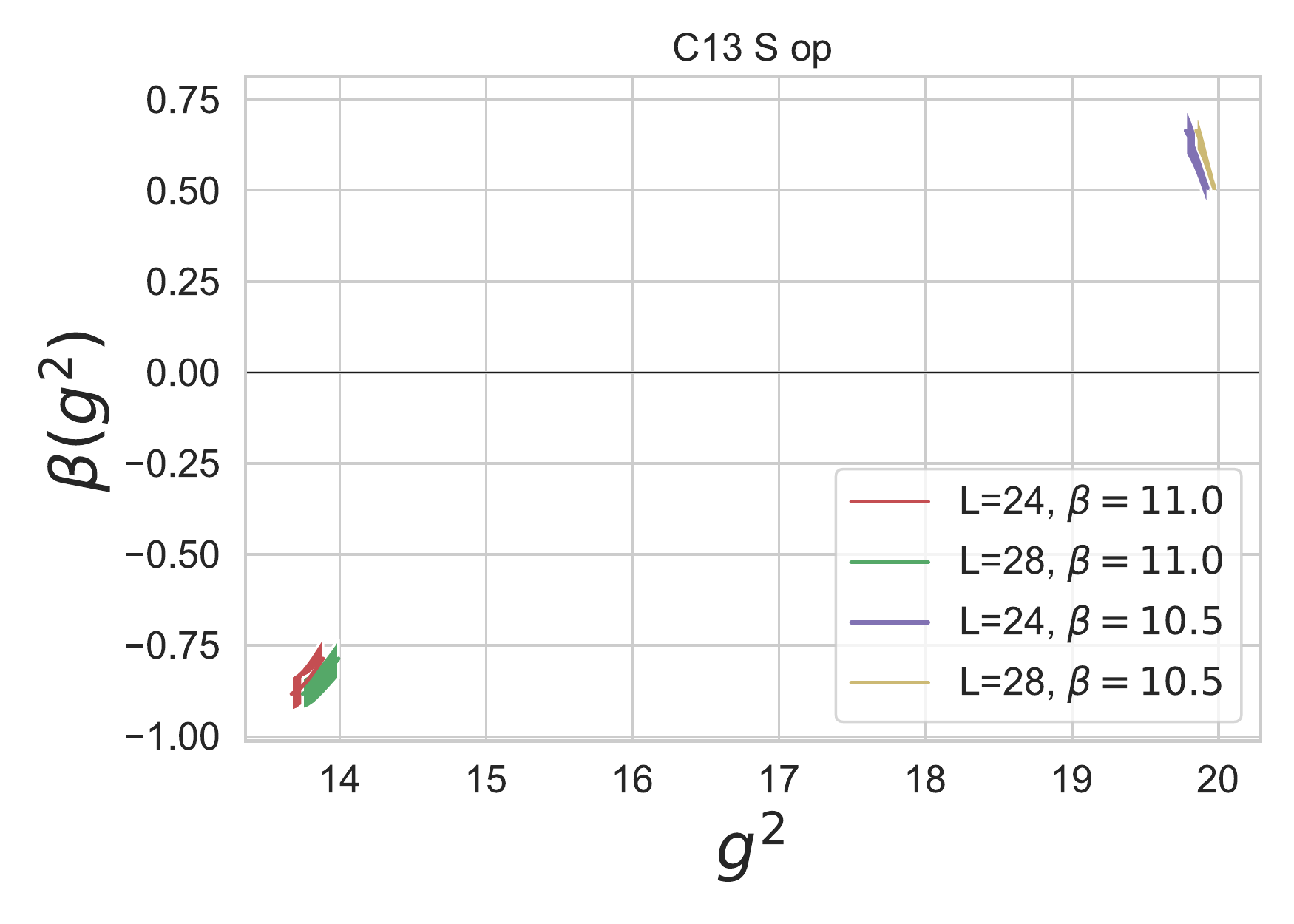}
\end{center}
\begin{quotation}
\caption{\label{fig:vol} Volume comparison for the S operator measured in C13 flow in the $\beta=11.0$
and~10.5 ensembles.  The volumes are $24^3\times48$ and $28^3\times56$.
Top: Gradient flow coupling $\ggf$ as a function of the lattice flow time $t/a^2$.
Bottom: $\beta(g^2)$ \vs~$\ggf$ at flow times $2.4\leq t/a^2\leq3.2$.
}
\end{quotation}
\vspace*{-4ex}
\end{figure}

In \Fig{fig:vol} we examine the effect of increasing the lattice volume.
We have generated two $28^3\times56$ ensembles---at $\b=11$ which is
a weak (bare) coupling and at $\b=10.5$ which is a strong coupling.
In the top panel, we show the flows at these two bare couplings
for the two volumes, again using C13 flow and the S operator.
The change in $\ggf$ and in its derivative
due to changing the lattice volume is evidently small.
The bottom panel shows the effect on the raw beta function.
For both bare couplings, an increase in $L/a$ from~24 to~28 has the effect
of shifting $\ggf$ upward by about~0.1, leaving the $\beta$ function
otherwise unchanged.  We expect the extrapolation to infinite volume
to be linear in $(a/L)^4$.  It can be easily checked that
the increase $L/a=24 \to 28$ takes $(a/L)^4$ halfway to the limit.
Hence, the expected change in $\beta(\ggf)$ is a horizontal shift
by twice the amount seen in \Fig{fig:vol}, with no qualitative change
in the overall shape, including the existence of a fixed point.  In the rest
of this section we concentrate on our eight $24^3\times48$ ensembles.

\subsection{\label{interpolation} Interpolation}

In order to determine the continuum limit of $\beta(\ggf)$
we will extrapolate $t/a^2 \to \infty$ at fixed $\ggf$.
In a theory with a rapidly running coupling the graph of the raw data
for $\beta(\ggf)$, Fig.~\ref{fig:Wil-raw} or~\ref{fig:C13-raw},
would yield several ensembles that give different values
for $\beta(\ggf)$ at any fixed coupling $\ggf$ \cite{Hasenfratz:2019hpg,Peterson:2021lvb}.
We would then read off the corresponding flow time $t/a^2$
and $\beta(\ggf)$ for each ensemble, and take $t/a^2 \to \infty$.

\begin{figure*}[t]
\begin{center}
\includegraphics*[width=0.45\textwidth]{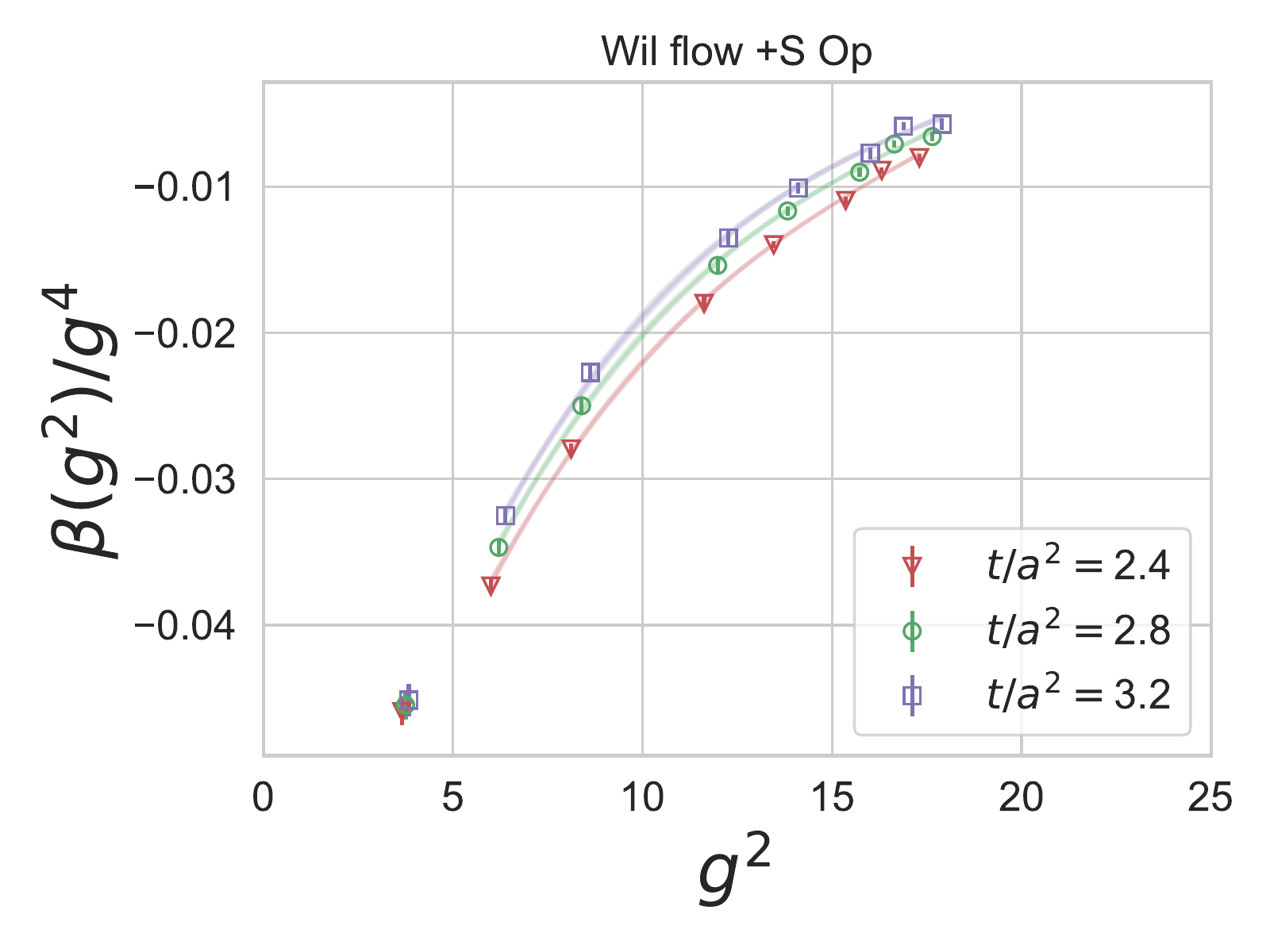}
\includegraphics*[width=0.45\textwidth]{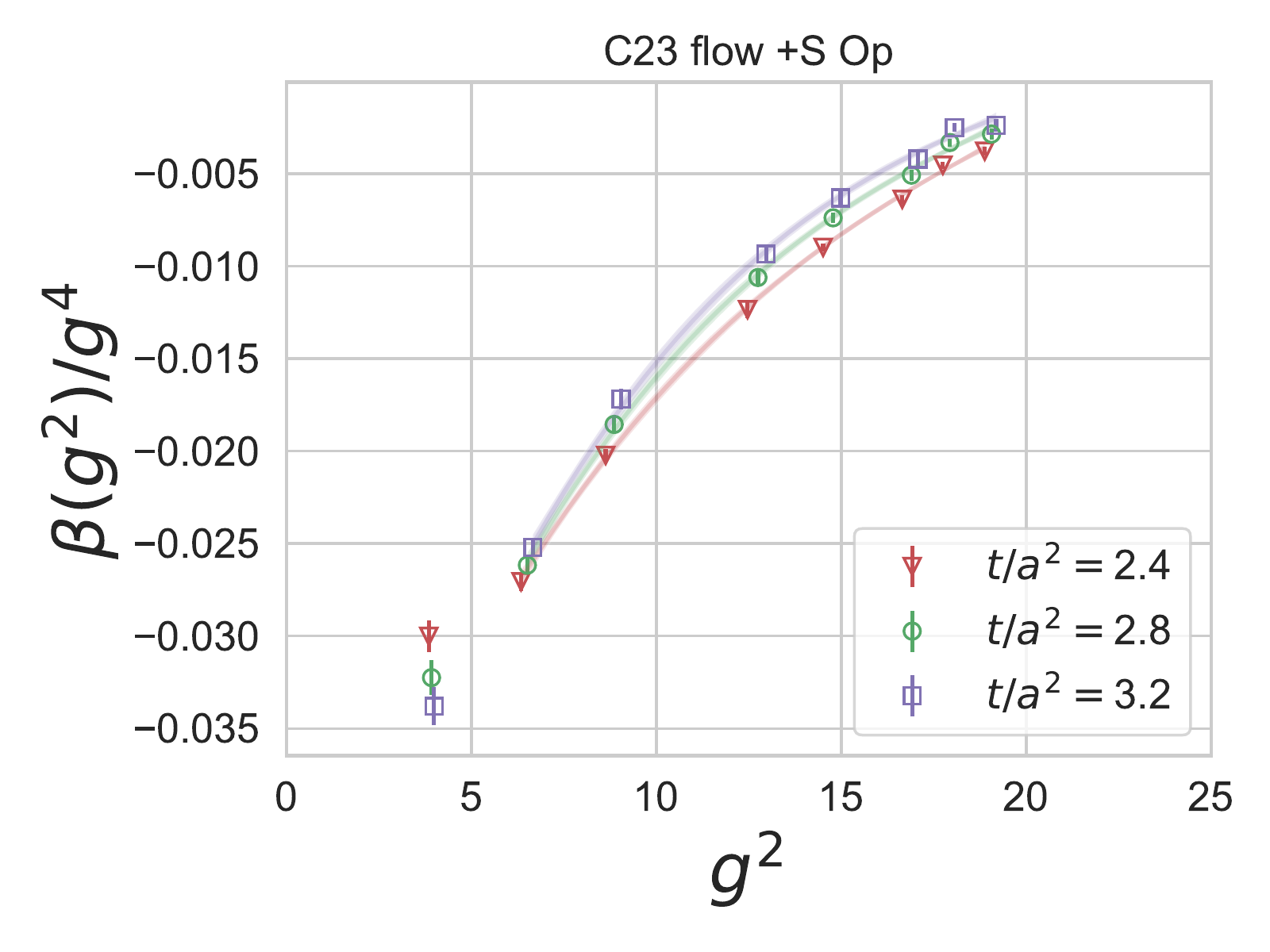}
\\
\includegraphics*[width=0.45\textwidth]{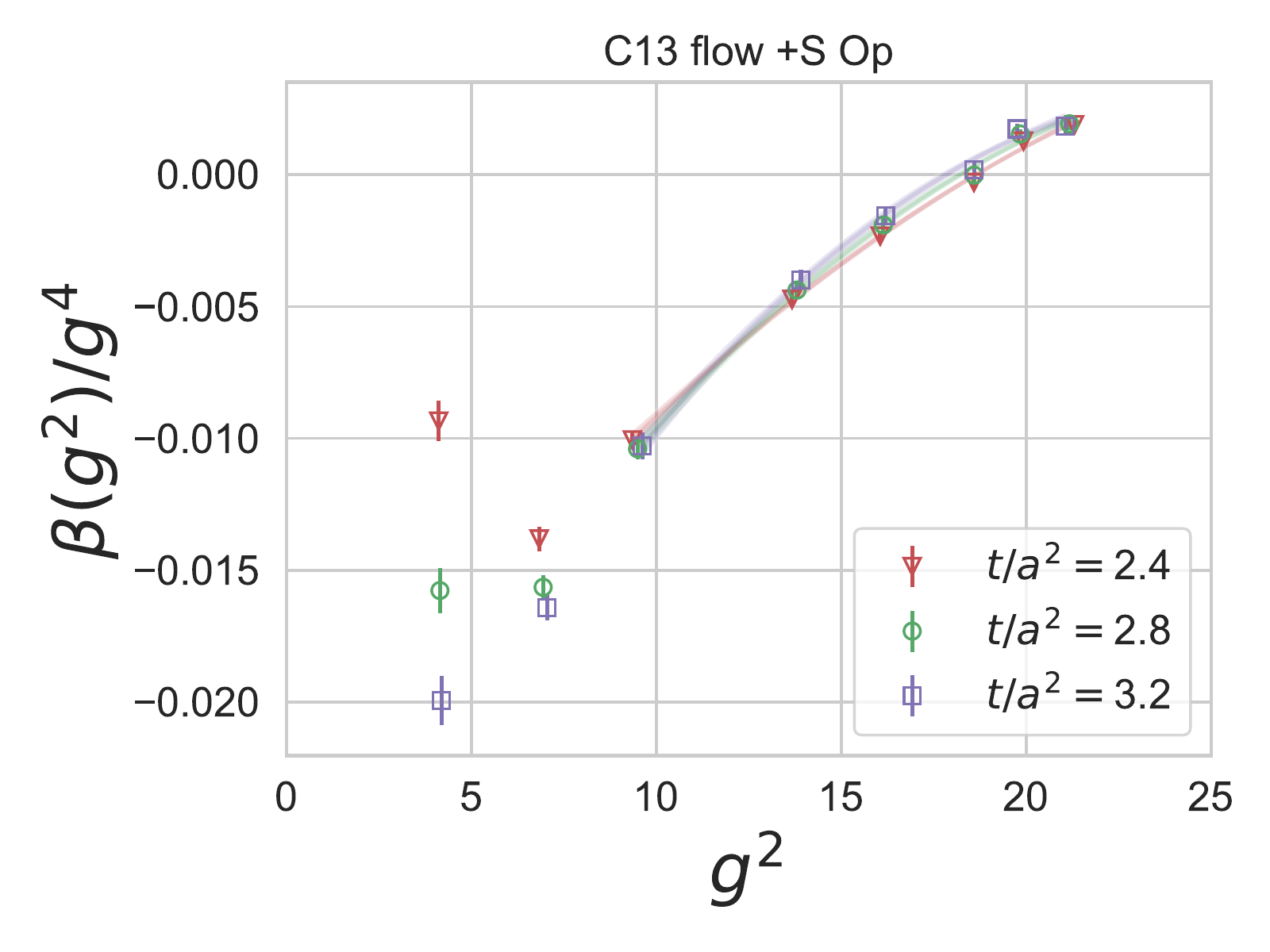}
\end{center}
\caption{\label{interpolation-W2448} Interpolation of $\beta(\ggf)/g^4$ \vs~$\ggf$ for selected values of $t/a^2$
in Wilson flow (top left), C23 flow (top right), and C13 flow (bottom).
The coupling $\ggf$ is defined by the S operator,
measured in the eight $24^3\times48$ ensembles.
In each plot, moving from left to right corresponds to
stronger GF {\em and\/} bare couplings: the leftmost points come
from $\beta=13$ and the rightmost from $\beta=10.4$.
The curves connect only the points that were included in the interpolation.
}
\end{figure*}

Since our theory runs slowly, each ensemble covers only a small range
of $\ggf$ and hence we have no overlaps between ensembles at any value
of the coupling.  Therefore, we have to interpolate $\beta(\ggf)$ \vs\ $\ggf$
at fixed flow time---see \Fig{interpolation-W2448}.
For a given flow time $t/a^2$, we identify the $(g^2,\beta(g^2))$ pair
on each ensemble.  This gives us eight data points on each fixed-$t$ curve.
To make vertical slices in $\beta(\ggf)$,
we interpolate between ensembles at each fixed $t$.
In weak coupling, $\beta(g^2)\propto g^4$,
and hence we use a polynomial interpolation even in strong coupling,
$\beta(g^2)/g^4=c_0+c_1 g^2 + c_2 g^4 + \cdots$, as detailed below.
The curves plotted in \Fig{interpolation-W2448} show the interpolations
for the Wilson, C23 and C13 flows, using the S operator for the coupling.

Before performing each interpolation, we have to decide if we can use
all eight data points.  As can be seen in the figure,
the C13 and C23 flows show rapid change with $t/a^2$ at the weakest bare couplings
(the leftmost data points).  We interpret this as indication that these flows
require larger flow times in the weak coupling region to reach
the vicinity of the RT.
Interpolations with cubic, or even quartic, polynomials have poor $p$-values
if we include all 8 data points. Dropping the left-most data point $\beta=13.0$
is sufficient to raise the $p$-value above 10\% if we use a cubic
interpolating polynomial for $\beta(g^2)/g^4$.
Even though Wilson flow shows smaller cutoff effects at $\beta=13.0$,
for consistency we include only the seven ensembles at $\beta <13.0$  and use
a cubic interpolation for both the Wilson and C23 flows.
Moreover, for the C13 flow, it can be seen in the bottom panel of \Fig{interpolation-W2448} that the
$\beta=12.0$ point also shows large cutoff effects for the Symanzik operator.
Hence, for this flow we discard $\beta=12$
from the interpolation as well, and use a quadratic interpolating polynomial
to keep the same number of degrees of freedom in the interpolation
as for the other flows.

The curves plotted in \Fig{interpolation-W2448} connect only the data points that were included in the fits.
As usual in interpolating data, in later stages of the analysis
we will not use the interpolating curves outside the range of the data points that
they connect.

The Symanzik and C43 flows (not shown in the figure) follow an opposite trend,
exhibiting increasing cutoff effects in strong coupling.
For those flows we drop the data point at the strongest bare coupling, $\beta=10.4$.
Again, we use a cubic interpolating polynomial.

\begin{figure}[t]
\begin{center}
\includegraphics*[width=\columnwidth]{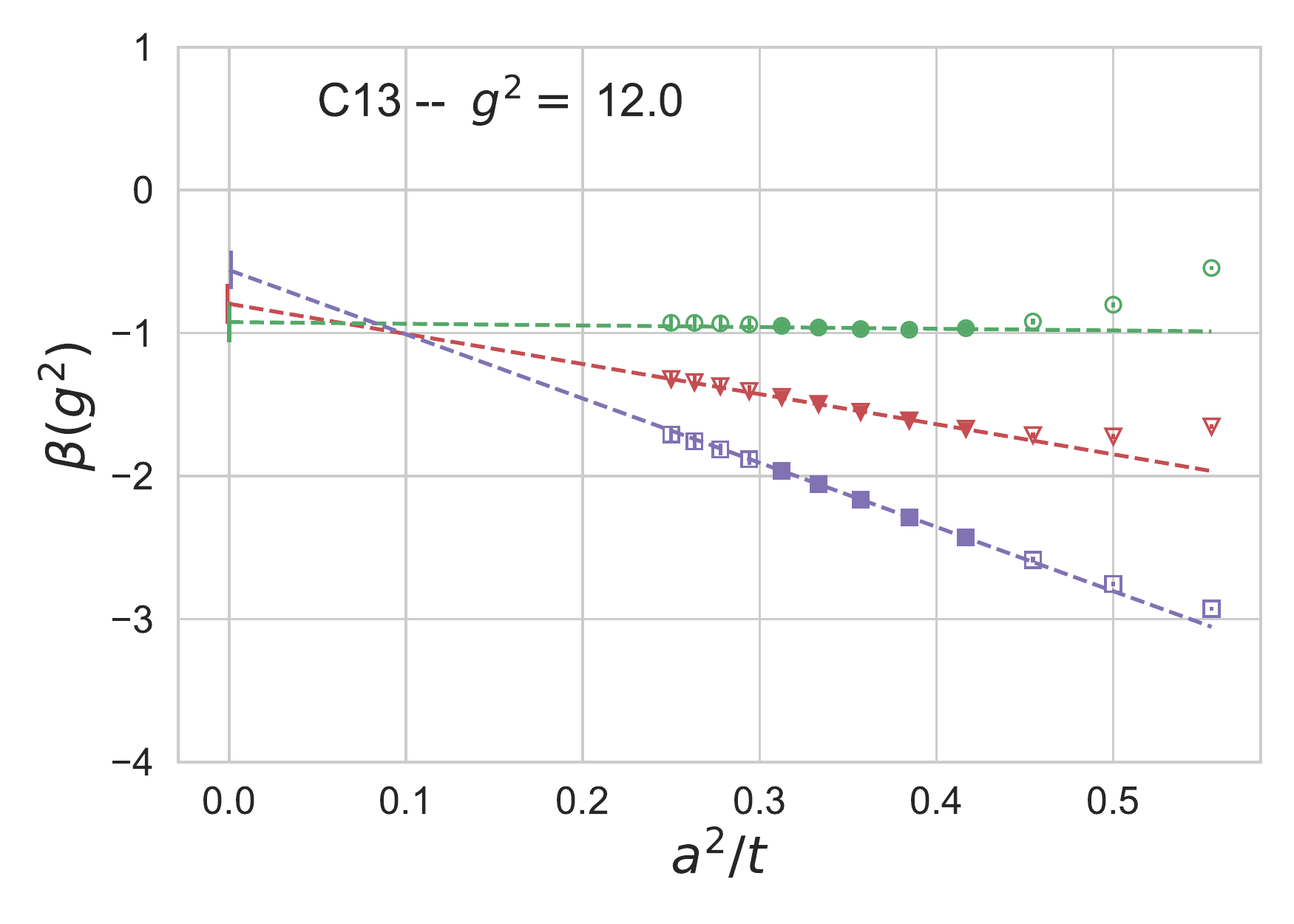}
\end{center}
\caption{\label{Cont-lim-tests-specimen} Extrapolations $a^2/t\to 0$ of $\beta(g^2)$ at $g^2=12.0$. Data are
(top to bottom) the S (green), W (red), and~C (purple) operators
measured in the C13 flow.  Data plotted with open symbols have been dropped
from the extrapolations.}
\end{figure}

\subsection{\label{contlim} Taking the continuum limit}
We now turn to the final stage of our analysis,
taking the continuum limit by extrapolating $t/a^2 \to \infty$.
Having constructed curves for $\beta(\ggf)$ for a selection of $t$ values,
we now consider the $\beta$ surface as a set of curves $\beta(a^2/t)$
for a selection of $\ggf$ values.  At any given physical coupling $\ggf$,
the beta functions extracted from our different discretizations (S, W, C)
must agree in the continuum limit; we will use this requirement
to impose cuts on the range of $g^2$ where each flow can be trusted.

We show an example of the extrapolation process
in \Fig{Cont-lim-tests-specimen}, which shows the extrapolation
of $\beta(\ggf)$ at $\ggf=12.0$ for the C13 flow,
for all three operators (S, W, C).  The difference between
the beta functions at finite lattice flow time $t/a^2$ obtained for any two operators should approach zero as
$(a^2/t)^\zeta$, where $\zeta$ is the scaling exponent of the
leading irrelevant operator.
At weak coupling one expects $\zeta=1+O(g^2)$, but in general the dependence on $g^2$ is not known.\footnote{A.~Hasenfratz and C.~T.~Peterson, in progress.
}
Our data do not allow us to resolve any statistically significant
deviations from $\zeta=1$, and so we will assume that all cutoff effects
scale linearly with $a^2/t$.

In \Fig{Cont-lim-tests-specimen} the data points plotted with open symbols
are not included in the extrapolation.  The smallest acceptable
flow time depends on $\ggf$---as well as, more generally,
on the flow---reflecting how close we are to the RT\@.
We also limit the maximum flow time to suppress finite volume effects.
The remaining data give adequate linear extrapolations.
We denote the extrapolated values by $\betaS,\betaW,\betaC$.

We determine the error of each extrapolation with a bootstrap analysis.
The closeness of the three extrapolated values indicates that
the three operators give consistent results in this case.
We will return to the precise condition for consistency shortly.

\begin{figure*}[t]
\begin{center}
\includegraphics*[width=0.31\textwidth]{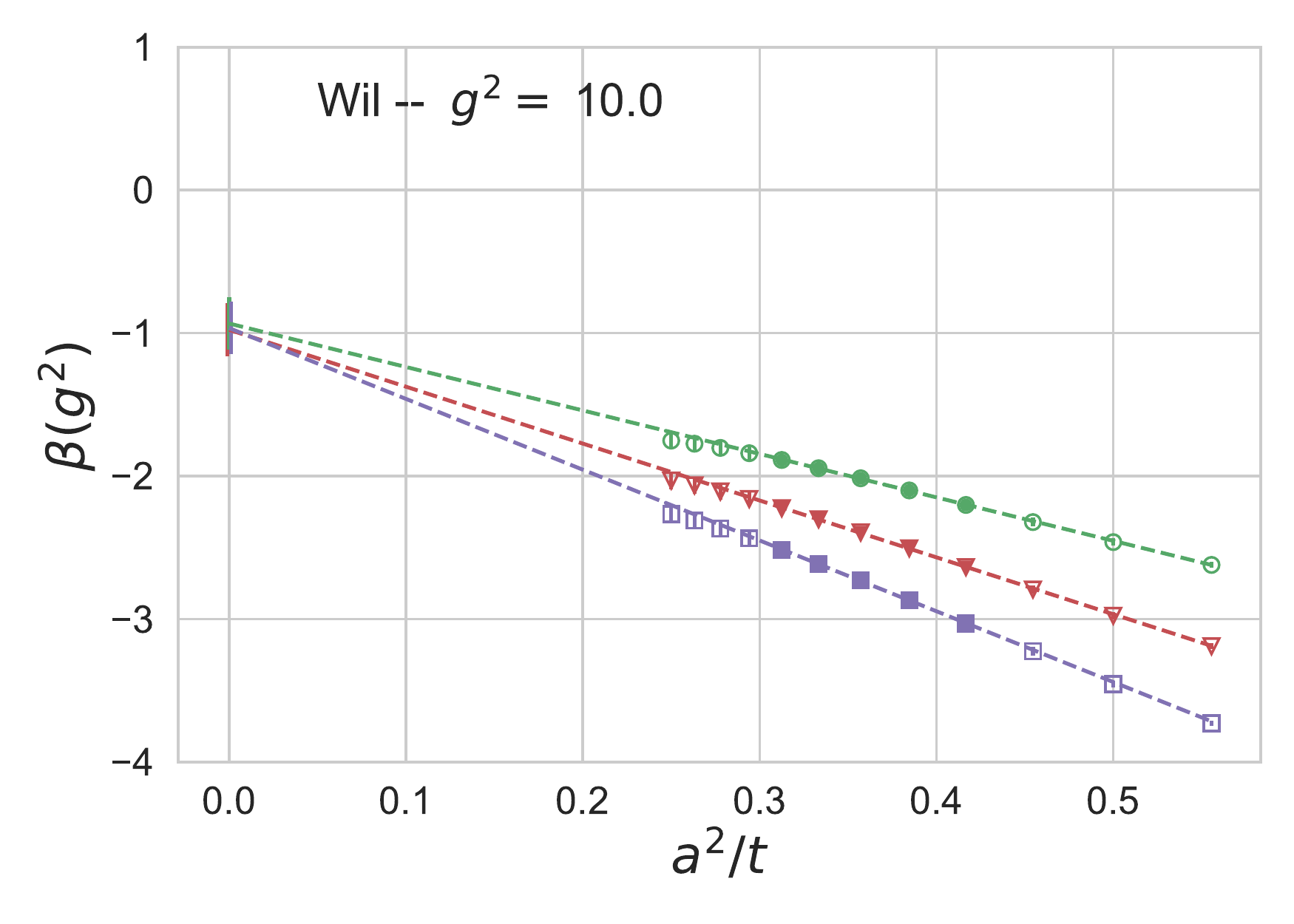}
\includegraphics*[width=0.31\textwidth]{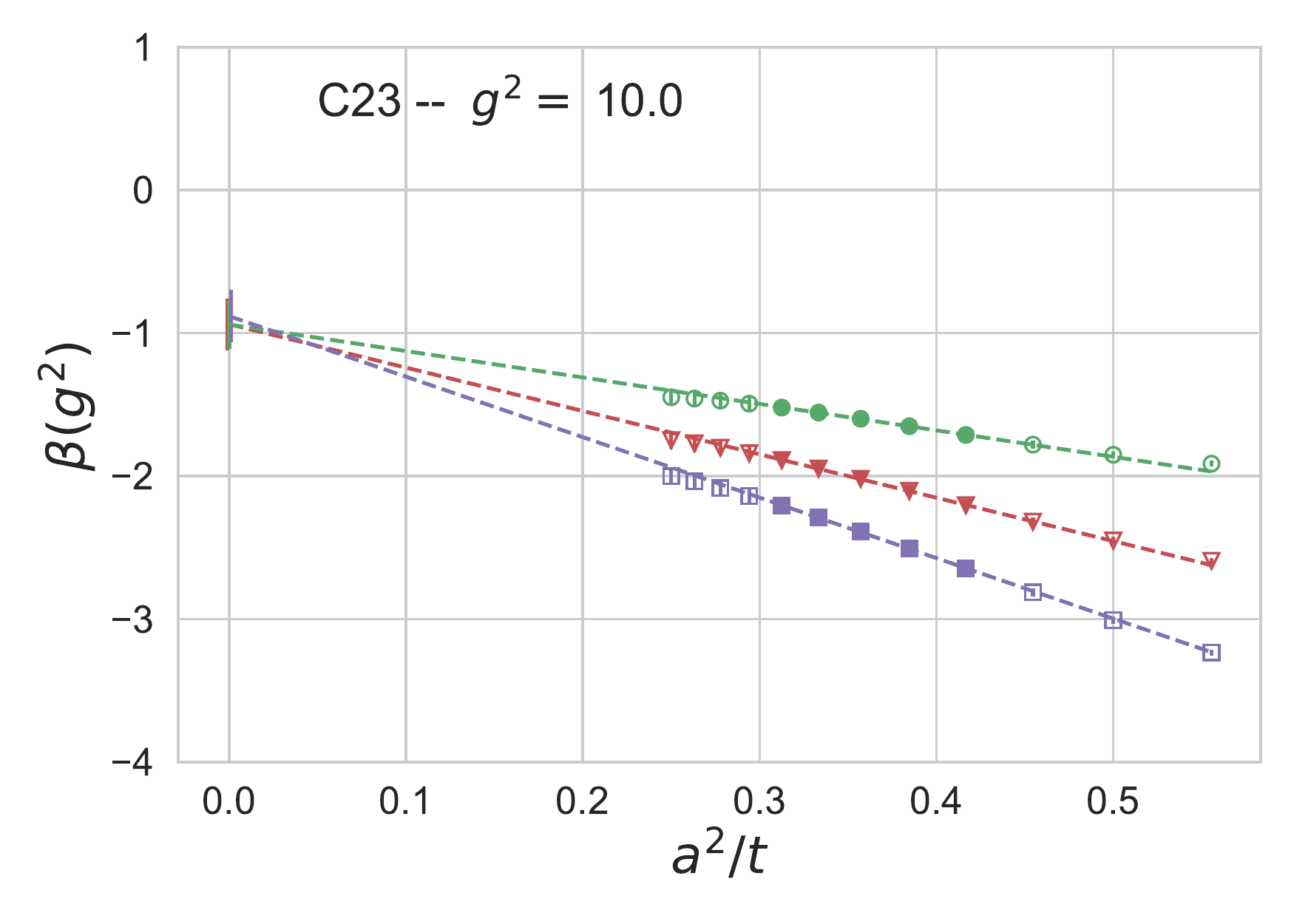}
\includegraphics*[width=0.31\textwidth]{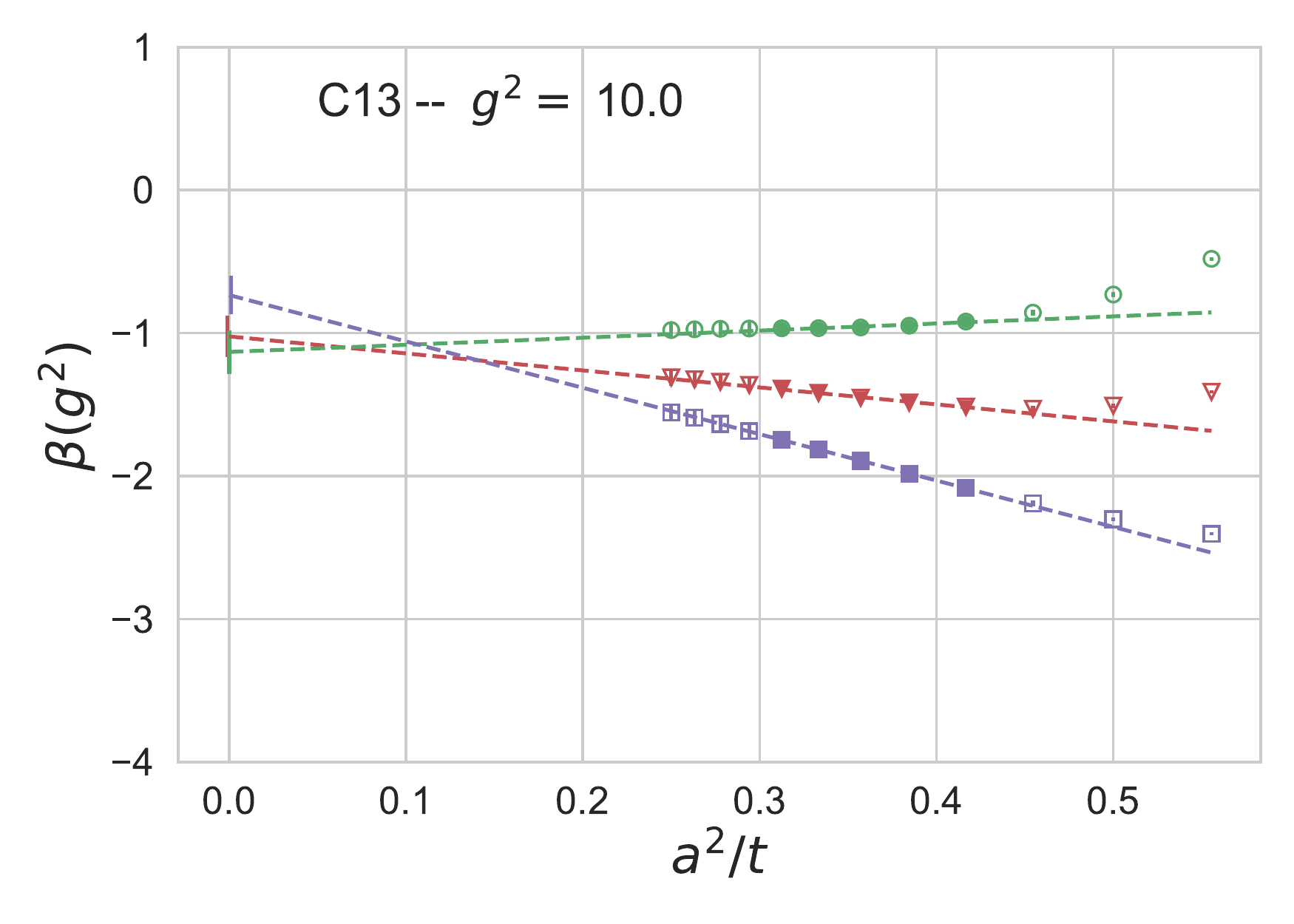}\\[.5ex]
\includegraphics*[width=0.31\textwidth]{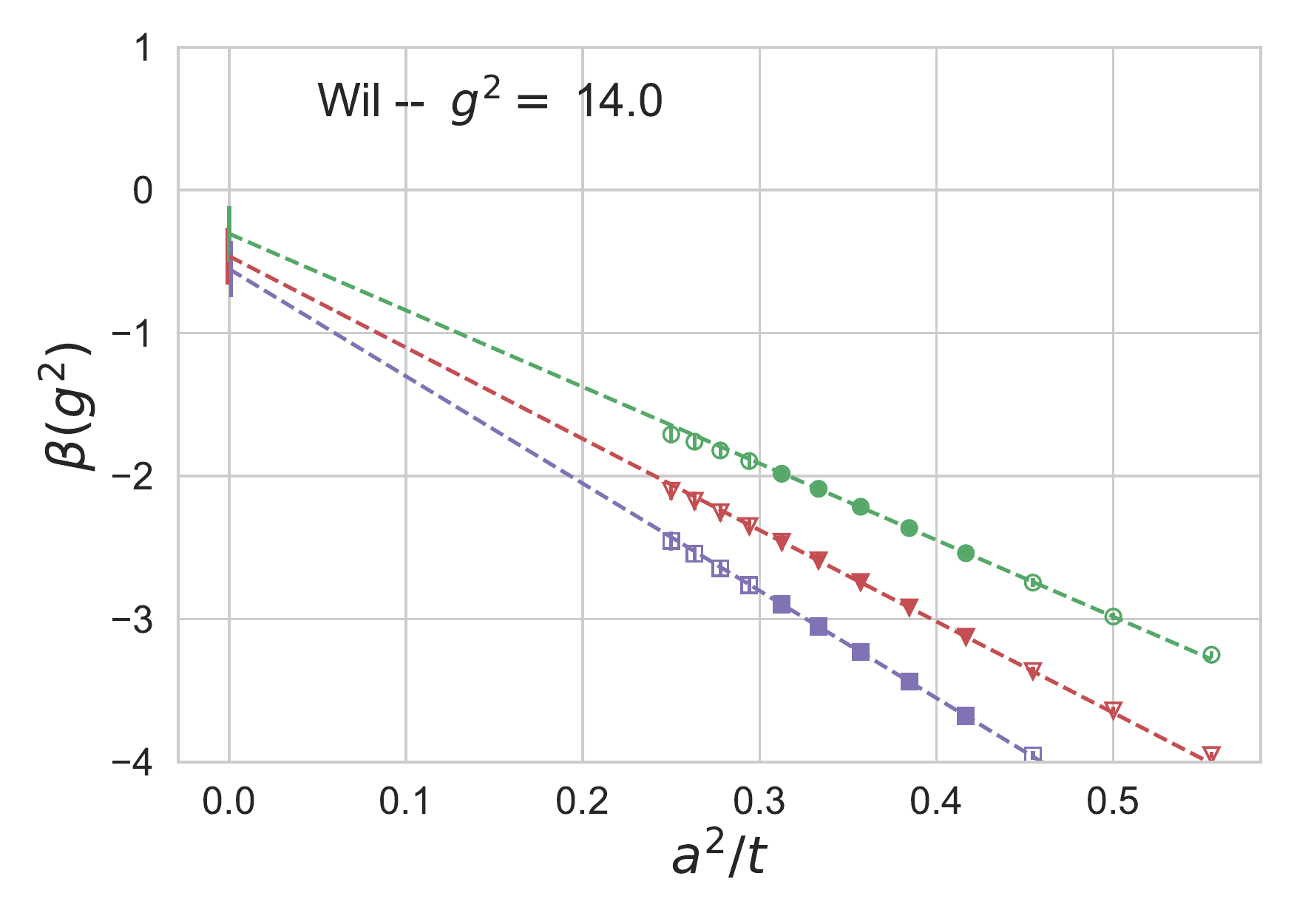}
\includegraphics*[width=0.31\textwidth]{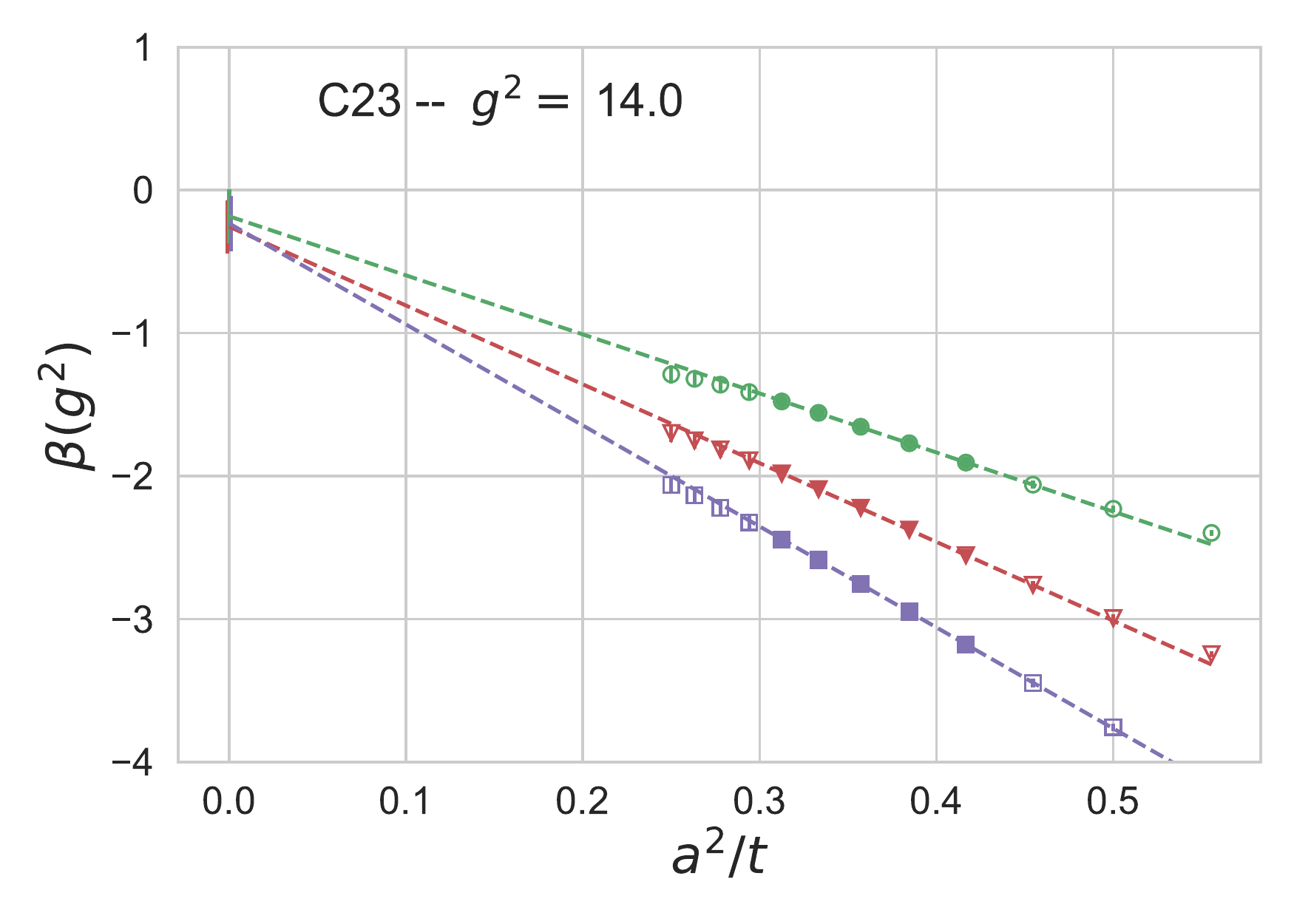}
\includegraphics*[width=0.31\textwidth]{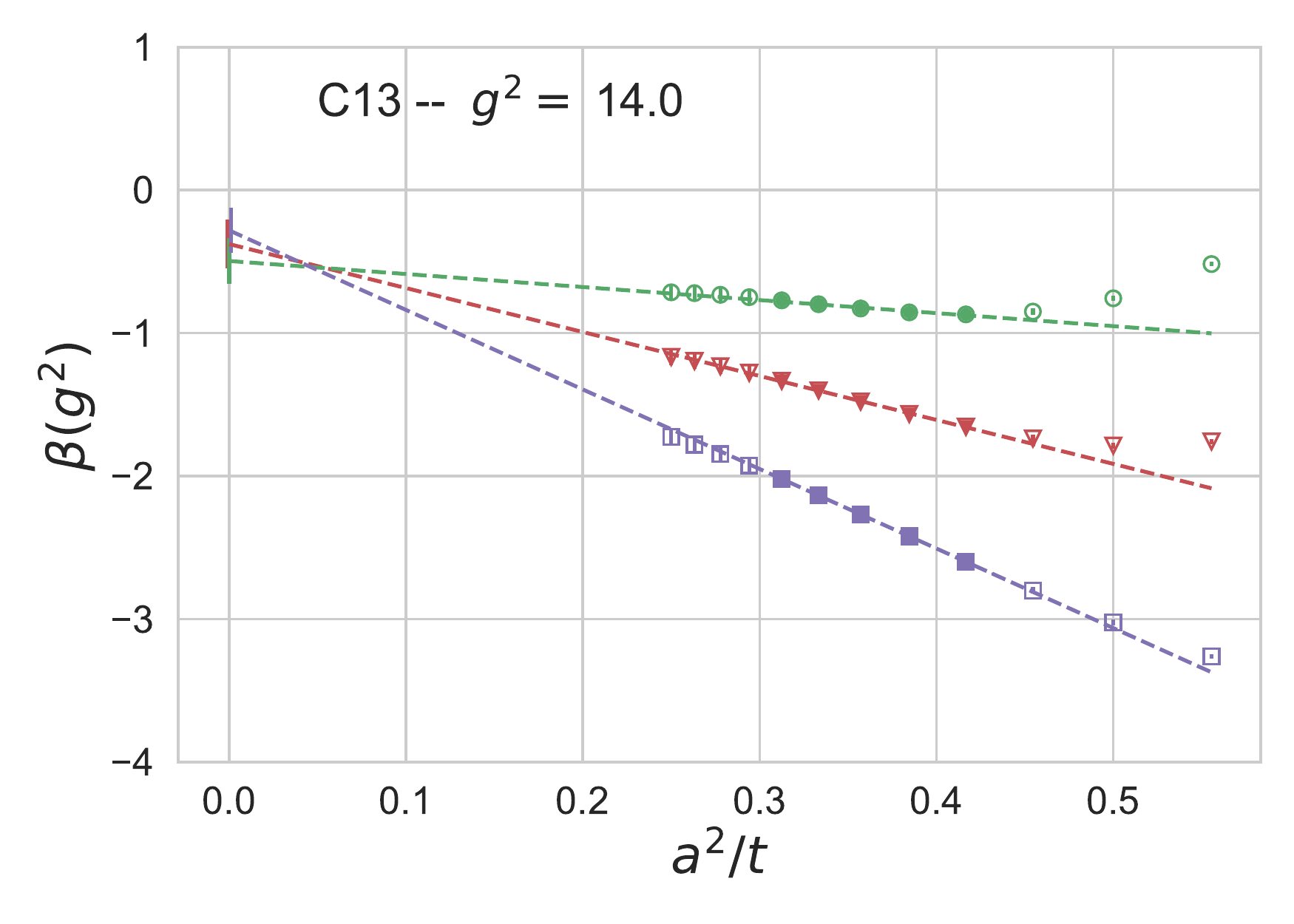}\\[.5ex]
\includegraphics*[width=0.31\textwidth]{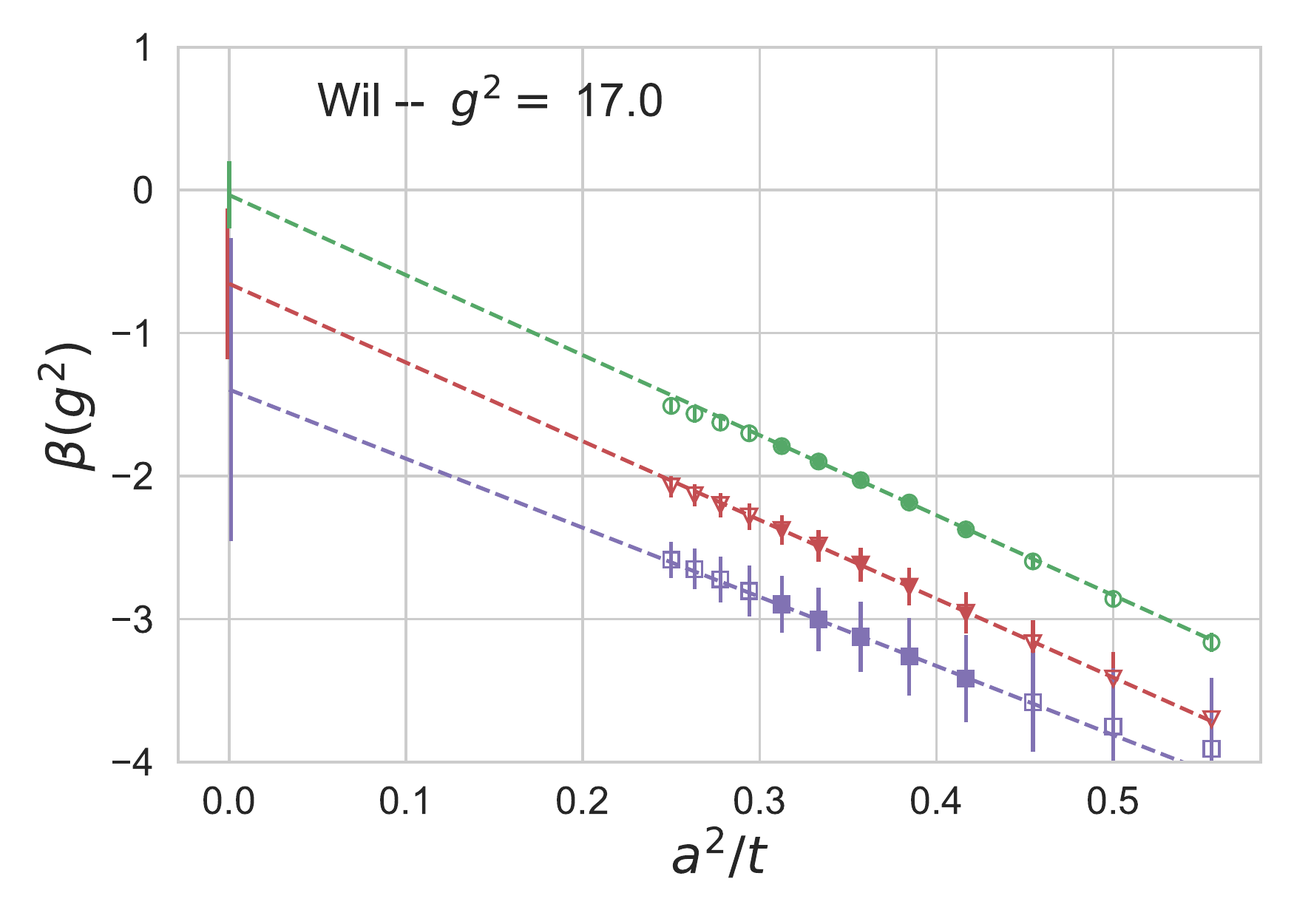}
\includegraphics*[width=0.31\textwidth]{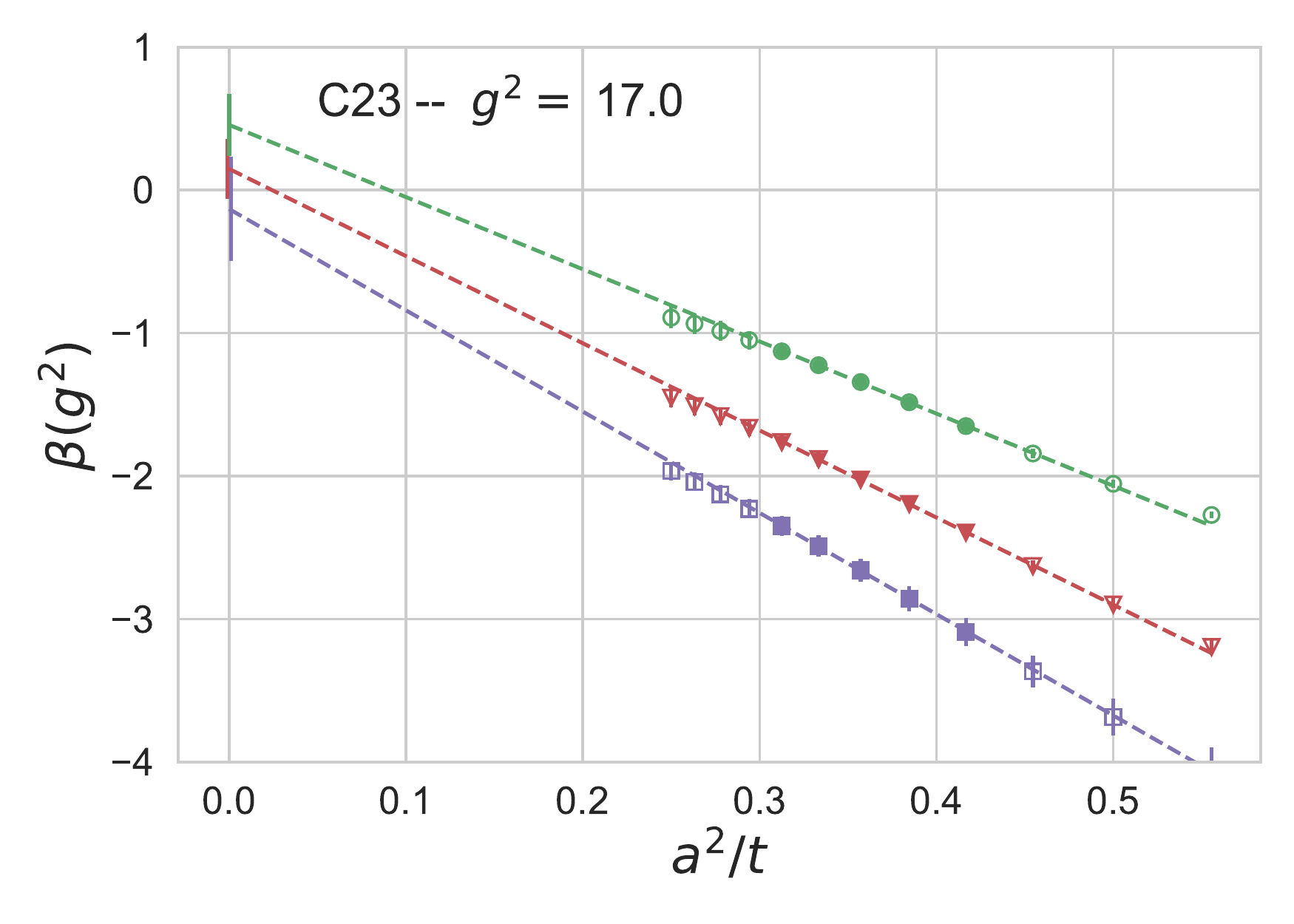}
\includegraphics*[width=0.31\textwidth]{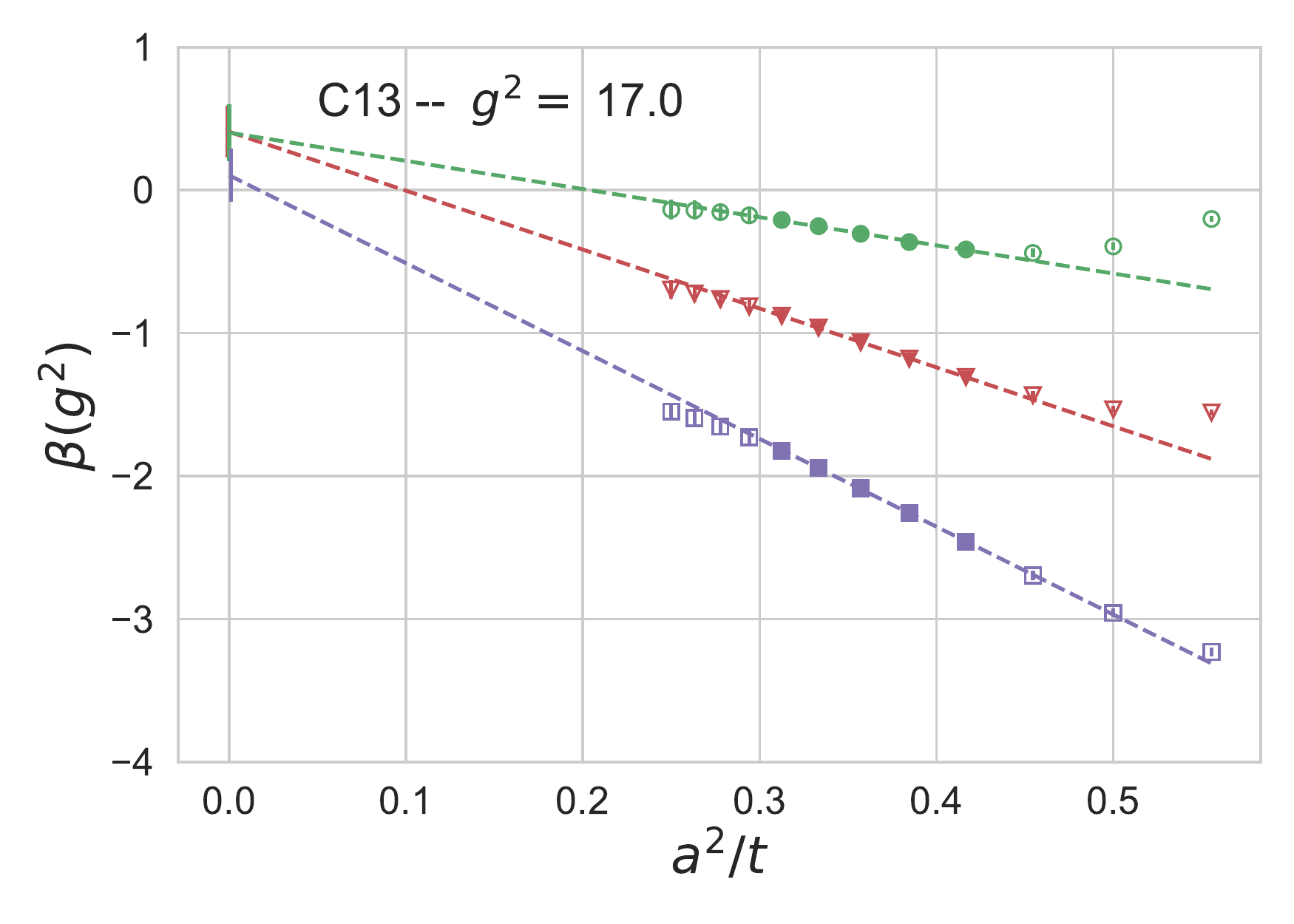}
\vskip -6ex
\end{center}
\caption{\label{Cont-lim-tests-long} Extrapolations $a^2/t\to 0$ of $\b(\ggf)$ for $\ggf=10.0$ (top), 14.0 (middle),
and~17.0 (bottom), which sample weak, intermediate, and strong couplings.
At each value of $\ggf$ we plot extrapolations for data
from Wilson flow (left), C23 flow (center), and C13 flow (right).
All data points come from flow times that belong to the interval $2.4\leq t/a^2\leq3.2$.
Color codes and symbols are the same as in \Fig{Cont-lim-tests-specimen}.
}
\end{figure*}

In \Fig{Cont-lim-tests-long} we show a compilation of continuum extrapolations.
The three rows of the figure show extrapolations at $\ggf=10.0,$ 14.0, and~17.0,
while the three columns correspond, from left to right, to Wilson, C23,
and C13 flows.  Each panel depicts the extrapolation process for all
three operators.  A general feature revealed in the figure is that the slope
of the extrapolation is always smallest for the S operator, with the
W operator coming next.  The C operator is last, having the largest slopes.
This signals that discretization effects are smallest for the S operator.
Combined with the other advantages of the S operator we have already discussed in \Sec{FV}, this naturally leads to the choice of the S operator
for our main result.

It is evident in \Fig{Cont-lim-tests-long} that there are cases where the
extrapolated values of the three operators are consistent with each other,
whereas for other cases they are not (for example, the panel in lower left). While these plots
give a general idea of the agreement of the extrapolations, they cannot
directly be used to determine the consistency of the extrapolations.
The reason is that data for the three operators are highly correlated.
In order to correctly assess the level of agreement between any
two extrapolations, their correlations must be taken into account.

\begin{figure*}[t]
\begin{center}
\includegraphics*[width=0.45\textwidth]{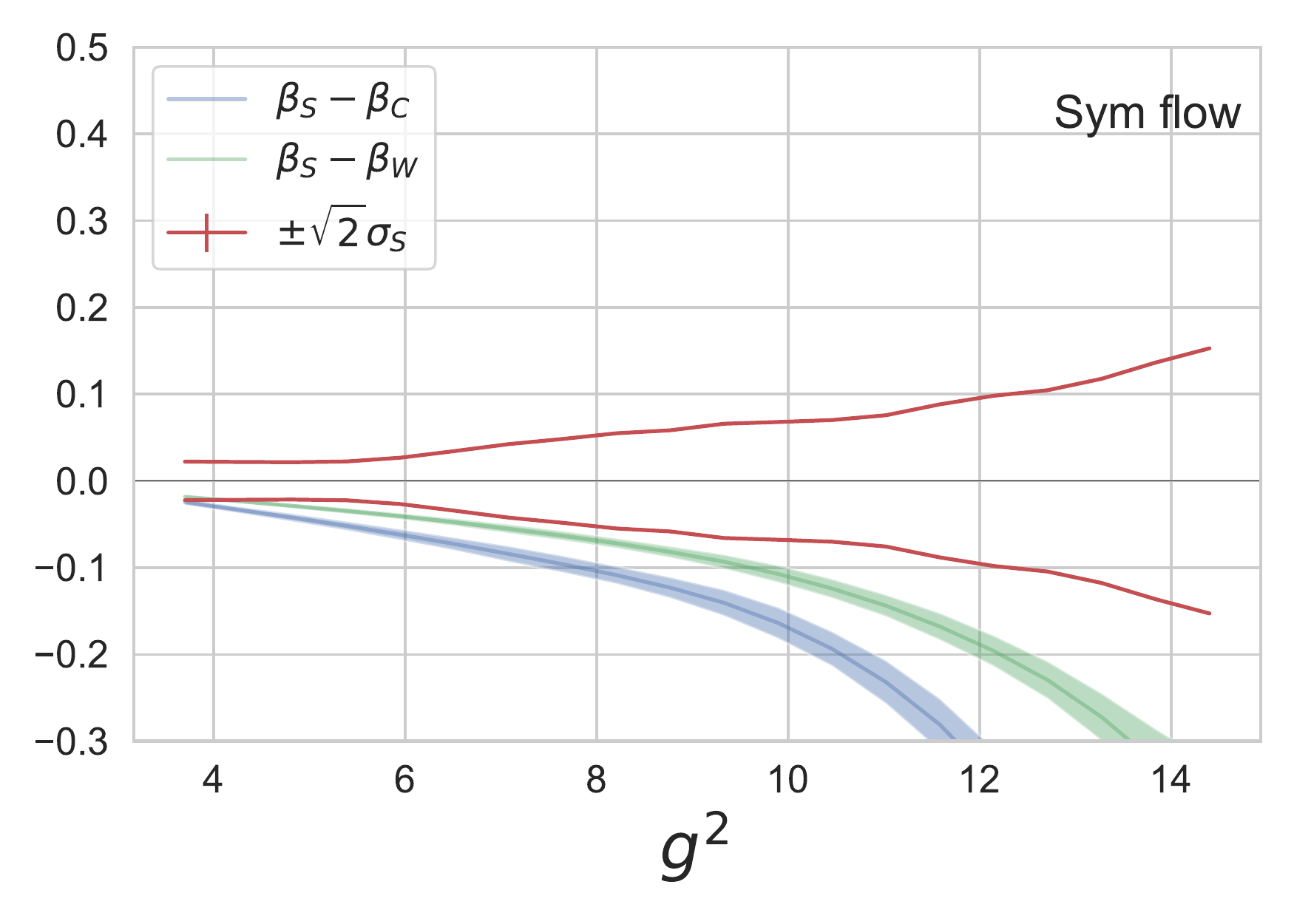}\\[.5ex]
\includegraphics*[width=0.45\textwidth]{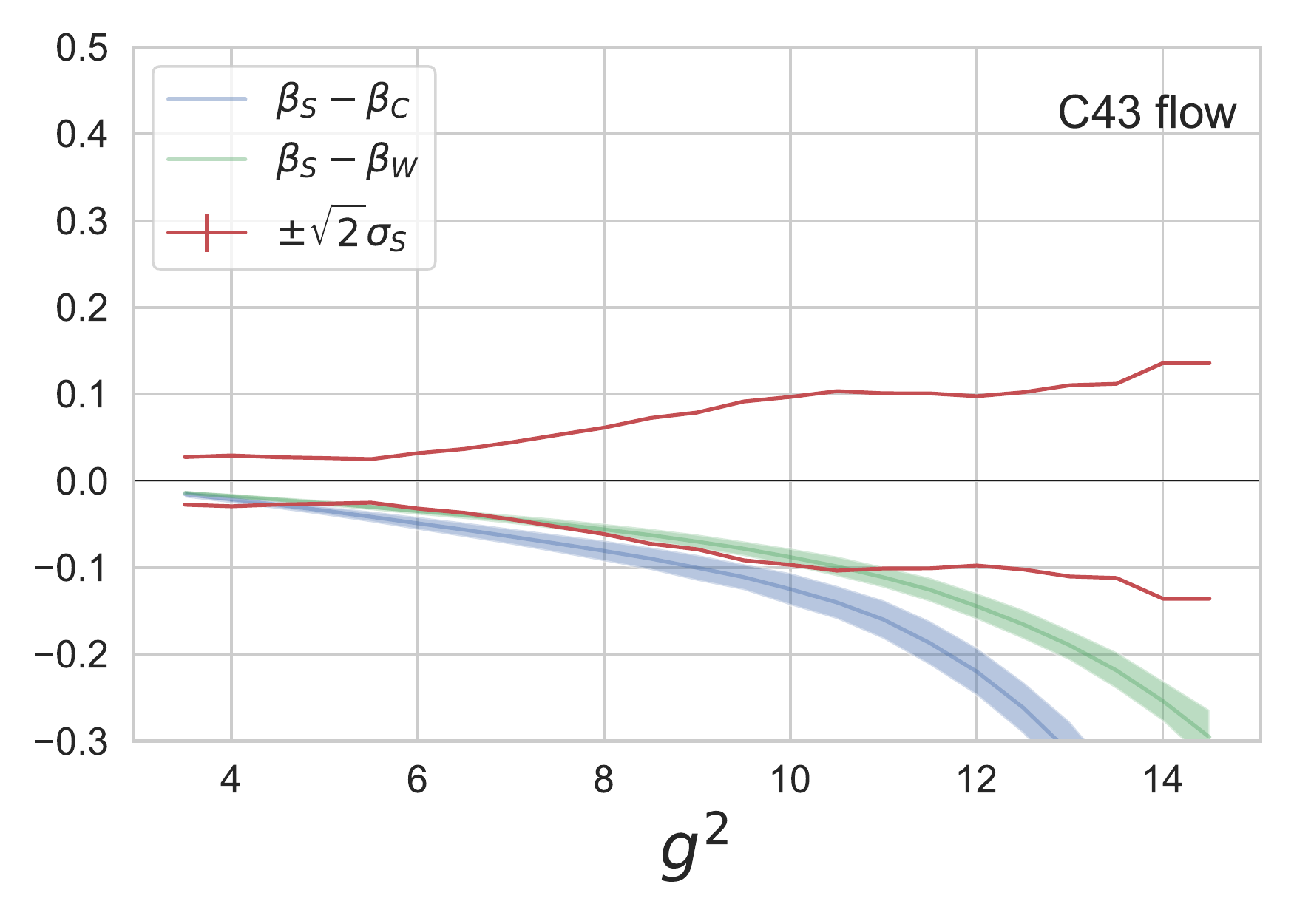}
\includegraphics*[width=0.45\textwidth]{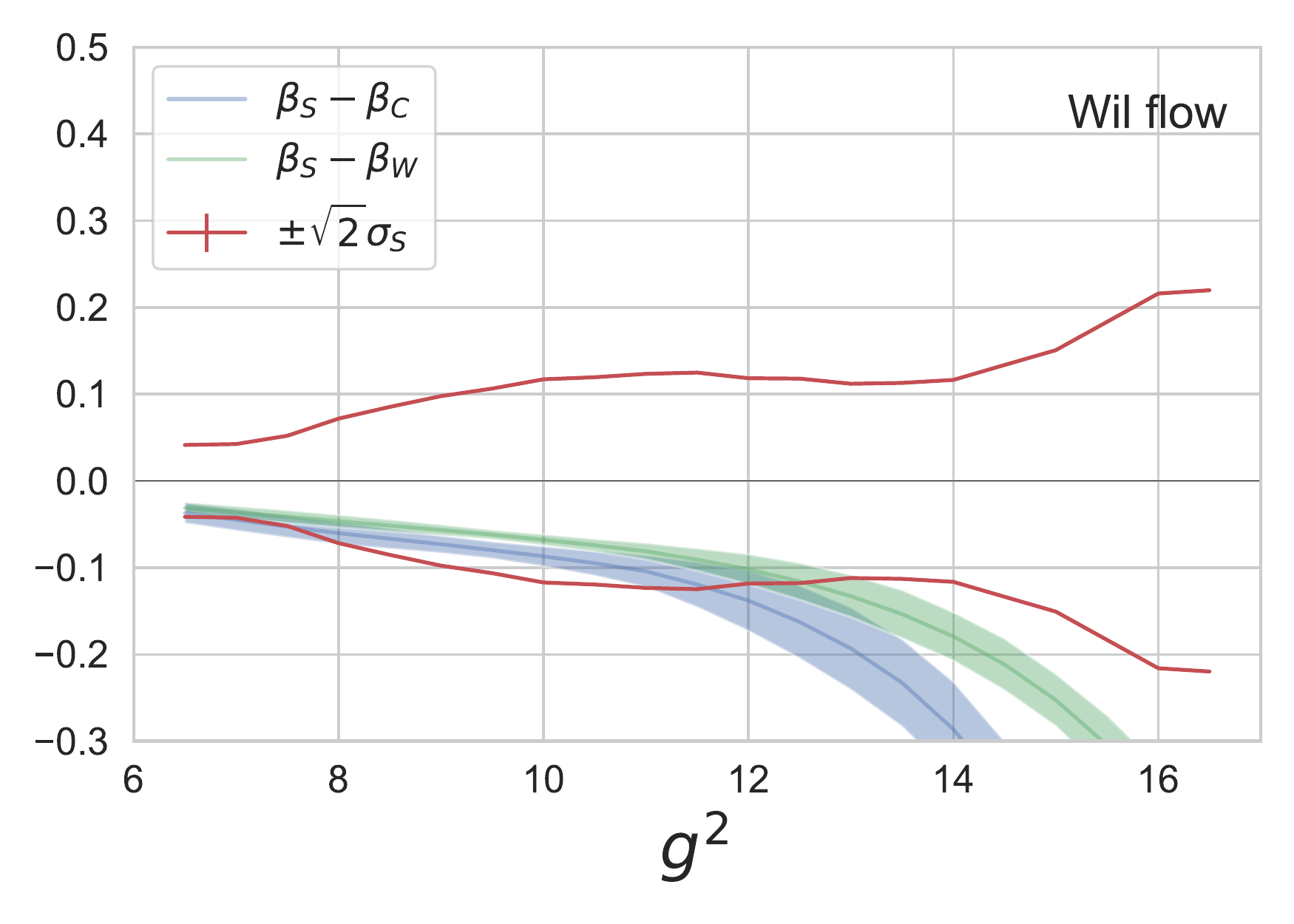}\\[.5ex]
\includegraphics*[width=0.45\textwidth]{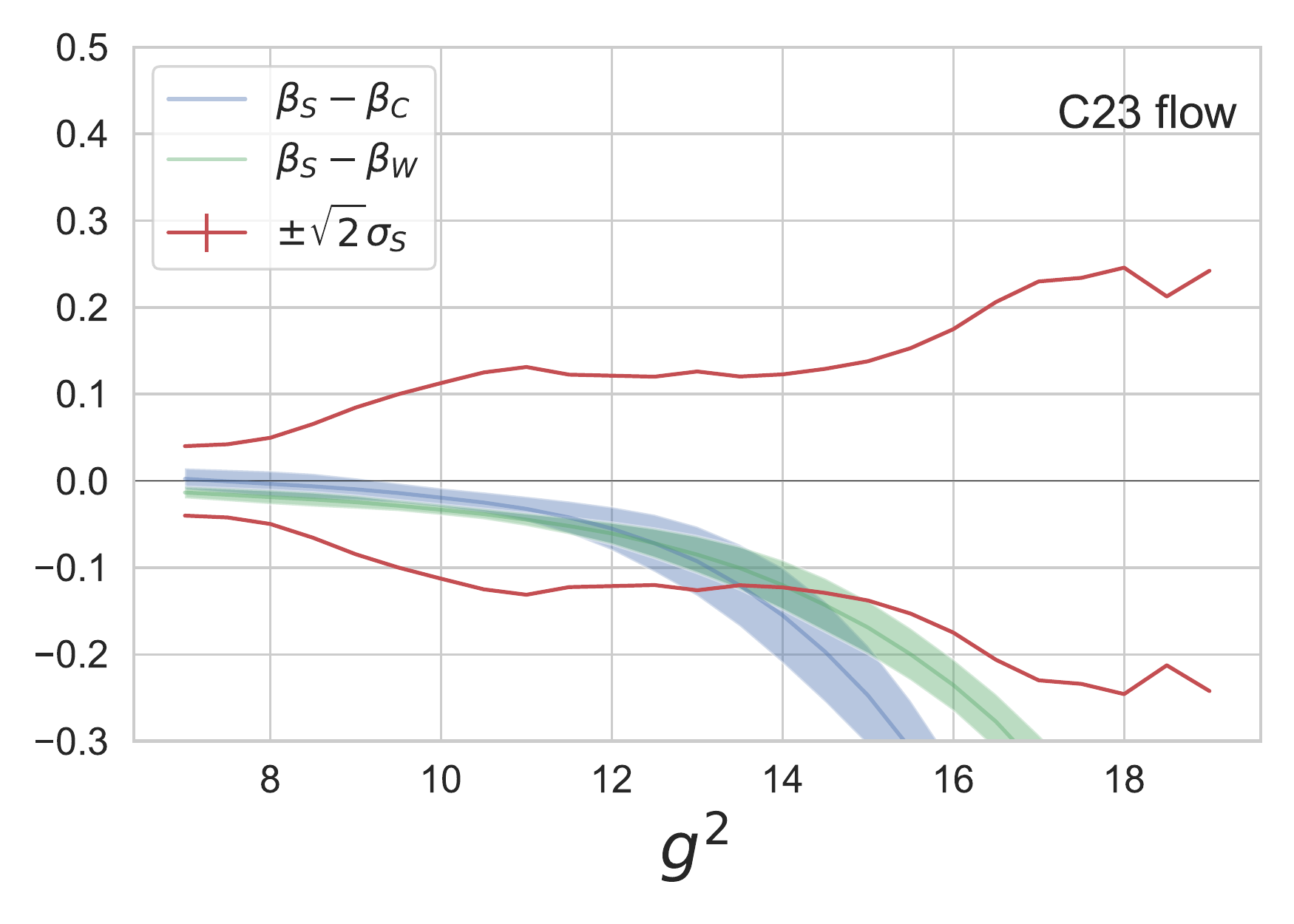}
\includegraphics*[width=0.45\textwidth]{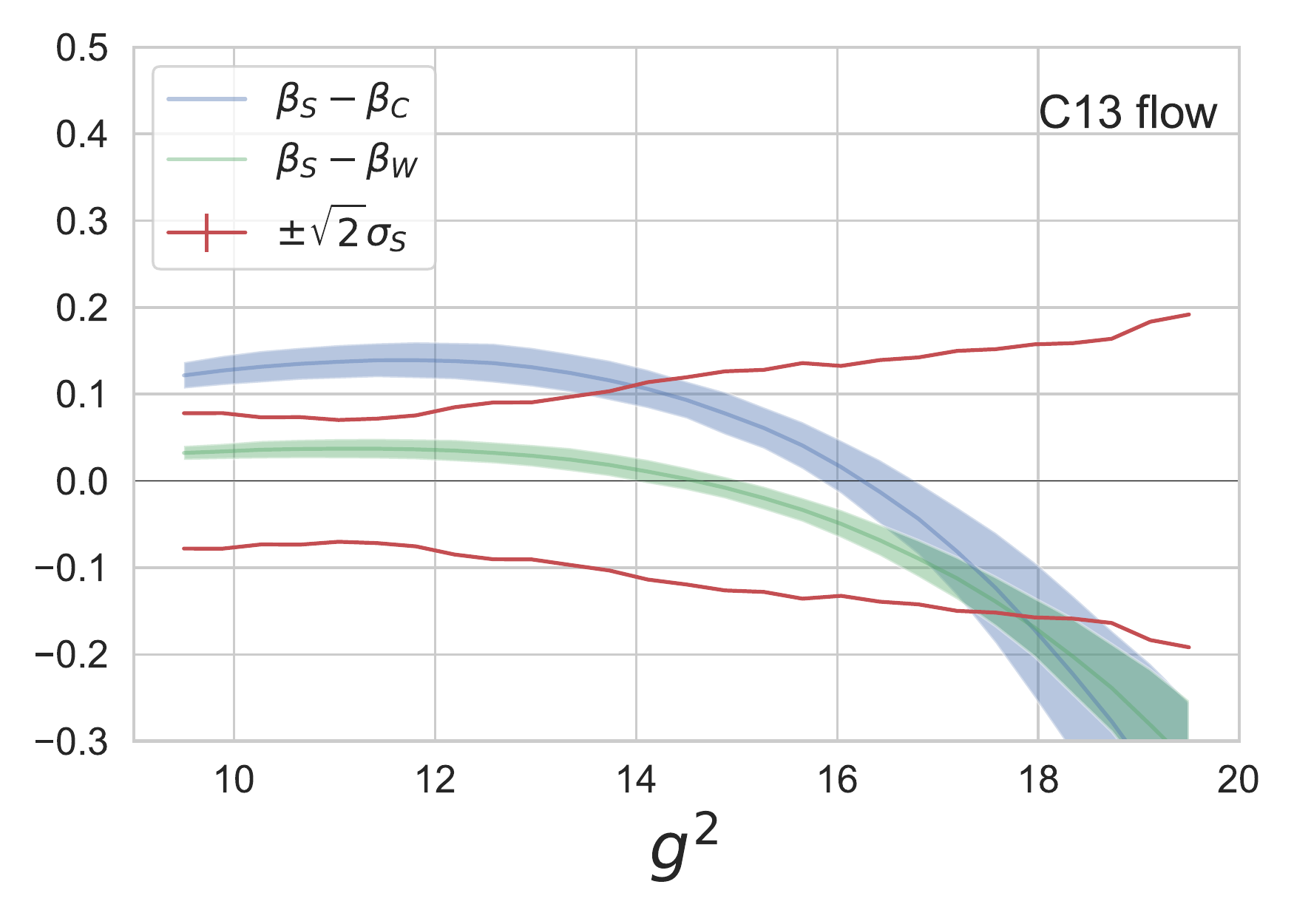}\\
\end{center}
\begin{quotation}
\caption{\label{fig:consistency} Comparing the differences $\betaS-\betaW$ (green) and $\betaS-\betaC$ (blue)
with the standard deviation $\sigmaS$ of the S operator,
at all physical couplings $\ggf$ for each of the five flows.
We require that the differences lie within
the curves $\pm\sqrt2\sigmaS$ (red).  The bounds of validity
$g^2_\text{min2}$ and $g^2_\text{max2}$ in Tables~\ref{tab:cuts_noC}
and~\ref{tab:cuts} have been set accordingly.
}
\end{quotation}
\vspace*{-3ex}
\end{figure*}

We therefore base the criteria for consistency on the plots
in \Fig{fig:consistency}, which include the effects of correlations.
Since we will be basing our main result on the S operator,
we plot red curves in each frame
to represent $\pm\sqrt2\sigmaS$, where $\sigmaS$ is the error
in $\betaS$.  The two bands in each frame are
the $\pm1\sigma$ bands for the correlated differences $\betaS-\betaW$
and $\betaS-\betaC$, which we also determine in the bootstrap analysis.
We comment that these two bands are much narrower than the span between the red curves; this reflects the strong correlations
among the three operators.

The range of $g^2$ values where a given flow can be trusted
is determined by the severity of discretization effects.
As we have seen (e.g., in Figs.~\ref{Cont-lim-tests-specimen}
and~\ref{Cont-lim-tests-long}),
the S operator shows the smallest discretization effects,
with the W operator coming next.
Hence we will determine this range by requiring consistency
between $\betaS$ and $\betaW$.
\Tab{tab:cuts_noC} shows the range of $\ggf$ in which
each set of flow data satisfies two constraints:
(1) $g^2$ lies within the range of interpolation in \Fig{interpolation-W2448},
resulting in an interval $[g^2_\text{min1},g^2_\text{max1}]$;
(2) the operators S and~W give consistent values for the extrapolation
to $a^2/t=0$, resulting in the smaller interval
$[g^2_\text{min2},g^2_\text{max2}]$.
The precise condition for consistency
is that the mean value of $\betaS-\betaW$,
represented by the green curves in \Fig{fig:consistency},
lies between the $\pm\sqrt2\sigmaS$ curves.

\begin{figure}[th]
\begin{center}
\includegraphics*[width=\columnwidth]{betafn_Sym_C43-Wil-C23-C13_noC.pdf}\\[.5ex]
\includegraphics*[width=\columnwidth]{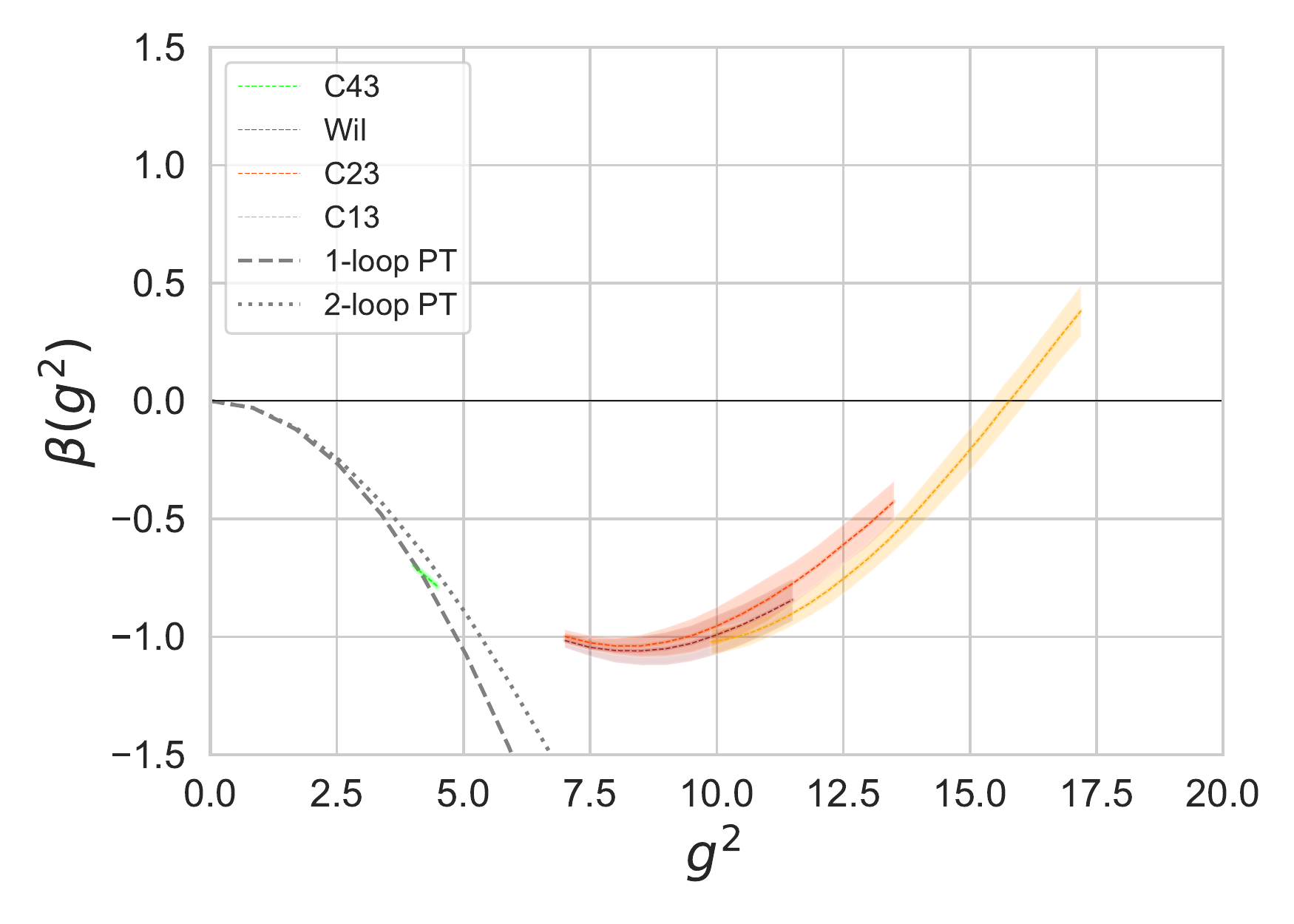}
\end{center}
\begin{quotation}
\caption{\label{betafn-matched2} The $\beta$ function obtained with five different GF transformations.
Both panels show the beta function derived from the S operator,
but the ranges in $\ggf$ allowed for each flow differ between them.
For the top panel (identical to \Fig{betafn-matched}),
only consistency between the S and W operators is required
(Table~\ref{tab:cuts_noC}). For the bottom panel,
all three operators are required to be consistent (Table~\ref{tab:cuts}).
In the bottom panel, there is a large gap between the $\ggf$ regions
of the C43 and Wilson flows---to bridge this gap a finer sequence of flows is needed.
Also, there is no region of validity for the Symanzik flow,
and thus it is not shown.
}
\end{quotation}
\vspace*{-3ex}
\end{figure}

The beta function resulting from this analysis is the final result
of this section.  It is displayed in \Fig{betafn-matched},
and replicated in the top panel of \Fig{betafn-matched2}.
There is a nice agreement between the predictions of the different flows
in their overlapping ranges.
On the weak-coupling side, our result connects smoothly to
perturbation theory.  At strong coupling, we find an IR fixed point at $g^2$
somewhere between 15.5 and~16.

For completeness, we also examine the consequences of requiring consistency
of all three operators.  In this case, both of the differences
$\betaS-\betaW$ and $\betaS-\betaC$
are required to lie between the $\pm\sqrt2\sigmaS$ curves, see \Tab{tab:cuts}.
This results in a smaller range of validity for each flow,
shown in the bottom panel of \Fig{betafn-matched2}\@.
It can be seen that now there is a gap between the coverage of the
C43 and Wilson flows; a set of flows with $c_p$ and $c_r$ chosen between
those of C43 and Wilson is needed to fill that gap.  We note that for C13 flow,
the differences $\betaS-\betaW$ and $\betaS-\betaC$
cross zero in the vicinity of the IR fixed point
(bottom right panel of \Fig{fig:consistency}).  In other words,
the differences among the beta functions extracted from the extrapolations
of the three operators are minimal near the fixed point.  This lends
further credibility to our determination of its existence and location.
\clearpage

\begin{table}[t]
\begin{ruledtabular}
\begin{tabular}{ccccc}
Flow     & $g^2_\text{min1}$ & $g^2_\text{max1}$ & $g^2_\text{min2}$ & $g^2_\text{max2}$
 \\ \hline
 Sym & 3.7 & 13.6 & 3.7 &  4.0 \\
 C43 & 3.7 & 14.4 & 3.7 & 11.0 \\
 Wil & 6.6 & 17.3 & 6.6 & 12.5 \\
 C23 & 6.8 & 18.9 & 6.8 & 14.0 \\
 C13 & 9.8 & 21.3 & 9.8 & 17.5 \\
\end{tabular}
\end{ruledtabular}
\caption{\label{tab:cuts_noC} Ranges of $\ggf$ in which each flow is included in the final result
for $\beta(\ggf)$.  $g^2_\text{min1}$ and $g^2_\text{max1}$ result
from limiting the ensembles included in the interpolations,
while $g^2_\text{min2}$ and $g^2_\text{max2}$ come from further demanding
consistency between the continuum extrapolations $\betaS$ and $\betaW$
(see \Fig{fig:consistency}).
Since the second requirement does not constrain the $g^2$ range any further on the weak-coupling side, the $g^2_\text{min1}$
and $g^2_\text{min2}$ columns are identical.
We quote $g^2_\text{max2}$ with a resolution of 0.5.
The last two columns are the ranges reflected in the upper panel of \Fig{betafn-matched2} (and in \Fig{betafn-matched}).
}
\end{table}

\begin{table}[ht]
\begin{ruledtabular}
\begin{tabular}{ccccc}
Flow     & $g^2_\text{min1}$ & $g^2_\text{max1}$ & $g^2_\text{min2}$ & $g^2_\text{max2}$
 \\ \hline
 Sym & 3.7 & 13.6 & (none) & (none) \\
 C43 & 3.7 & 14.4 &    3.7 &  4.5   \\
 Wil & 6.6 & 17.3 &    6.6 & 11.5   \\
 C23 & 6.8 & 18.9 &    6.8 & 13.5   \\
 C13 & 9.8 & 21.3 &   11.5 & 17.5   \\
\end{tabular}
\end{ruledtabular}
\caption{\label{tab:cuts} Same as Table~\ref{tab:cuts_noC}, but here the determination
of $g^2_\text{min2}$ and $g^2_\text{max2}$ is based on consistency
of $\betaS$, $\betaW$, and $\betaC$, with results reflected in the lower panel
of \Fig{betafn-matched2}.
}

\end{table}

\clearpage

\section{\label{sec:andim} Anomalous dimensions}

In this section we compute the mass anomalous dimensions for the fundamental
and sextet representations, as well as the anomalous dimensions of top-partner
chimera operators, as functions of the renormalized coupling $g^2$.
The anomalous dimensions in the infrared are then given by their values
at the fixed point  at $g^2 \simeq 15.5$.

The gradient flow equation for a fermion field
in the continuum is \cite{Luscher:2013cpa}
\begin{equation}
\label{flowf}
  \frac{\partial \c}{\partial t} = \D(t) \c \ ,
\end{equation}
where $\c$ is the flowed fermion field. $\D$ is the covariant laplacian,
constructed from the flowed gauge field at the same $t$.
Similarly to the gauge field, the initial condition is
\begin{equation}
\label{bcf}
\c(0) \big|_{t=0} =\j \ ,
\end{equation}
where $\j$ is the dynamical fermion field.

There is a technical issue in the application of the continuous RG method to operators made out of
fermion fields, described in Ref.~\cite{Carosso:2018bmz}.
Consider a two-point function of a flowed mesonic density $X'$ with an unflowed source $X$,
\begin{equation}
\label{meson}
\svev{X(0)\, X'(t)} \sim t^{-(d+\eta+\g)/2} \ .
\end{equation}
The scaling formula follows from the fact that here $X(0)$ is kept at $t=0$, while $X'(t)$ is constructed from
the flowed fermion fields at time $t$.  The exponent reflects
the classical dimension of the fermion bilinear, $d=3$;
the anomalous dimension of the elementary fermion field, $\eta/2$;
and the anomalous dimension of the meson operator, $\g$.
The reason for the appearance of $\eta$ is that the gradient flow,
unlike a blocking RG transformation, does not preserve
the normalization of the fermion kinetic term.

One way to eliminate $\eta$
is to divide $\svev{X(0)\, X'(t)}$ by the two-point function
of a conserved current, whose anomalous dimension vanishes.
We use the vector current, which scales according to (suppressing the vector index)
\begin{equation}
\label{vectorc}
\svev{V(0)\, V'(t)} \sim t^{-(d+\eta)/2} \ .
\end{equation}
Defining the ratio
\begin{equation}
\label{Rratio}
  R(t) = \frac{\svev{X(0)\, X'(t)}}{\svev{V(0)\, V'(t)}}\ ,
\end{equation}
we have
\begin{equation}
\label{R}
  R(t)  \sim t^{-\g/2} \ .
\end{equation}
Now $\g$ can be extracted from the logarithmic derivative,
\begin{equation}
\label{gammaX}
\g = -2\frac{t}{R}\,\frac{\partial R}{\partial t} \ .
\end{equation}
We choose the operators $X$ and $X'$ to lie in definite 3-volumes,
separated by euclidean time $x_4$.
We require $\sqrt{8t}\ll x_4$, meaning that the euclidean time separation
must be large compared to the smearing of the operators by the flow.
Once this condition is satisfied, the ratio $R(t)$
and the corresponding exponent $\gamma$ are expected to be independent of $x_4$.

\begin{figure}[t]
\begin{center}
\includegraphics*[width=\columnwidth]{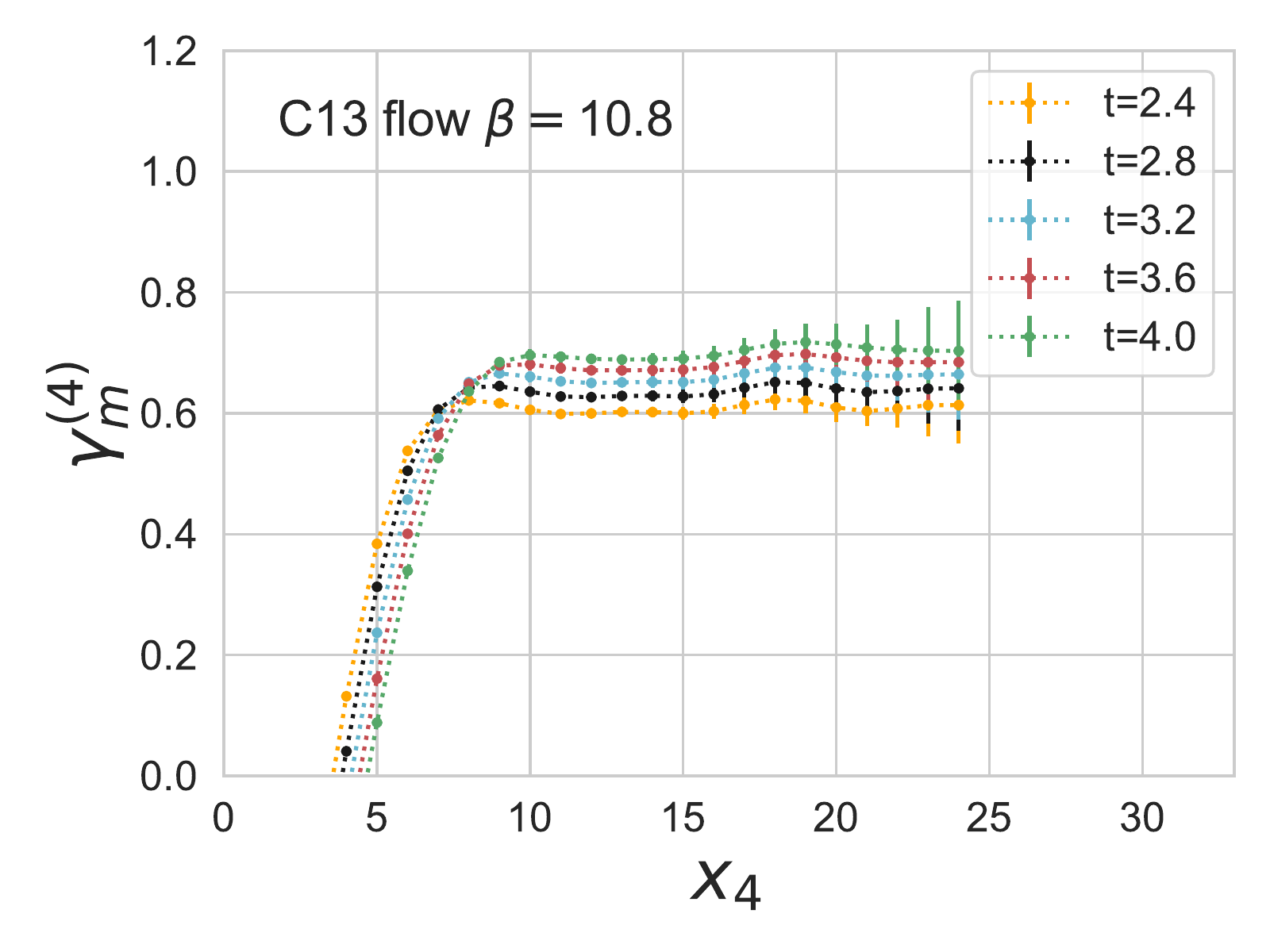}\\
\includegraphics*[width=\columnwidth]{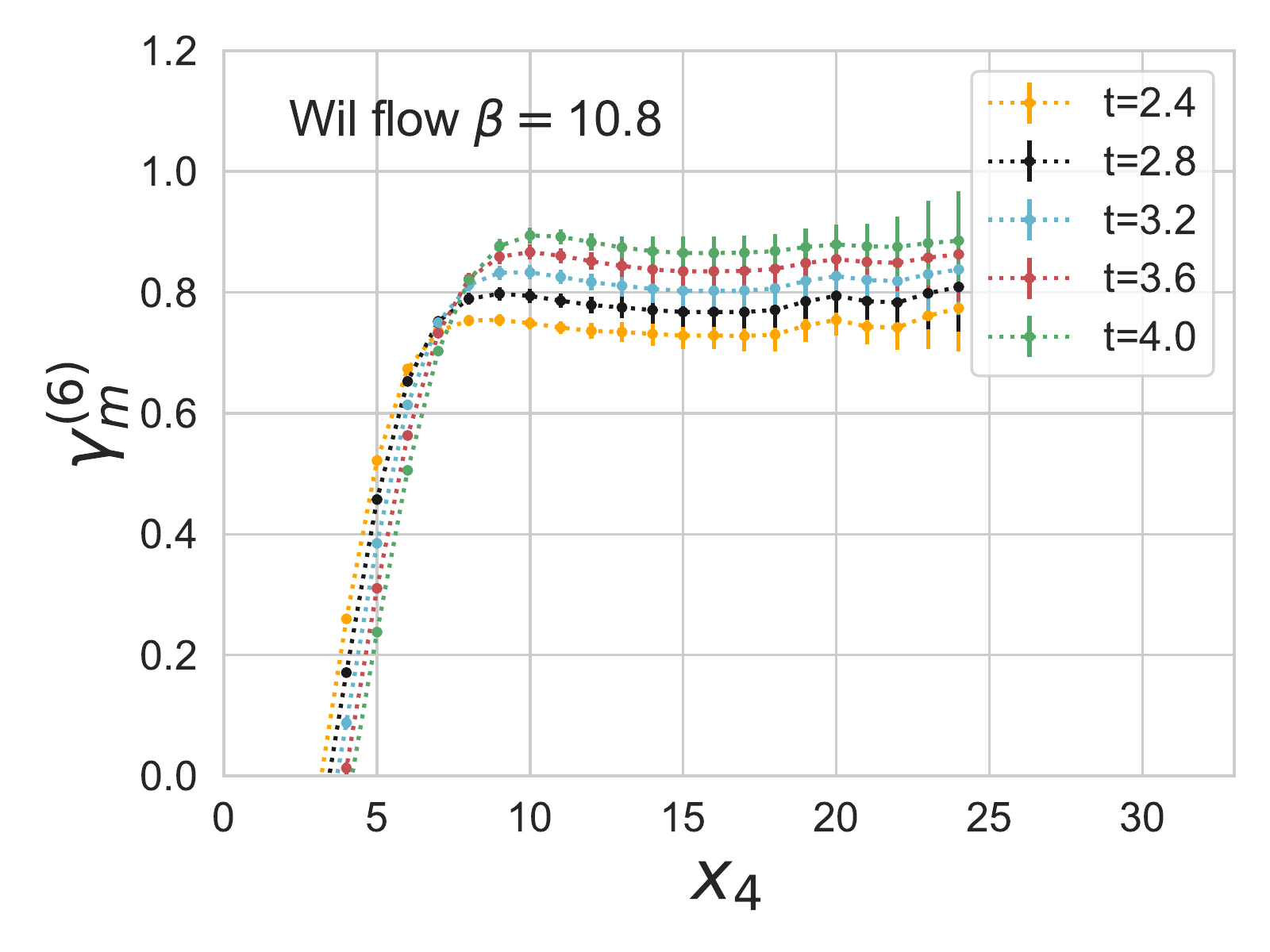}
\vskip -6ex
\end{center}
\caption{\label{gamma-vs-x4} Examples of $x_4$ plateaus. Both panels show raw results for the
mass anomalous dimension at $\b=10.8$. Top: C13 flow,
fundamental representation.  Bottom: Wilson flow,
sextet representation.
}
\end{figure}

\subsection{\label{gammam} Mass anomalous dimensions}

In the lattice calculation, for each representation we construct zero-momentum
meson correlation functions,
\begin{equation}
\label{mesoncorr}
C_X(x_4,t) = \svev{X(x_4^0)\, X'(x_4^0+x_4,t)} \ ,
\end{equation}
where $X=S$ (scalar) or $P$ (pseudoscalar).%
\footnote{We verified that the scalar and pseudoscalar correlators agree within error,
as expected when the chiral limit is taken in finite volume.}
The source $X$ is placed at euclidean time $x_4^0$,
while the sink $X'$ is separated by time $x_4$ from the source.
To shorten autocorrelation times, we shift the source plane $x_4^0$
between successive lattice configurations of each ensemble.
After fixing to Coulomb gauge, we construct the source $X$
using gaussian-smeared fermion fields with smearing radius $R_0=6$.
We solve the Dirac equation using the conjugate gradient algorithm,
and then flow the solution by integrating the fermion flow equation
following \rcite{Luscher:2013cpa}.
We use point sinks, projected to $\vec{p}=0$.  As for the gauge field,
flowed correlation functions are recorded at interval $\D t=0.1$.

Having similarly computed the flowed two-point function of the vector current,
we use \Eq{gammaX} to determine the mass anomalous dimension $\g_m$
as a function of the euclidean time separation $x_4$
and the flow time $t$.  Examples of raw results for $\g_m$
are shown in \Fig{gamma-vs-x4} for both representations.
A new ingredient, clearly visible in the figure,
is the $x_4$ plateaus.  This provides a practical criterion for
satisfying the condition $\sqrt{8t}\ll x_4$, and hence
we extract the anomalous dimensions using values of $x_4$ inside the plateau,
averaging over the range $11\leq x_4\leq14$.

The rest of the calculation follows the same steps as for
the beta function.  We interpolate the raw results to obtain
$\g_m$ as a function of $\ggf$, and then use the interpolations $\g_m(t,\ggf)$
to take the continuum limit $a^2/t \to 0$ at fixed $\ggf$.

Final results for the mass anomalous dimensions are shown in \Fig{gamma-cont} above.
In the weak-coupling region, the mass anomalous dimensions
agree with one-loop perturbation theory,
\begin{equation}
\label{gamma1loop}
\g_m = \frac{6 \ggf C_2}{16 \pi^2} \ ,
\end{equation}
where $C_2$ is the quadratic Casimir operator: $C_2 = 15/8$ for the
fundamental representation and $C_2 = 5/2$ for the sextet representation.
At larger couplings, the calculated mass anomalous dimensions
move below the one-loop result.
At the IR fixed point we have $\g_m^{(4)}\simeq 0.75$
for the fundamental representation and $\g_m^{(6)} \simeq 1.0$
for the sextet representation.  Both are quite large, suggesting that while
the 4+4 system is IR conformal, it is  not too far from the conformal sill.

These results, like the beta function presented above, are based on data obtained on lattices of size $24^3\times48$.  We have carried out comparisons to the two $28^3\times56$ ensembles, parallel to the analysis of $\beta(\ggf)$ in Sec.~\ref{FV}.
In a plot along the lines of Fig.~\ref{fig:vol}, we find of course the same shift in
$\ggf$, but, again, no 
change in the data beyond the statistical error bars.
Hence we expect $\gamma_m(\ggf)$ to move horizontally with the beta function as we approach the infinite-volume limit, with no other change.
We obtain the same behavior for the chimera anomalous dimensions, presented below.

\subsection{\label{andimchim} Chimera anomalous dimensions}

Many different chimera operators can be used to create a top-partner state
\cite{Golterman:2015zwa,Golterman:2017vdj}.
We consider operators of the lowest possible mass dimension, which are
three-fermion operators with no derivatives.
To write them, we introduce Dirac fermions for the sextet representation,
$\c_{ABi}$, $i=1,\ldots,4$, where $i$ is a flavor index
while $A,B=1,\ldots,4$, are the hypercolor indices,
which are antisymmetrized.
The fundamental representation fermions are denoted $\j_{Aa}$,
where $a=1,\ldots,4$ is also a flavor index.
The three-fermion chimera operators are
\begin{subequations}
\label{chimop}
\begin{eqnarray}
\label{chimopa}
  B_{iab}^{IJ} &=& \e_{ABCD}\, P_I\, \c_{ABi}\,
  \left( \j_{Ca}^T \,C P_J\, \j_{Db} \right) ,
\\
\label{chimopg}
  B^I_{iab\m} &=& \e_{ABCD}\, P_I\, \g_\m \c_{ABi}
  \left(\j^T_{Ca} C \g_\m \g_5 \j_{Db} \right) ,\hspace{5ex}
\end{eqnarray}
\end{subequations}
where the labels $I,J$ take the values $R,L$.
Here $C$ is the charge-conjugation matrix and $P_{R,L}=(1\pm\g_5)/2$.
In \Eq{chimopg} there is no summation over $\m$.

Projecting these operators onto the quantum numbers of the top quark
depends on the details of the embedding of the Standard Model symmetries
in the hypercolor (M6 or M11) model.
For the M6 model, see Refs.~\cite{Golterman:2015zwa,Golterman:2017vdj}.
This will not concern us here.

We construct chimera two-point functions following closely
\rcite{Ayyar:2018glg}.  All the chimera correlators we consider
have the general form
\begin{equation}
\label{chimcorr}
C_\pm(x_4,t) = {\rm Tr} \svev{B(x_4^0+x_4,t)\, \bar\Lambda(x_4^0) P_\pm} \ ,
\end{equation}
where $P_\pm = \half(1\pm \g_4)$ are parity projectors.  The source operator
$\bar\Lambda$ is a quark-model creation operator for the top partner,
which is kept at $t=0$ (for details, see \rcite{Ayyar:2018glg}).
The sink operator $B(x_4^0+x_4,t)$ is one of the operators in \Eq{chimop}.
We use the same flowed fermion fields as for the mesons.
As before, the sink is projected onto zero spatial momentum.

The correlation functions~(\ref{chimcorr}) are related by
discrete (lattice) symmetries.  First, the $B^I_{iab\m}$ operators are related
by hypercubic rotations;
since we calculate correlation functions with $x_4$ dependence,
we separate the $\mu=4$ operator from $\mu=1,2,3$
and lump the latter into a ``space'' component, viz.,
\begin{subequations}
\label{Bgamma}
\begin{eqnarray}
\label{Bgammaa}
B^{sI}_{iab} &=& \frac{1}{3} \sum_{k=1}^3 B^I_{iabk} \ ,
\\
\label{Bgammab}
B^{tI}_{iab} &=& B^I_{iab4} \ .
\end{eqnarray}
\end{subequations}
In addition, the correlation functions~(\ref{chimcorr}) are related
by the usual continuum discrete symmetries.
Parity flips the chiral projectors, and so it takes, for example,
$B^{RR}(\vec{p}=0,x_4) \leftrightarrow B^{LL}(\vec{p}=0,x_4)$.
Euclidean time reversal likewise
flips the parity projectors, and transforms the time component
according to $x_4 \leftrightarrow T-x_4$, where $T$ is the temporal
extent of the lattice.\footnote{
The product of euclidean time reversal and parity acts on a generic fermion field as $\j(x)\to\g_5\j(-x)$ in infinite volume.
We improve statistics by averaging over
quartets of correlators related by parity and time reversal.}

The outcome is that there are only three independent
anomalous dimensions.\footnote{%
  Naturally, the anomalous dimensions are independent of flavor indices.
}
We denote them $\g^{RR}_{ch}$ for $B^{RR}$ and $B^{LL}$;
$\g^{RL}_{ch}$ for $B^{RL}$ and $B^{LR}$; and $\g^s_{ch}$ for $B^{sI}$ and $B^{tI}$.

Since each chimera operator consists of two fundamental and one sextet fermion,
we normalize the flowed chimera two-point functions by constructing ratios
[cf.~\Eq{R}]
\begin{equation}
\label{chimR}
R_\pm(x_4,t) = \frac{C_\pm(x_4,t)}{C_V^{(4)}(x_4,t) \sqrt{C_V^{(6)}(x_4,t)}} \ .
\end{equation}
Here $C_V^{(4)}(x_4,t)$ and $C_V^{(6)}(x_4,t)$ are the two-point functions
of the fundamental and sextet vector currents.
We then use \Eq{gammaX} as before to extract the anomalous dimension.

\begin{figure*}[t]
\begin{center}
\includegraphics*[width=0.45\textwidth]{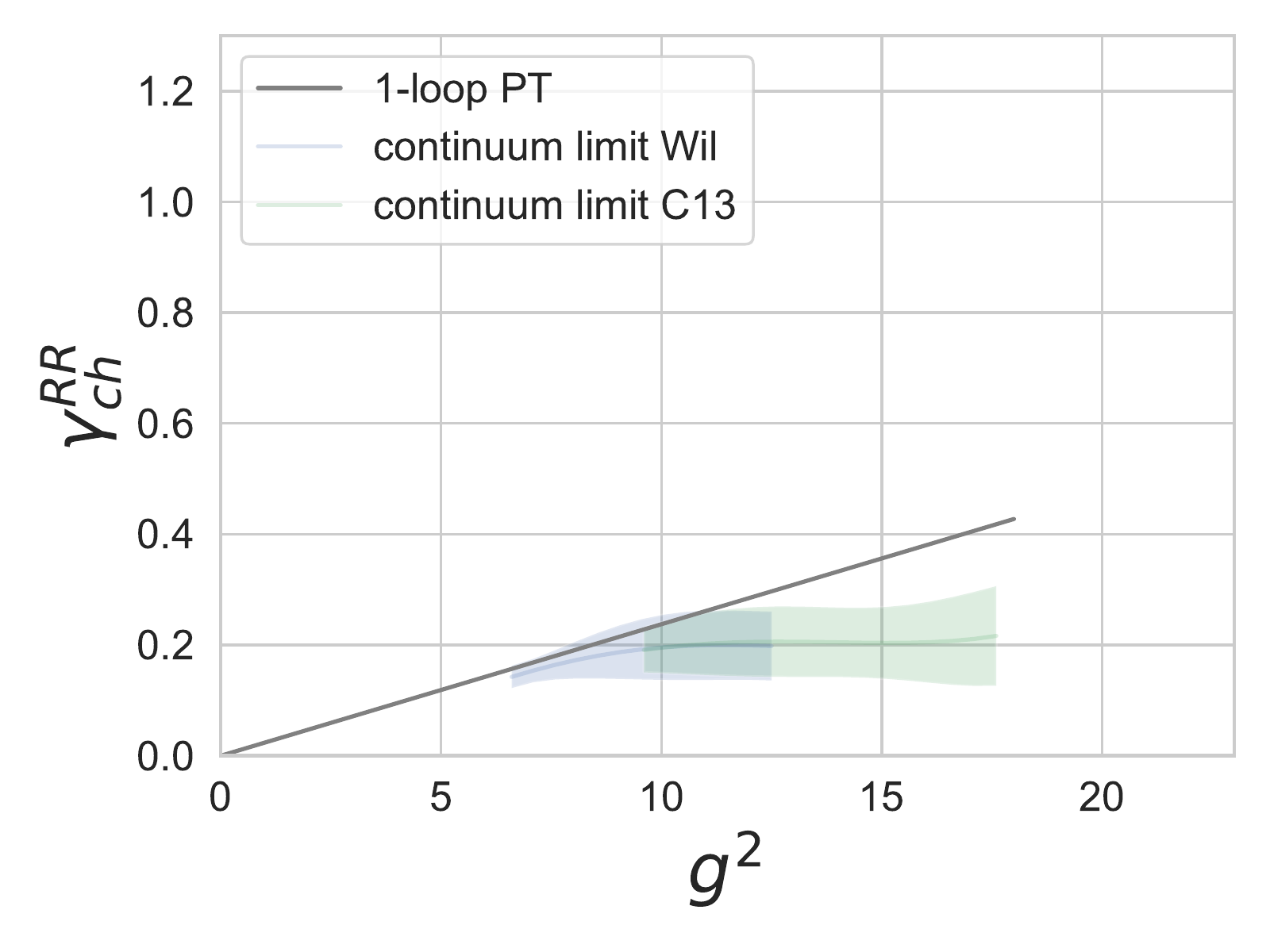}
\includegraphics*[width=0.45\textwidth]{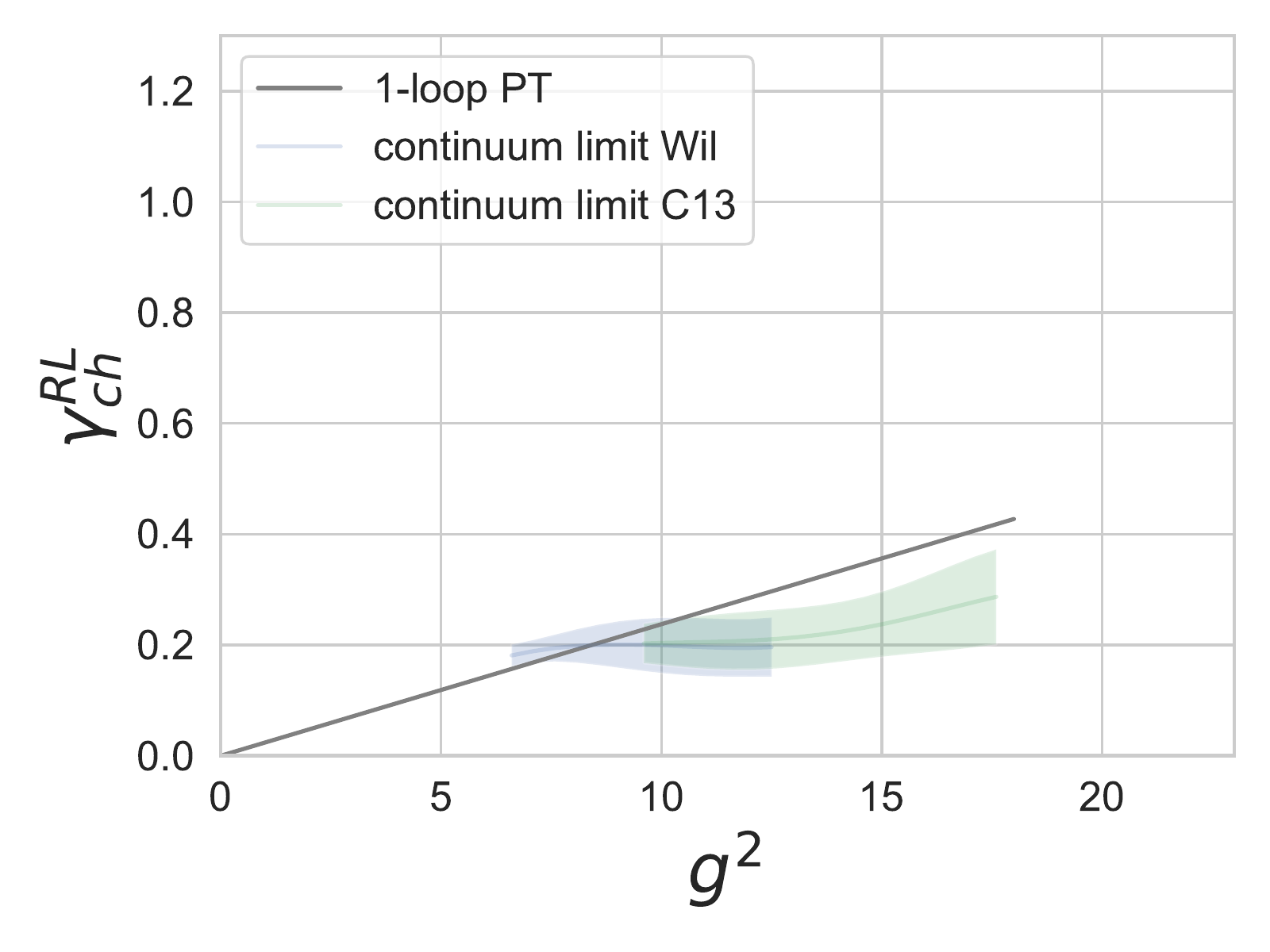} \\[0.5ex]
\includegraphics*[width=0.45\textwidth]{gammafn_gammas.pdf}
\end{center}
\caption{\label{andim-chimera2}
Anomalous dimensions of chimera operators.
Top left: $\g^{RR}_{ch}$.  Top right: $\g^{RL}_{ch}$.
Bottom: $\g^s_{ch}$ (same plot as in \Fig{andim-chimera}).
}
\end{figure*}

Figure~\ref{andim-chimera2} shows our results for the three independent
anomalous dimensions $\g^{RR}_{ch}$, $\g^{RL}_{ch}$ and $\g^s_{ch}$.
The general behavior of each of them closely resembles that of
the mass anomalous dimensions.
At our weakest couplings, the anomalous dimensions agree with one-loop
perturbation theory, whereas for larger couplings,
the calculated anomalous dimensions fall below the one-loop curve.
In addition, $\g^{RR}_{ch}$ and $\g^{RL}_{ch}$ agree within error,
which is not required by symmetry.

The one-loop predictions are given by $\g = c g^2 / (16\p^2)$,
where $c=15/4$ for both $\g^{RR}_{ch}$ and $\g^{RL}_{ch}$,
and $c=15/2$ for $\g^s_{ch}$ \cite{DeGrand:2015yna} (see also \cite{BuarqueFranzosi:2019eee}).
These anomalous dimensions happen to satisfy
\begin{equation}
\label{grat}
\g^{RR}_{ch}:\g^{RL}_{ch}:\g^s_{ch}:\g_m^{(4)}:\g_m^{(6)} = 1:1:2:3:4 \ .
\end{equation}
Thus, while at one loop $\g^s_{ch}$ is twice $\g^{RR}_{ch}$ or $\g^{RL}_{ch}$,
it is still smaller than both of the mass anomalous dimensions.
This pattern persists in our non-perturbative results.
Indeed, the ratios~(\ref{grat}) are roughly preserved near the fixed point
as well, where $\g^{RR}_{ch}\approx \g^{RL}_{ch}\simeq 0.25$
and $\g^s_{ch}\simeq 0.5$.

\section{\label{sec:conc} Conclusions}
As reported in this paper, we have employed lattice techniques to study the 4+4 model,
an SU(4) gauge theory with four Dirac fermions in the fundamental representation and with eight Majorana fermions in the sextet representation.
We used the continuous RG method, based on a gradient flow,
to calculate the beta function and a number of important anomalous dimensions.
Our lattice action includes a set of Pauli--Villars fields,
which enable us to reach otherwise inaccessible strong couplings.
We find that the 4+4 model has an infrared-stable fixed point at $g^2 \simeq 15.5$.

We have dealt carefully with the continuum limit.
Moreover, we have argued that finite-volume effects on both the beta function $\beta(\ggf)$ and the anomalous dimensions are limited to a common horizontal shift of a few percent, which leaves the anomalous dimensions at the infrared fixed point unchanged.

The M6 and M11 models of the Ferretti--Karateev list
\cite{Ferretti:2013kya,Belyaev:2016ftv} can be reached from the 4+4 model
by giving large masses to a suitable subset of the fermions,
while the fermion fields needed for the Ferretti--Karateev model itself
are kept massless or given much smaller masses.  Assuming that
the Ferretti--Karateev model is below the conformal sill, it follows that
the decoupling of the heavy fermions will trigger chiral symmetry breaking
and confinement.  The hypercolor scale $\LHC$ thus follows
the heavy mass scale \cite{Brower:2015owo,Hasenfratz:2016gut,Witzel:2019oej,Appelquist:2020xua}.

We also find that the mass anomalous dimensions of both representations
are quite large at the fixed point, even reaching $\g_m^{(6)}\simeq1$
for the sextet representation.
The anomalous dimensions of all top-partner chimera operators are smaller,
reaching 0.5 or less.  If one chooses masses for the fermions of the 4+4 model,
as just discussed, to reach the M6 or the M11 model, the anomalous dimensions
that govern the running of operators from the EHC scale down to the HC scale
will still be controlled by the fixed point.
The critical value of the chimera anomalous dimension,
where suppression by a power of $\LHC/\LEHC$ is eliminated,
is $\g = 2$ (see for example \rcite{Panico:2015jxa}).
Our results, then, indicate that
these anomalous dimensions may not be large enough
for a phenomenologically successful composite Higgs model.

\begin{acknowledgments}
Computations for this work were carried out on facilities
of the USQCD Collaboration, which are funded by the Office of Science
of the U.S.~Department of Energy.
A.H. and E.N.\ acknowledge support by DOE grant DE-SC0010005.
The work of B.S.\ and Y.S.\ was supported by the Israel Science Foundation
under grant No.~1429/21.

\end{acknowledgments}

\appendix*

\section{Lattice matters \label{latt}}

\subsection{\label{Slatt} Lattice action and simulation code}
As in our previous work on the 2+2 model, we use a Wilson-clover fermion action, with a gauge-invariant kinetic term for each fermion species.
We use normalized hypercubic (nHYP) smeared gauge links for the fundamental representation
\cite{Hasenfratz:2001hp,Hasenfratz:2007rf};
gauge links for the sextet representation are constructed from these smeared links.
The clover coefficient is set equal to unity for both fermion species
\cite{Bernard:1999kc,Shamir:2010cq}.
The gauge field action has the form $\b S_\text{plaq}+\g S_\text{NDS}$,
where $S_\text{plaq}$ is the usual plaquette action.
$S_\text{NDS}$ is the nHYP dislocation suppressing action,
a smeared action designed to reduce gauge-field roughness that would create
large fermion forces in the molecular dynamics evolution \cite{DeGrand:2014rwa}.
We hold the ratio $\g/\b=1/125$ fixed \cite{Ayyar:2017qdf}.

As discussed in \Sec{sec:GF}, in this work we have added a set of Pauli--Villars (PV) fields, which allow us to reach much further into strong renormalized coupling \cite{Hasenfratz:2021zsl}.
The PV action is the same Wilson-clover action as for the fermions, but the PV fields have opposite statistics.
The ratio of PV to fermion fields is $3:1$,
that is, we have 12 PV fields in each representation.
The bare mass of the PV fields is kept at $am_0=1$ ($\kappa=1/10$)
in order that they decouple in the continuum limit.
This can be gauged by the mass of the ghost pion made out of
two PV fields, which we find to be roughly equal to 2 in lattice units.
This means that the effective gauge action induced by integrating over
the PV fields is essentially local. For comparison, we note that
the physical pions in both representations come out to have masses
dictated by the volume, $m_\pi\simeq\pi/L$, where the lattice size
is $L=24a$ or~$28a$.

Our multi-representation code is a derivative of the MILC code \cite{MILC}.
We use three nested update levels with one Hasenbush preconditioning mass.
The innermost level (\verb+level=0+) contains the gauge update.
Pseudo\-fermion actions are simulated at \verb+level=1+, which takes care
of the upper end of the fermion spectrum, and at \verb+level=2+,
the outermost level, which takes care of the lower end of this spectrum.
The PV fields are simulated at \verb+level=1+, as is $S_\text{NDS}$.

\subsection{\label{app:ensembles} Ensembles}
For each value of the gauge coupling $\beta$, we determine the critical point $K_c=(\k_{4}^c,\k_{6}^c)$ by tuning the fermion masses to zero
for both representations, as calculated from the axial Ward identities (AWI).
We require the actual AWI masses to be $|m_q| \lesssim 0.002$,
as we have seen that these masses are small enough to have
negligible effect on the flow.  The ensembles used in this work
are shown in \Tab{table:ensembles}.
In practice, we tuned to $K_c$ for each of our $24^3\times48$ ensembles.
In the $28^3\times56$ ensembles we kept the same $(\k_4,\k_6)$ as for the
smaller volume and verified that the AWI masses are practically unchanged.

\begin{table*}[ht]
\begin{ruledtabular}
\begin{tabular}{ccllcll}
$L/a$ & $\beta$ & $\ph{X}\kappa_4$ & $\ph{X}\kappa_6$ & lattices & $\ph{XX}m_{q4}$ & $\ph{XX}m_{q6}$ \\
\hline
24       & 10.4         & 0.12805 & 0.12905 & 35 & $\ph{-}0.0017(1)$ & $-0.0003(1)$  \\
24       & 10.5         & 0.12787 & 0.12876 & 25 & $\ph{-}0.0013(1)$ & $\ph{-}0.0011(1)$   \\
24       & 10.6         & 0.12775 & 0.12855 & 35 & $-0.0007(1)$      & $\ph{-}0.0005(1)$   \\
24       & 10.8         & 0.12750 & 0.12804 & 25 & $-0.0028(1)$      & $\ph{-}0.0040(1)$   \\
24       & 11.0         & 0.12717 & 0.12777 & 25 & $-0.0010(1)$      & $\ph{-}0.0015(1)$   \\
24       & 11.5         & 0.12670 & 0.12724 & 25 & $-0.0012(1)$      & $-0.0009(1)$  \\
24       & 12.0         & 0.12633 & 0.12680 & 45 & $\ph{-}0.00154(3)$& $\ph{-}0.00139(3)$  \\
24       & 13.0         & 0.12600 & 0.12640 & 45 & $\ph{-}0.00141(1)$& $\ph{-}0.00010(1)$  \\
\hline
28       & 10.5         & 0.12787 & 0.12876 & 25& $\ph{-}0.0018(1)$ & $\ph{-}0.0015(1)$  \\
28       & 11.0         & 0.12717 & 0.12777 & 25& $-0.00096(4)$     & $\ph{-}0.0016(1)$  \\
\end{tabular}
\end{ruledtabular}
\caption{\label{table:ensembles}
Parameters and fermion masses for the ensembles.
The volume is $L^3\times T$, with $T=2L$.
The fifth column gives the number of configurations that were used in the flow analysis.
We saved a configuration every 5 trajectories (after equilibration).
$m_{q4}$ and $m_{q6}$ are the AWI masses of the
fundamental and sextet representations, respectively.
}
\end{table*}

\bibliography{4+4paper2c_post}

\begin{thebibliography}{55}%
\makeatletter
\providecommand \@ifxundefined [1]{%
 \@ifx{#1\undefined}
}%
\providecommand \@ifnum [1]{%
 \ifnum #1\expandafter \@firstoftwo
 \else \expandafter \@secondoftwo
 \fi
}%
\providecommand \@ifx [1]{%
 \ifx #1\expandafter \@firstoftwo
 \else \expandafter \@secondoftwo
 \fi
}%
\providecommand \natexlab [1]{#1}%
\providecommand \enquote  [1]{``#1''}%
\providecommand \bibnamefont  [1]{#1}%
\providecommand \bibfnamefont [1]{#1}%
\providecommand \citenamefont [1]{#1}%
\providecommand \href@noop [0]{\@secondoftwo}%
\providecommand \href [0]{\begingroup \@sanitize@url \@href}%
\providecommand \@href[1]{\@@startlink{#1}\@@href}%
\providecommand \@@href[1]{\endgroup#1\@@endlink}%
\providecommand \@sanitize@url [0]{\catcode `\\12\catcode `\$12\catcode
  `\&12\catcode `\#12\catcode `\^12\catcode `\_12\catcode `\%12\relax}%
\providecommand \@@startlink[1]{}%
\providecommand \@@endlink[0]{}%
\providecommand \url  [0]{\begingroup\@sanitize@url \@url }%
\providecommand \@url [1]{\endgroup\@href {#1}{\urlprefix }}%
\providecommand \urlprefix  [0]{URL }%
\providecommand \Eprint [0]{\href }%
\providecommand \doibase [0]{https://doi.org/}%
\providecommand \selectlanguage [0]{\@gobble}%
\providecommand \bibinfo  [0]{\@secondoftwo}%
\providecommand \bibfield  [0]{\@secondoftwo}%
\providecommand \translation [1]{[#1]}%
\providecommand \BibitemOpen [0]{}%
\providecommand \bibitemStop [0]{}%
\providecommand \bibitemNoStop [0]{.\EOS\space}%
\providecommand \EOS [0]{\spacefactor3000\relax}%
\providecommand \BibitemShut  [1]{\csname bibitem#1\endcsname}%
\let\auto@bib@innerbib\@empty
\bibitem [{\citenamefont {Georgi}\ and\ \citenamefont
  {Kaplan}(1984)}]{Georgi:1984af}%
  \BibitemOpen
  \bibfield  {author} {\bibinfo {author} {\bibfnamefont {H.}~\bibnamefont
  {Georgi}}\ and\ \bibinfo {author} {\bibfnamefont {D.~B.}\ \bibnamefont
  {Kaplan}},\ }\bibfield  {title} {\bibinfo {title} {{Composite Higgs and
  Custodial SU(2)}},\ }\href {https://doi.org/10.1016/0370-2693(84)90341-1}
  {\bibfield  {journal} {\bibinfo  {journal} {Phys. Lett.}\ }\textbf {\bibinfo
  {volume} {145B}},\ \bibinfo {pages} {216} (\bibinfo {year}
  {1984})}\BibitemShut {NoStop}%
\bibitem [{\citenamefont {Dugan}\ \emph {et~al.}(1985)\citenamefont {Dugan},
  \citenamefont {Georgi},\ and\ \citenamefont {Kaplan}}]{Dugan:1984hq}%
  \BibitemOpen
  \bibfield  {author} {\bibinfo {author} {\bibfnamefont {M.~J.}\ \bibnamefont
  {Dugan}}, \bibinfo {author} {\bibfnamefont {H.}~\bibnamefont {Georgi}},\ and\
  \bibinfo {author} {\bibfnamefont {D.~B.}\ \bibnamefont {Kaplan}},\ }\bibfield
   {title} {\bibinfo {title} {{Anatomy of a Composite Higgs Model}},\ }\href
  {https://doi.org/10.1016/0550-3213(85)90221-4} {\bibfield  {journal}
  {\bibinfo  {journal} {Nucl. Phys.}\ }\textbf {\bibinfo {volume} {B254}},\
  \bibinfo {pages} {299} (\bibinfo {year} {1985})}\BibitemShut {NoStop}%
\bibitem [{\citenamefont {Kaplan}(1991)}]{Kaplan:1991dc}%
  \BibitemOpen
  \bibfield  {author} {\bibinfo {author} {\bibfnamefont {D.~B.}\ \bibnamefont
  {Kaplan}},\ }\bibfield  {title} {\bibinfo {title} {{Flavor at SSC energies: A
  New mechanism for dynamically generated fermion masses}},\ }\href
  {https://doi.org/10.1016/S0550-3213(05)80021-5} {\bibfield  {journal}
  {\bibinfo  {journal} {Nucl. Phys.}\ }\textbf {\bibinfo {volume} {B365}},\
  \bibinfo {pages} {259} (\bibinfo {year} {1991})}\BibitemShut {NoStop}%
\bibitem [{\citenamefont {Contino}(2011)}]{Contino:2010rs}%
  \BibitemOpen
  \bibfield  {author} {\bibinfo {author} {\bibfnamefont {R.}~\bibnamefont
  {Contino}},\ }\bibfield  {title} {\bibinfo {title} {{The Higgs as a Composite
  Nambu-Goldstone Boson}},\ }in\ \href
  {https://doi.org/10.1142/9789814327183_0005} {\emph {\bibinfo {booktitle}
  {{Physics of the large and the small, TASI 09, proceedings of the Theoretical
  Advanced Study Institute in Elementary Particle Physics, Boulder, Colorado,
  USA, 1-26 June 2009}}}}\ (\bibinfo {year} {2011})\ pp.\ \bibinfo {pages}
  {235--306},\ \Eprint {https://arxiv.org/abs/1005.4269} {arXiv:1005.4269
  [hep-ph]} \BibitemShut {NoStop}%
\bibitem [{\citenamefont {Bellazzini}\ \emph {et~al.}(2014)\citenamefont
  {Bellazzini}, \citenamefont {Cs{\'a}ki},\ and\ \citenamefont
  {Serra}}]{Bellazzini:2014yua}%
  \BibitemOpen
  \bibfield  {author} {\bibinfo {author} {\bibfnamefont {B.}~\bibnamefont
  {Bellazzini}}, \bibinfo {author} {\bibfnamefont {C.}~\bibnamefont
  {Cs{\'a}ki}},\ and\ \bibinfo {author} {\bibfnamefont {J.}~\bibnamefont
  {Serra}},\ }\bibfield  {title} {\bibinfo {title} {{Composite Higgses}},\
  }\href {https://doi.org/10.1140/epjc/s10052-014-2766-x} {\bibfield  {journal}
  {\bibinfo  {journal} {Eur. Phys. J.}\ }\textbf {\bibinfo {volume} {C74}},\
  \bibinfo {pages} {2766} (\bibinfo {year} {2014})},\ \Eprint
  {https://arxiv.org/abs/1401.2457} {arXiv:1401.2457 [hep-ph]} \BibitemShut
  {NoStop}%
\bibitem [{\citenamefont {Panico}\ and\ \citenamefont
  {Wulzer}(2016)}]{Panico:2015jxa}%
  \BibitemOpen
  \bibfield  {author} {\bibinfo {author} {\bibfnamefont {G.}~\bibnamefont
  {Panico}}\ and\ \bibinfo {author} {\bibfnamefont {A.}~\bibnamefont
  {Wulzer}},\ }\bibfield  {title} {\bibinfo {title} {{The Composite
  Nambu-Goldstone Higgs}},\ }\href@noop {} {\bibfield  {journal} {\bibinfo
  {journal} {Lect. Notes Phys.}\ }\textbf {\bibinfo {volume} {913}} (\bibinfo
  {year} {2016})},\ \Eprint {https://arxiv.org/abs/1506.01961}
  {arXiv:1506.01961 [hep-ph]} \BibitemShut {NoStop}%
\bibitem [{\citenamefont {Ferretti}\ and\ \citenamefont
  {Karateev}(2014)}]{Ferretti:2013kya}%
  \BibitemOpen
  \bibfield  {author} {\bibinfo {author} {\bibfnamefont {G.}~\bibnamefont
  {Ferretti}}\ and\ \bibinfo {author} {\bibfnamefont {D.}~\bibnamefont
  {Karateev}},\ }\bibfield  {title} {\bibinfo {title} {{Fermionic UV
  completions of Composite Higgs models}},\ }\href
  {https://doi.org/10.1007/JHEP03(2014)077} {\bibfield  {journal} {\bibinfo
  {journal} {JHEP}\ }\textbf {\bibinfo {volume} {03}},\ \bibinfo {pages}
  {077}},\ \Eprint {https://arxiv.org/abs/1312.5330} {arXiv:1312.5330 [hep-ph]}
  \BibitemShut {NoStop}%
\bibitem [{\citenamefont {Ferretti}(2014)}]{Ferretti:2014qta}%
  \BibitemOpen
  \bibfield  {author} {\bibinfo {author} {\bibfnamefont {G.}~\bibnamefont
  {Ferretti}},\ }\bibfield  {title} {\bibinfo {title} {{UV Completions of
  Partial Compositeness: The Case for a SU(4) Gauge Group}},\ }\href
  {https://doi.org/10.1007/JHEP06(2014)142} {\bibfield  {journal} {\bibinfo
  {journal} {JHEP}\ }\textbf {\bibinfo {volume} {06}},\ \bibinfo {pages}
  {142}},\ \Eprint {https://arxiv.org/abs/1404.7137} {arXiv:1404.7137 [hep-ph]}
  \BibitemShut {NoStop}%
\bibitem [{\citenamefont {Ferretti}(2016)}]{Ferretti:2016upr}%
  \BibitemOpen
  \bibfield  {author} {\bibinfo {author} {\bibfnamefont {G.}~\bibnamefont
  {Ferretti}},\ }\bibfield  {title} {\bibinfo {title} {{Gauge theories of
  Partial Compositeness: Scenarios for Run-II of the LHC}},\ }\href
  {https://doi.org/10.1007/JHEP06(2016)107} {\bibfield  {journal} {\bibinfo
  {journal} {JHEP}\ }\textbf {\bibinfo {volume} {06}},\ \bibinfo {pages}
  {107}},\ \Eprint {https://arxiv.org/abs/1604.06467} {arXiv:1604.06467
  [hep-ph]} \BibitemShut {NoStop}%
\bibitem [{\citenamefont {Belyaev}\ \emph {et~al.}(2017)\citenamefont
  {Belyaev}, \citenamefont {Cacciapaglia}, \citenamefont {Cai}, \citenamefont
  {Ferretti}, \citenamefont {Flacke}, \citenamefont {Parolini},\ and\
  \citenamefont {Serodio}}]{Belyaev:2016ftv}%
  \BibitemOpen
  \bibfield  {author} {\bibinfo {author} {\bibfnamefont {A.}~\bibnamefont
  {Belyaev}}, \bibinfo {author} {\bibfnamefont {G.}~\bibnamefont
  {Cacciapaglia}}, \bibinfo {author} {\bibfnamefont {H.}~\bibnamefont {Cai}},
  \bibinfo {author} {\bibfnamefont {G.}~\bibnamefont {Ferretti}}, \bibinfo
  {author} {\bibfnamefont {T.}~\bibnamefont {Flacke}}, \bibinfo {author}
  {\bibfnamefont {A.}~\bibnamefont {Parolini}},\ and\ \bibinfo {author}
  {\bibfnamefont {H.}~\bibnamefont {Serodio}},\ }\bibfield  {title} {\bibinfo
  {title} {{Di-boson signatures as Standard Candles for Partial
  Compositeness}},\ }\href {https://doi.org/10.1007/JHEP01(2017)094} {\bibfield
   {journal} {\bibinfo  {journal} {JHEP}\ }\textbf {\bibinfo {volume} {01}},\
  \bibinfo {pages} {094}},\ \Eprint {https://arxiv.org/abs/1610.06591}
  {arXiv:1610.06591 [hep-ph]} \BibitemShut {NoStop}%
\bibitem [{\citenamefont {Vecchi}(2017)}]{Vecchi:2015fma}%
  \BibitemOpen
  \bibfield  {author} {\bibinfo {author} {\bibfnamefont {L.}~\bibnamefont
  {Vecchi}},\ }\bibfield  {title} {\bibinfo {title} {{A dangerous irrelevant
  UV-completion of the composite Higgs}},\ }\href
  {https://doi.org/10.1007/JHEP02(2017)094} {\bibfield  {journal} {\bibinfo
  {journal} {JHEP}\ }\textbf {\bibinfo {volume} {02}},\ \bibinfo {pages}
  {094}},\ \Eprint {https://arxiv.org/abs/1506.00623} {arXiv:1506.00623
  [hep-ph]} \BibitemShut {NoStop}%
\bibitem [{\citenamefont {Cacciapaglia}\ \emph {et~al.}(2021)\citenamefont
  {Cacciapaglia}, \citenamefont {Vatani},\ and\ \citenamefont
  {Zhang}}]{Cacciapaglia:2020jvj}%
  \BibitemOpen
  \bibfield  {author} {\bibinfo {author} {\bibfnamefont {G.}~\bibnamefont
  {Cacciapaglia}}, \bibinfo {author} {\bibfnamefont {S.}~\bibnamefont
  {Vatani}},\ and\ \bibinfo {author} {\bibfnamefont {C.}~\bibnamefont
  {Zhang}},\ }\bibfield  {title} {\bibinfo {title} {{The Techni-Pati-Salam
  Composite Higgs}},\ }\href {https://doi.org/10.1103/PhysRevD.103.055001}
  {\bibfield  {journal} {\bibinfo  {journal} {Phys. Rev. D}\ }\textbf {\bibinfo
  {volume} {103}},\ \bibinfo {pages} {055001} (\bibinfo {year} {2021})},\
  \Eprint {https://arxiv.org/abs/2005.12302} {arXiv:2005.12302 [hep-ph]}
  \BibitemShut {NoStop}%
\bibitem [{\citenamefont {Bando}\ \emph {et~al.}(1988)\citenamefont {Bando},
  \citenamefont {Kugo},\ and\ \citenamefont {Yamawaki}}]{Bando:1987br}%
  \BibitemOpen
  \bibfield  {author} {\bibinfo {author} {\bibfnamefont {M.}~\bibnamefont
  {Bando}}, \bibinfo {author} {\bibfnamefont {T.}~\bibnamefont {Kugo}},\ and\
  \bibinfo {author} {\bibfnamefont {K.}~\bibnamefont {Yamawaki}},\ }\bibfield
  {title} {\bibinfo {title} {{Nonlinear Realization and Hidden Local
  Symmetries}},\ }\href {https://doi.org/10.1016/0370-1573(88)90019-1}
  {\bibfield  {journal} {\bibinfo  {journal} {Phys. Rept.}\ }\textbf {\bibinfo
  {volume} {164}},\ \bibinfo {pages} {217} (\bibinfo {year}
  {1988})}\BibitemShut {NoStop}%
\bibitem [{\citenamefont {Hill}\ and\ \citenamefont
  {Simmons}(2003)}]{Hill:2002ap}%
  \BibitemOpen
  \bibfield  {author} {\bibinfo {author} {\bibfnamefont {C.~T.}\ \bibnamefont
  {Hill}}\ and\ \bibinfo {author} {\bibfnamefont {E.~H.}\ \bibnamefont
  {Simmons}},\ }\bibfield  {title} {\bibinfo {title} {{Strong Dynamics and
  Electroweak Symmetry Breaking}},\ }\href
  {https://doi.org/10.1016/S0370-1573(03)00140-6} {\bibfield  {journal}
  {\bibinfo  {journal} {Phys. Rept.}\ }\textbf {\bibinfo {volume} {381}},\
  \bibinfo {pages} {235} (\bibinfo {year} {2003})},\ \bibinfo {note} {[Erratum:
  Phys.Rept. 390, 553--554 (2004)]},\ \Eprint
  {https://arxiv.org/abs/hep-ph/0203079} {arXiv:hep-ph/0203079} \BibitemShut
  {NoStop}%
\bibitem [{\citenamefont {Kaplan}\ \emph {et~al.}(2009)\citenamefont {Kaplan},
  \citenamefont {Lee}, \citenamefont {Son},\ and\ \citenamefont
  {Stephanov}}]{Kaplan:2009kr}%
  \BibitemOpen
  \bibfield  {author} {\bibinfo {author} {\bibfnamefont {D.~B.}\ \bibnamefont
  {Kaplan}}, \bibinfo {author} {\bibfnamefont {J.-W.}\ \bibnamefont {Lee}},
  \bibinfo {author} {\bibfnamefont {D.~T.}\ \bibnamefont {Son}},\ and\ \bibinfo
  {author} {\bibfnamefont {M.~A.}\ \bibnamefont {Stephanov}},\ }\bibfield
  {title} {\bibinfo {title} {{Conformality Lost}},\ }\href
  {https://doi.org/10.1103/PhysRevD.80.125005} {\bibfield  {journal} {\bibinfo
  {journal} {Phys. Rev. D}\ }\textbf {\bibinfo {volume} {80}},\ \bibinfo
  {pages} {125005} (\bibinfo {year} {2009})},\ \Eprint
  {https://arxiv.org/abs/0905.4752} {arXiv:0905.4752 [hep-th]} \BibitemShut
  {NoStop}%
\bibitem [{\citenamefont {Brower}\ \emph {et~al.}(2016)\citenamefont {Brower},
  \citenamefont {Hasenfratz}, \citenamefont {Rebbi}, \citenamefont {Weinberg},\
  and\ \citenamefont {Witzel}}]{Brower:2015owo}%
  \BibitemOpen
  \bibfield  {author} {\bibinfo {author} {\bibfnamefont {R.~C.}\ \bibnamefont
  {Brower}}, \bibinfo {author} {\bibfnamefont {A.}~\bibnamefont {Hasenfratz}},
  \bibinfo {author} {\bibfnamefont {C.}~\bibnamefont {Rebbi}}, \bibinfo
  {author} {\bibfnamefont {E.}~\bibnamefont {Weinberg}},\ and\ \bibinfo
  {author} {\bibfnamefont {O.}~\bibnamefont {Witzel}},\ }\bibfield  {title}
  {\bibinfo {title} {{Composite Higgs model at a conformal fixed point}},\
  }\href {https://doi.org/10.1103/PhysRevD.93.075028} {\bibfield  {journal}
  {\bibinfo  {journal} {Phys. Rev. D}\ }\textbf {\bibinfo {volume} {93}},\
  \bibinfo {pages} {075028} (\bibinfo {year} {2016})},\ \Eprint
  {https://arxiv.org/abs/1512.02576} {arXiv:1512.02576 [hep-ph]} \BibitemShut
  {NoStop}%
\bibitem [{\citenamefont {Hasenfratz}\ \emph {et~al.}(2017)\citenamefont
  {Hasenfratz}, \citenamefont {Rebbi},\ and\ \citenamefont
  {Witzel}}]{Hasenfratz:2016gut}%
  \BibitemOpen
  \bibfield  {author} {\bibinfo {author} {\bibfnamefont {A.}~\bibnamefont
  {Hasenfratz}}, \bibinfo {author} {\bibfnamefont {C.}~\bibnamefont {Rebbi}},\
  and\ \bibinfo {author} {\bibfnamefont {O.}~\bibnamefont {Witzel}},\
  }\bibfield  {title} {\bibinfo {title} {{Large scale separation and resonances
  within LHC range from a prototype BSM model}},\ }\href
  {https://doi.org/10.1016/j.physletb.2017.07.058} {\bibfield  {journal}
  {\bibinfo  {journal} {Phys. Lett. B}\ }\textbf {\bibinfo {volume} {773}},\
  \bibinfo {pages} {86} (\bibinfo {year} {2017})},\ \Eprint
  {https://arxiv.org/abs/1609.01401} {arXiv:1609.01401 [hep-ph]} \BibitemShut
  {NoStop}%
\bibitem [{\citenamefont {Witzel}\ and\ \citenamefont
  {Hasenfratz}(2019)}]{Witzel:2019oej}%
  \BibitemOpen
  \bibfield  {author} {\bibinfo {author} {\bibfnamefont {O.}~\bibnamefont
  {Witzel}}\ and\ \bibinfo {author} {\bibfnamefont {A.}~\bibnamefont
  {Hasenfratz}} (\bibinfo {collaboration} {Lattice Strong Dynamics}),\
  }\bibfield  {title} {\bibinfo {title} {{Constructing a composite Higgs model
  with built-in large separation of scales}},\ }\href
  {https://doi.org/10.22323/1.363.0115} {\bibfield  {journal} {\bibinfo
  {journal} {PoS}\ }\textbf {\bibinfo {volume} {LATTICE2019}},\ \bibinfo
  {pages} {115} (\bibinfo {year} {2019})},\ \Eprint
  {https://arxiv.org/abs/1912.12255} {arXiv:1912.12255 [hep-lat]} \BibitemShut
  {NoStop}%
\bibitem [{\citenamefont {Appelquist}\ \emph {et~al.}(2021)\citenamefont
  {Appelquist} \emph {et~al.}}]{Appelquist:2020xua}%
  \BibitemOpen
  \bibfield  {author} {\bibinfo {author} {\bibfnamefont {T.}~\bibnamefont
  {Appelquist}} \emph {et~al.} (\bibinfo {collaboration} {Lattice Strong
  Dynamics}),\ }\bibfield  {title} {\bibinfo {title} {{Near-conformal dynamics
  in a chirally broken system}},\ }\href
  {https://doi.org/10.1103/PhysRevD.103.014504} {\bibfield  {journal} {\bibinfo
   {journal} {Phys. Rev. D}\ }\textbf {\bibinfo {volume} {103}},\ \bibinfo
  {pages} {014504} (\bibinfo {year} {2021})},\ \Eprint
  {https://arxiv.org/abs/2007.01810} {arXiv:2007.01810 [hep-ph]} \BibitemShut
  {NoStop}%
\bibitem [{\citenamefont {Ayyar}\ \emph
  {et~al.}(2018{\natexlab{a}})\citenamefont {Ayyar}, \citenamefont {DeGrand},
  \citenamefont {Golterman}, \citenamefont {Hackett}, \citenamefont {Jay},
  \citenamefont {Neil}, \citenamefont {Shamir},\ and\ \citenamefont
  {Svetitsky}}]{Ayyar:2017qdf}%
  \BibitemOpen
  \bibfield  {author} {\bibinfo {author} {\bibfnamefont {V.}~\bibnamefont
  {Ayyar}}, \bibinfo {author} {\bibfnamefont {T.}~\bibnamefont {DeGrand}},
  \bibinfo {author} {\bibfnamefont {M.}~\bibnamefont {Golterman}}, \bibinfo
  {author} {\bibfnamefont {D.~C.}\ \bibnamefont {Hackett}}, \bibinfo {author}
  {\bibfnamefont {W.~I.}\ \bibnamefont {Jay}}, \bibinfo {author} {\bibfnamefont
  {E.~T.}\ \bibnamefont {Neil}}, \bibinfo {author} {\bibfnamefont
  {Y.}~\bibnamefont {Shamir}},\ and\ \bibinfo {author} {\bibfnamefont
  {B.}~\bibnamefont {Svetitsky}},\ }\bibfield  {title} {\bibinfo {title}
  {{Spectroscopy of SU(4) composite Higgs theory with two distinct fermion
  representations}},\ }\href {https://doi.org/10.1103/PhysRevD.97.074505}
  {\bibfield  {journal} {\bibinfo  {journal} {Phys. Rev.}\ }\textbf {\bibinfo
  {volume} {D97}},\ \bibinfo {pages} {074505} (\bibinfo {year}
  {2018}{\natexlab{a}})},\ \Eprint {https://arxiv.org/abs/1710.00806}
  {arXiv:1710.00806 [hep-lat]} \BibitemShut {NoStop}%
\bibitem [{\citenamefont {Ayyar}\ \emph
  {et~al.}(2018{\natexlab{b}})\citenamefont {Ayyar}, \citenamefont {Degrand},
  \citenamefont {Hackett}, \citenamefont {Jay}, \citenamefont {Neil},
  \citenamefont {Shamir},\ and\ \citenamefont {Svetitsky}}]{Ayyar:2018zuk}%
  \BibitemOpen
  \bibfield  {author} {\bibinfo {author} {\bibfnamefont {V.}~\bibnamefont
  {Ayyar}}, \bibinfo {author} {\bibfnamefont {T.}~\bibnamefont {Degrand}},
  \bibinfo {author} {\bibfnamefont {D.~C.}\ \bibnamefont {Hackett}}, \bibinfo
  {author} {\bibfnamefont {W.~I.}\ \bibnamefont {Jay}}, \bibinfo {author}
  {\bibfnamefont {E.~T.}\ \bibnamefont {Neil}}, \bibinfo {author}
  {\bibfnamefont {Y.}~\bibnamefont {Shamir}},\ and\ \bibinfo {author}
  {\bibfnamefont {B.}~\bibnamefont {Svetitsky}},\ }\bibfield  {title} {\bibinfo
  {title} {{Baryon spectrum of SU(4) composite Higgs theory with two distinct
  fermion representations}},\ }\href
  {https://doi.org/10.1103/PhysRevD.97.114505} {\bibfield  {journal} {\bibinfo
  {journal} {Phys. Rev.}\ }\textbf {\bibinfo {volume} {D97}},\ \bibinfo {pages}
  {114505} (\bibinfo {year} {2018}{\natexlab{b}})},\ \Eprint
  {https://arxiv.org/abs/1801.05809} {arXiv:1801.05809 [hep-ph]} \BibitemShut
  {NoStop}%
\bibitem [{\citenamefont {Ayyar}\ \emph
  {et~al.}(2018{\natexlab{c}})\citenamefont {Ayyar}, \citenamefont {DeGrand},
  \citenamefont {Hackett}, \citenamefont {Jay}, \citenamefont {Neil},
  \citenamefont {Shamir},\ and\ \citenamefont {Svetitsky}}]{Ayyar:2018ppa}%
  \BibitemOpen
  \bibfield  {author} {\bibinfo {author} {\bibfnamefont {V.}~\bibnamefont
  {Ayyar}}, \bibinfo {author} {\bibfnamefont {T.}~\bibnamefont {DeGrand}},
  \bibinfo {author} {\bibfnamefont {D.~C.}\ \bibnamefont {Hackett}}, \bibinfo
  {author} {\bibfnamefont {W.~I.}\ \bibnamefont {Jay}}, \bibinfo {author}
  {\bibfnamefont {E.~T.}\ \bibnamefont {Neil}}, \bibinfo {author}
  {\bibfnamefont {Y.}~\bibnamefont {Shamir}},\ and\ \bibinfo {author}
  {\bibfnamefont {B.}~\bibnamefont {Svetitsky}},\ }\bibfield  {title} {\bibinfo
  {title} {{Finite-temperature phase structure of SU(4) gauge theory with
  multiple fermion representations}},\ }\href
  {https://doi.org/10.1103/PhysRevD.97.114502} {\bibfield  {journal} {\bibinfo
  {journal} {Phys. Rev.}\ }\textbf {\bibinfo {volume} {D97}},\ \bibinfo {pages}
  {114502} (\bibinfo {year} {2018}{\natexlab{c}})},\ \Eprint
  {https://arxiv.org/abs/1802.09644} {arXiv:1802.09644 [hep-lat]} \BibitemShut
  {NoStop}%
\bibitem [{\citenamefont {Ayyar}\ \emph
  {et~al.}(2018{\natexlab{d}})\citenamefont {Ayyar}, \citenamefont {DeGrand},
  \citenamefont {Hackett}, \citenamefont {Jay}, \citenamefont {Neil},
  \citenamefont {Shamir},\ and\ \citenamefont {Svetitsky}}]{Ayyar:2018glg}%
  \BibitemOpen
  \bibfield  {author} {\bibinfo {author} {\bibfnamefont {V.}~\bibnamefont
  {Ayyar}}, \bibinfo {author} {\bibfnamefont {T.}~\bibnamefont {DeGrand}},
  \bibinfo {author} {\bibfnamefont {D.~C.}\ \bibnamefont {Hackett}}, \bibinfo
  {author} {\bibfnamefont {W.~I.}\ \bibnamefont {Jay}}, \bibinfo {author}
  {\bibfnamefont {E.~T.}\ \bibnamefont {Neil}}, \bibinfo {author}
  {\bibfnamefont {Y.}~\bibnamefont {Shamir}},\ and\ \bibinfo {author}
  {\bibfnamefont {B.}~\bibnamefont {Svetitsky}},\ }\bibfield  {title} {\bibinfo
  {title} {{Partial compositeness from baryon matrix elements on the
  lattice}},\ }\href@noop {} {\  (\bibinfo {year} {2018}{\natexlab{d}})},\
  \Eprint {https://arxiv.org/abs/1812.02727} {arXiv:1812.02727 [hep-ph]}
  \BibitemShut {NoStop}%
\bibitem [{\citenamefont {Ayyar}\ \emph {et~al.}(2019)\citenamefont {Ayyar},
  \citenamefont {Golterman}, \citenamefont {Hackett}, \citenamefont {Jay},
  \citenamefont {Neil}, \citenamefont {Shamir},\ and\ \citenamefont
  {Svetitsky}}]{Ayyar:2019exp}%
  \BibitemOpen
  \bibfield  {author} {\bibinfo {author} {\bibfnamefont {V.}~\bibnamefont
  {Ayyar}}, \bibinfo {author} {\bibfnamefont {M.~F.}\ \bibnamefont
  {Golterman}}, \bibinfo {author} {\bibfnamefont {D.~C.}\ \bibnamefont
  {Hackett}}, \bibinfo {author} {\bibfnamefont {W.}~\bibnamefont {Jay}},
  \bibinfo {author} {\bibfnamefont {E.~T.}\ \bibnamefont {Neil}}, \bibinfo
  {author} {\bibfnamefont {Y.}~\bibnamefont {Shamir}},\ and\ \bibinfo {author}
  {\bibfnamefont {B.}~\bibnamefont {Svetitsky}},\ }\bibfield  {title} {\bibinfo
  {title} {{Radiative Contribution to the Composite-Higgs Potential in a
  Two-Representation Lattice Model}},\ }\href
  {https://doi.org/10.1103/PhysRevD.99.094504} {\bibfield  {journal} {\bibinfo
  {journal} {Phys. Rev. D}\ }\textbf {\bibinfo {volume} {99}},\ \bibinfo
  {pages} {094504} (\bibinfo {year} {2019})},\ \Eprint
  {https://arxiv.org/abs/1903.02535} {arXiv:1903.02535 [hep-lat]} \BibitemShut
  {NoStop}%
\bibitem [{\citenamefont {Golterman}\ \emph {et~al.}(2021)\citenamefont
  {Golterman}, \citenamefont {Jay}, \citenamefont {Neil}, \citenamefont
  {Shamir},\ and\ \citenamefont {Svetitsky}}]{Golterman:2020pyx}%
  \BibitemOpen
  \bibfield  {author} {\bibinfo {author} {\bibfnamefont {M.}~\bibnamefont
  {Golterman}}, \bibinfo {author} {\bibfnamefont {W.~I.}\ \bibnamefont {Jay}},
  \bibinfo {author} {\bibfnamefont {E.~T.}\ \bibnamefont {Neil}}, \bibinfo
  {author} {\bibfnamefont {Y.}~\bibnamefont {Shamir}},\ and\ \bibinfo {author}
  {\bibfnamefont {B.}~\bibnamefont {Svetitsky}},\ }\bibfield  {title} {\bibinfo
  {title} {{Low-energy constant $L_{10}$ in a two-representation lattice
  theory}},\ }\href {https://doi.org/10.1103/PhysRevD.103.074509} {\bibfield
  {journal} {\bibinfo  {journal} {Phys. Rev. D}\ }\textbf {\bibinfo {volume}
  {103}},\ \bibinfo {pages} {074509} (\bibinfo {year} {2021})},\ \Eprint
  {https://arxiv.org/abs/2010.01920} {arXiv:2010.01920 [hep-lat]} \BibitemShut
  {NoStop}%
\bibitem [{\citenamefont {Cossu}\ \emph {et~al.}(2019)\citenamefont {Cossu},
  \citenamefont {Del~Debbio}, \citenamefont {Panero},\ and\ \citenamefont
  {Preti}}]{Cossu:2019hse}%
  \BibitemOpen
  \bibfield  {author} {\bibinfo {author} {\bibfnamefont {G.}~\bibnamefont
  {Cossu}}, \bibinfo {author} {\bibfnamefont {L.}~\bibnamefont {Del~Debbio}},
  \bibinfo {author} {\bibfnamefont {M.}~\bibnamefont {Panero}},\ and\ \bibinfo
  {author} {\bibfnamefont {D.}~\bibnamefont {Preti}},\ }\bibfield  {title}
  {\bibinfo {title} {{Strong dynamics with matter in multiple representations:
  $\mathrm {SU}(4)$ gauge theory with fundamental and sextet fermions}},\
  }\href {https://doi.org/10.1140/epjc/s10052-019-7137-1} {\bibfield  {journal}
  {\bibinfo  {journal} {Eur. Phys. J. C}\ }\textbf {\bibinfo {volume} {79}},\
  \bibinfo {pages} {638} (\bibinfo {year} {2019})},\ \Eprint
  {https://arxiv.org/abs/1904.08885} {arXiv:1904.08885 [hep-lat]} \BibitemShut
  {NoStop}%
\bibitem [{\citenamefont {Del~Debbio}\ \emph {et~al.}(2023)\citenamefont
  {Del~Debbio}, \citenamefont {Lupo}, \citenamefont {Panero},\ and\
  \citenamefont {Tantalo}}]{DelDebbio:2022qgu}%
  \BibitemOpen
  \bibfield  {author} {\bibinfo {author} {\bibfnamefont {L.}~\bibnamefont
  {Del~Debbio}}, \bibinfo {author} {\bibfnamefont {A.}~\bibnamefont {Lupo}},
  \bibinfo {author} {\bibfnamefont {M.}~\bibnamefont {Panero}},\ and\ \bibinfo
  {author} {\bibfnamefont {N.}~\bibnamefont {Tantalo}},\ }\bibfield  {title}
  {\bibinfo {title} {{Multi-representation dynamics of SU(4) composite Higgs
  models: chiral limit and spectral reconstructions}},\ }\href
  {https://doi.org/10.1140/epjc/s10052-023-11363-8} {\bibfield  {journal}
  {\bibinfo  {journal} {Eur. Phys. J. C}\ }\textbf {\bibinfo {volume} {83}},\
  \bibinfo {pages} {220} (\bibinfo {year} {2023})},\ \Eprint
  {https://arxiv.org/abs/2211.09581} {arXiv:2211.09581 [hep-lat]} \BibitemShut
  {NoStop}%
\bibitem [{\citenamefont {Bennett}\ \emph {et~al.}(2018)\citenamefont
  {Bennett}, \citenamefont {Hong}, \citenamefont {Lee}, \citenamefont {Lin},
  \citenamefont {Lucini}, \citenamefont {Piai},\ and\ \citenamefont
  {Vadacchino}}]{Bennett:2017kga}%
  \BibitemOpen
  \bibfield  {author} {\bibinfo {author} {\bibfnamefont {E.}~\bibnamefont
  {Bennett}}, \bibinfo {author} {\bibfnamefont {D.~K.}\ \bibnamefont {Hong}},
  \bibinfo {author} {\bibfnamefont {J.-W.}\ \bibnamefont {Lee}}, \bibinfo
  {author} {\bibfnamefont {C.~J.~D.}\ \bibnamefont {Lin}}, \bibinfo {author}
  {\bibfnamefont {B.}~\bibnamefont {Lucini}}, \bibinfo {author} {\bibfnamefont
  {M.}~\bibnamefont {Piai}},\ and\ \bibinfo {author} {\bibfnamefont
  {D.}~\bibnamefont {Vadacchino}},\ }\bibfield  {title} {\bibinfo {title}
  {{Sp(4) gauge theory on the lattice: towards SU(4)/Sp(4) composite Higgs (and
  beyond)}},\ }\href {https://doi.org/10.1007/JHEP03(2018)185} {\bibfield
  {journal} {\bibinfo  {journal} {JHEP}\ }\textbf {\bibinfo {volume} {03}},\
  \bibinfo {pages} {185}},\ \Eprint {https://arxiv.org/abs/1712.04220}
  {arXiv:1712.04220 [hep-lat]} \BibitemShut {NoStop}%
\bibitem [{\citenamefont {Bennett}\ \emph {et~al.}(2019)\citenamefont
  {Bennett}, \citenamefont {Hong}, \citenamefont {Lee}, \citenamefont {Lin},
  \citenamefont {Lucini}, \citenamefont {Piai},\ and\ \citenamefont
  {Vadacchino}}]{Bennett:2019jzz}%
  \BibitemOpen
  \bibfield  {author} {\bibinfo {author} {\bibfnamefont {E.}~\bibnamefont
  {Bennett}}, \bibinfo {author} {\bibfnamefont {D.~K.}\ \bibnamefont {Hong}},
  \bibinfo {author} {\bibfnamefont {J.-W.}\ \bibnamefont {Lee}}, \bibinfo
  {author} {\bibfnamefont {C.~J.~D.}\ \bibnamefont {Lin}}, \bibinfo {author}
  {\bibfnamefont {B.}~\bibnamefont {Lucini}}, \bibinfo {author} {\bibfnamefont
  {M.}~\bibnamefont {Piai}},\ and\ \bibinfo {author} {\bibfnamefont
  {D.}~\bibnamefont {Vadacchino}},\ }\bibfield  {title} {\bibinfo {title}
  {{Sp(4) gauge theories on the lattice: $N_f=2$ dynamical fundamental
  fermions}},\ }\href {https://doi.org/10.1007/JHEP12(2019)053} {\bibfield
  {journal} {\bibinfo  {journal} {JHEP}\ }\textbf {\bibinfo {volume} {12}},\
  \bibinfo {pages} {053}},\ \Eprint {https://arxiv.org/abs/1909.12662}
  {arXiv:1909.12662 [hep-lat]} \BibitemShut {NoStop}%
\bibitem [{\citenamefont {Bennett}\ \emph {et~al.}(2022)\citenamefont
  {Bennett}, \citenamefont {Hong}, \citenamefont {Hsiao}, \citenamefont {Lee},
  \citenamefont {Lin}, \citenamefont {Lucini}, \citenamefont {Mesiti},
  \citenamefont {Piai},\ and\ \citenamefont {Vadacchino}}]{Bennett:2022yfa}%
  \BibitemOpen
  \bibfield  {author} {\bibinfo {author} {\bibfnamefont {E.}~\bibnamefont
  {Bennett}}, \bibinfo {author} {\bibfnamefont {D.~K.}\ \bibnamefont {Hong}},
  \bibinfo {author} {\bibfnamefont {H.}~\bibnamefont {Hsiao}}, \bibinfo
  {author} {\bibfnamefont {J.-W.}\ \bibnamefont {Lee}}, \bibinfo {author}
  {\bibfnamefont {C.~J.~D.}\ \bibnamefont {Lin}}, \bibinfo {author}
  {\bibfnamefont {B.}~\bibnamefont {Lucini}}, \bibinfo {author} {\bibfnamefont
  {M.}~\bibnamefont {Mesiti}}, \bibinfo {author} {\bibfnamefont
  {M.}~\bibnamefont {Piai}},\ and\ \bibinfo {author} {\bibfnamefont
  {D.}~\bibnamefont {Vadacchino}},\ }\bibfield  {title} {\bibinfo {title}
  {{Lattice studies of the Sp(4) gauge theory with two fundamental and three
  antisymmetric Dirac fermions}},\ }\href
  {https://doi.org/10.1103/PhysRevD.106.014501} {\bibfield  {journal} {\bibinfo
   {journal} {Phys. Rev. D}\ }\textbf {\bibinfo {volume} {106}},\ \bibinfo
  {pages} {014501} (\bibinfo {year} {2022})},\ \Eprint
  {https://arxiv.org/abs/2202.05516} {arXiv:2202.05516 [hep-lat]} \BibitemShut
  {NoStop}%
\bibitem [{\citenamefont {Hsiao}\ \emph {et~al.}(2022)\citenamefont {Hsiao},
  \citenamefont {Bennett}, \citenamefont {Hong}, \citenamefont {Lee},
  \citenamefont {Lin}, \citenamefont {Lucini}, \citenamefont {Piai},\ and\
  \citenamefont {Vadacchino}}]{Hsiao:2022kxf}%
  \BibitemOpen
  \bibfield  {author} {\bibinfo {author} {\bibfnamefont {H.}~\bibnamefont
  {Hsiao}}, \bibinfo {author} {\bibfnamefont {E.}~\bibnamefont {Bennett}},
  \bibinfo {author} {\bibfnamefont {D.~K.}\ \bibnamefont {Hong}}, \bibinfo
  {author} {\bibfnamefont {J.-W.}\ \bibnamefont {Lee}}, \bibinfo {author}
  {\bibfnamefont {C.~J.~D.}\ \bibnamefont {Lin}}, \bibinfo {author}
  {\bibfnamefont {B.}~\bibnamefont {Lucini}}, \bibinfo {author} {\bibfnamefont
  {M.}~\bibnamefont {Piai}},\ and\ \bibinfo {author} {\bibfnamefont
  {D.}~\bibnamefont {Vadacchino}},\ }\bibfield  {title} {\bibinfo {title}
  {{Spectroscopy of chimera baryons in a Sp(4) lattice gauge theory}},\
  }\href@noop {} {\  (\bibinfo {year} {2022})},\ \Eprint
  {https://arxiv.org/abs/2211.03955} {arXiv:2211.03955 [hep-lat]} \BibitemShut
  {NoStop}%
\bibitem [{\citenamefont {Bennett}\ \emph {et~al.}(2023)\citenamefont
  {Bennett}, \citenamefont {Holligan}, \citenamefont {Hong}, \citenamefont
  {Hsiao}, \citenamefont {Lee}, \citenamefont {Lin}, \citenamefont {Lucini},
  \citenamefont {Mesiti}, \citenamefont {Piai},\ and\ \citenamefont
  {Vadacchino}}]{Bennett:2023wjw}%
  \BibitemOpen
  \bibfield  {author} {\bibinfo {author} {\bibfnamefont {E.}~\bibnamefont
  {Bennett}}, \bibinfo {author} {\bibfnamefont {J.}~\bibnamefont {Holligan}},
  \bibinfo {author} {\bibfnamefont {D.~K.}\ \bibnamefont {Hong}}, \bibinfo
  {author} {\bibfnamefont {H.}~\bibnamefont {Hsiao}}, \bibinfo {author}
  {\bibfnamefont {J.-W.}\ \bibnamefont {Lee}}, \bibinfo {author} {\bibfnamefont
  {C.~J.~D.}\ \bibnamefont {Lin}}, \bibinfo {author} {\bibfnamefont
  {B.}~\bibnamefont {Lucini}}, \bibinfo {author} {\bibfnamefont
  {M.}~\bibnamefont {Mesiti}}, \bibinfo {author} {\bibfnamefont
  {M.}~\bibnamefont {Piai}},\ and\ \bibinfo {author} {\bibfnamefont
  {D.}~\bibnamefont {Vadacchino}},\ }\bibfield  {title} {\bibinfo {title}
  {{Sp($2N$) Lattice Gauge Theories and Extensions of the Standard Model of
  Particle Physics}}\ }(\bibinfo {year} {2023})\ \Eprint
  {https://arxiv.org/abs/2304.01070} {arXiv:2304.01070 [hep-lat]} \BibitemShut
  {NoStop}%
\bibitem [{\citenamefont {Kim}\ \emph {et~al.}(2020)\citenamefont {Kim},
  \citenamefont {Hong},\ and\ \citenamefont {Lee}}]{Kim:2020yvr}%
  \BibitemOpen
  \bibfield  {author} {\bibinfo {author} {\bibfnamefont {B.~S.}\ \bibnamefont
  {Kim}}, \bibinfo {author} {\bibfnamefont {D.~K.}\ \bibnamefont {Hong}},\ and\
  \bibinfo {author} {\bibfnamefont {J.-W.}\ \bibnamefont {Lee}},\ }\bibfield
  {title} {\bibinfo {title} {{Into the conformal window: Multirepresentation
  gauge theories}},\ }\href {https://doi.org/10.1103/PhysRevD.101.056008}
  {\bibfield  {journal} {\bibinfo  {journal} {Phys. Rev. D}\ }\textbf {\bibinfo
  {volume} {101}},\ \bibinfo {pages} {056008} (\bibinfo {year} {2020})},\
  \Eprint {https://arxiv.org/abs/2001.02690} {arXiv:2001.02690 [hep-ph]}
  \BibitemShut {NoStop}%
\bibitem [{\citenamefont {Hasenfratz}\ and\ \citenamefont
  {Witzel}(2020)}]{Hasenfratz:2019hpg}%
  \BibitemOpen
  \bibfield  {author} {\bibinfo {author} {\bibfnamefont {A.}~\bibnamefont
  {Hasenfratz}}\ and\ \bibinfo {author} {\bibfnamefont {O.}~\bibnamefont
  {Witzel}},\ }\bibfield  {title} {\bibinfo {title} {{Continuous
  renormalization group $\beta$ function from lattice simulations}},\ }\href
  {https://doi.org/10.1103/PhysRevD.101.034514} {\bibfield  {journal} {\bibinfo
   {journal} {Phys. Rev. D}\ }\textbf {\bibinfo {volume} {101}},\ \bibinfo
  {pages} {034514} (\bibinfo {year} {2020})},\ \Eprint
  {https://arxiv.org/abs/1910.06408} {arXiv:1910.06408 [hep-lat]} \BibitemShut
  {NoStop}%
\bibitem [{\citenamefont {Hasenfratz}\ and\ \citenamefont
  {Witzel}(2019)}]{Hasenfratz:2019puu}%
  \BibitemOpen
  \bibfield  {author} {\bibinfo {author} {\bibfnamefont {A.}~\bibnamefont
  {Hasenfratz}}\ and\ \bibinfo {author} {\bibfnamefont {O.}~\bibnamefont
  {Witzel}},\ }\bibfield  {title} {\bibinfo {title} {{Continuous $\beta$
  function for the SU(3) gauge systems with two and twelve fundamental
  flavors}},\ }\href {https://doi.org/10.22323/1.363.0094} {\bibfield
  {journal} {\bibinfo  {journal} {PoS}\ }\textbf {\bibinfo {volume}
  {LATTICE2019}},\ \bibinfo {pages} {094} (\bibinfo {year} {2019})},\ \Eprint
  {https://arxiv.org/abs/1911.11531} {arXiv:1911.11531 [hep-lat]} \BibitemShut
  {NoStop}%
\bibitem [{\citenamefont {Peterson}\ \emph {et~al.}(2022)\citenamefont
  {Peterson}, \citenamefont {Hasenfratz}, \citenamefont {van Sickle},\ and\
  \citenamefont {Witzel}}]{Peterson:2021lvb}%
  \BibitemOpen
  \bibfield  {author} {\bibinfo {author} {\bibfnamefont {C.~T.}\ \bibnamefont
  {Peterson}}, \bibinfo {author} {\bibfnamefont {A.}~\bibnamefont
  {Hasenfratz}}, \bibinfo {author} {\bibfnamefont {J.}~\bibnamefont {van
  Sickle}},\ and\ \bibinfo {author} {\bibfnamefont {O.}~\bibnamefont
  {Witzel}},\ }\bibfield  {title} {\bibinfo {title} {{Determination of the
  continuous $\beta$ function of SU(3) Yang-Mills theory}},\ }\href
  {https://doi.org/10.22323/1.396.0174} {\bibfield  {journal} {\bibinfo
  {journal} {PoS}\ }\textbf {\bibinfo {volume} {LATTICE2021}},\ \bibinfo
  {pages} {174} (\bibinfo {year} {2022})},\ \Eprint
  {https://arxiv.org/abs/2109.09720} {arXiv:2109.09720 [hep-lat]} \BibitemShut
  {NoStop}%
\bibitem [{\citenamefont {Hasenfratz}\ \emph {et~al.}(2023)\citenamefont
  {Hasenfratz}, \citenamefont {Peterson}, \citenamefont {van Sickle},\ and\
  \citenamefont {Witzel}}]{Hasenfratz:2023bok}%
  \BibitemOpen
  \bibfield  {author} {\bibinfo {author} {\bibfnamefont {A.}~\bibnamefont
  {Hasenfratz}}, \bibinfo {author} {\bibfnamefont {C.~T.}\ \bibnamefont
  {Peterson}}, \bibinfo {author} {\bibfnamefont {J.}~\bibnamefont {van
  Sickle}},\ and\ \bibinfo {author} {\bibfnamefont {O.}~\bibnamefont
  {Witzel}},\ }\bibfield  {title} {\bibinfo {title} {{$\Lambda$ parameter of
  the SU(3) Yang-Mills theory from the continuous $\beta$ function}},\
  }\href@noop {} {\  (\bibinfo {year} {2023})},\ \Eprint
  {https://arxiv.org/abs/2303.00704} {arXiv:2303.00704 [hep-lat]} \BibitemShut
  {NoStop}%
\bibitem [{\citenamefont {Fodor}\ \emph {et~al.}(2018)\citenamefont {Fodor},
  \citenamefont {Holland}, \citenamefont {Kuti}, \citenamefont {Nogradi},\ and\
  \citenamefont {Wong}}]{Fodor:2017die}%
  \BibitemOpen
  \bibfield  {author} {\bibinfo {author} {\bibfnamefont {Z.}~\bibnamefont
  {Fodor}}, \bibinfo {author} {\bibfnamefont {K.}~\bibnamefont {Holland}},
  \bibinfo {author} {\bibfnamefont {J.}~\bibnamefont {Kuti}}, \bibinfo {author}
  {\bibfnamefont {D.}~\bibnamefont {Nogradi}},\ and\ \bibinfo {author}
  {\bibfnamefont {C.~H.}\ \bibnamefont {Wong}},\ }\bibfield  {title} {\bibinfo
  {title} {{A new method for the beta function in the chiral symmetry broken
  phase}},\ }\href {https://doi.org/10.1051/epjconf/201817508027} {\bibfield
  {journal} {\bibinfo  {journal} {EPJ Web Conf.}\ }\textbf {\bibinfo {volume}
  {175}},\ \bibinfo {pages} {08027} (\bibinfo {year} {2018})},\ \Eprint
  {https://arxiv.org/abs/1711.04833} {arXiv:1711.04833 [hep-lat]} \BibitemShut
  {NoStop}%
\bibitem [{\citenamefont {L{\"u}scher}(2010)}]{Luscher:2010iy}%
  \BibitemOpen
  \bibfield  {author} {\bibinfo {author} {\bibfnamefont {M.}~\bibnamefont
  {L{\"u}scher}},\ }\bibfield  {title} {\bibinfo {title} {{Properties and uses
  of the Wilson flow in lattice QCD}},\ }\href
  {https://doi.org/10.1007/JHEP08(2010)071, 10.1007/JHEP03(2014)092} {\bibfield
   {journal} {\bibinfo  {journal} {JHEP}\ }\textbf {\bibinfo {volume} {08}},\
  \bibinfo {pages} {071}},\ \bibinfo {note} {[Erratum: JHEP 03, 092 (2014)]},\
  \Eprint {https://arxiv.org/abs/1006.4518} {arXiv:1006.4518 [hep-lat]}
  \BibitemShut {NoStop}%
\bibitem [{\citenamefont {Fodor}\ \emph {et~al.}(2012)\citenamefont {Fodor},
  \citenamefont {Holland}, \citenamefont {Kuti}, \citenamefont {Nogradi},\ and\
  \citenamefont {Wong}}]{Fodor:2012td}%
  \BibitemOpen
  \bibfield  {author} {\bibinfo {author} {\bibfnamefont {Z.}~\bibnamefont
  {Fodor}}, \bibinfo {author} {\bibfnamefont {K.}~\bibnamefont {Holland}},
  \bibinfo {author} {\bibfnamefont {J.}~\bibnamefont {Kuti}}, \bibinfo {author}
  {\bibfnamefont {D.}~\bibnamefont {Nogradi}},\ and\ \bibinfo {author}
  {\bibfnamefont {C.~H.}\ \bibnamefont {Wong}},\ }\bibfield  {title} {\bibinfo
  {title} {{The Yang-Mills gradient flow in finite volume}},\ }\href
  {https://doi.org/10.1007/JHEP11(2012)007} {\bibfield  {journal} {\bibinfo
  {journal} {JHEP}\ }\textbf {\bibinfo {volume} {11}},\ \bibinfo {pages}
  {007}},\ \Eprint {https://arxiv.org/abs/1208.1051} {arXiv:1208.1051
  [hep-lat]} \BibitemShut {NoStop}%
\bibitem [{\citenamefont {L{\"u}scher}(2013)}]{Luscher:2013cpa}%
  \BibitemOpen
  \bibfield  {author} {\bibinfo {author} {\bibfnamefont {M.}~\bibnamefont
  {L{\"u}scher}},\ }\bibfield  {title} {\bibinfo {title} {{Chiral symmetry and
  the Yang--Mills gradient flow}},\ }\href
  {https://doi.org/10.1007/JHEP04(2013)123} {\bibfield  {journal} {\bibinfo
  {journal} {JHEP}\ }\textbf {\bibinfo {volume} {04}},\ \bibinfo {pages}
  {123}},\ \Eprint {https://arxiv.org/abs/1302.5246} {arXiv:1302.5246
  [hep-lat]} \BibitemShut {NoStop}%
\bibitem [{\citenamefont {Carosso}\ \emph {et~al.}(2018)\citenamefont
  {Carosso}, \citenamefont {Hasenfratz},\ and\ \citenamefont
  {Neil}}]{Carosso:2018bmz}%
  \BibitemOpen
  \bibfield  {author} {\bibinfo {author} {\bibfnamefont {A.}~\bibnamefont
  {Carosso}}, \bibinfo {author} {\bibfnamefont {A.}~\bibnamefont
  {Hasenfratz}},\ and\ \bibinfo {author} {\bibfnamefont {E.~T.}\ \bibnamefont
  {Neil}},\ }\bibfield  {title} {\bibinfo {title} {{Nonperturbative
  Renormalization of Operators in Near-Conformal Systems Using Gradient
  Flows}},\ }\href {https://doi.org/10.1103/PhysRevLett.121.201601} {\bibfield
  {journal} {\bibinfo  {journal} {Phys. Rev. Lett.}\ }\textbf {\bibinfo
  {volume} {121}},\ \bibinfo {pages} {201601} (\bibinfo {year} {2018})},\
  \Eprint {https://arxiv.org/abs/1806.01385} {arXiv:1806.01385 [hep-lat]}
  \BibitemShut {NoStop}%
\bibitem [{\citenamefont {Hasenfratz}\ \emph {et~al.}(2021)\citenamefont
  {Hasenfratz}, \citenamefont {Shamir},\ and\ \citenamefont
  {Svetitsky}}]{Hasenfratz:2021zsl}%
  \BibitemOpen
  \bibfield  {author} {\bibinfo {author} {\bibfnamefont {A.}~\bibnamefont
  {Hasenfratz}}, \bibinfo {author} {\bibfnamefont {Y.}~\bibnamefont {Shamir}},\
  and\ \bibinfo {author} {\bibfnamefont {B.}~\bibnamefont {Svetitsky}},\
  }\bibfield  {title} {\bibinfo {title} {{Taming lattice artifacts with
  Pauli-Villars fields}},\ }\href {https://doi.org/10.1103/PhysRevD.104.074509}
  {\bibfield  {journal} {\bibinfo  {journal} {Phys. Rev. D}\ }\textbf {\bibinfo
  {volume} {104}},\ \bibinfo {pages} {074509} (\bibinfo {year} {2021})},\
  \Eprint {https://arxiv.org/abs/2109.02790} {arXiv:2109.02790 [hep-lat]}
  \BibitemShut {NoStop}%
\bibitem [{\citenamefont {Ramos}\ and\ \citenamefont
  {Sint}(2016)}]{Ramos:2015baa}%
  \BibitemOpen
  \bibfield  {author} {\bibinfo {author} {\bibfnamefont {A.}~\bibnamefont
  {Ramos}}\ and\ \bibinfo {author} {\bibfnamefont {S.}~\bibnamefont {Sint}},\
  }\bibfield  {title} {\bibinfo {title} {{Symanzik improvement of the gradient
  flow in lattice gauge theories}},\ }\href
  {https://doi.org/10.1140/epjc/s10052-015-3831-9} {\bibfield  {journal}
  {\bibinfo  {journal} {Eur. Phys. J. C}\ }\textbf {\bibinfo {volume} {76}},\
  \bibinfo {pages} {15} (\bibinfo {year} {2016})},\ \Eprint
  {https://arxiv.org/abs/1508.05552} {arXiv:1508.05552 [hep-lat]} \BibitemShut
  {NoStop}%
\bibitem [{\citenamefont {Hasenfratz}\ and\ \citenamefont
  {Witzel}(2021)}]{Hasenfratz:2020vta}%
  \BibitemOpen
  \bibfield  {author} {\bibinfo {author} {\bibfnamefont {A.}~\bibnamefont
  {Hasenfratz}}\ and\ \bibinfo {author} {\bibfnamefont {O.}~\bibnamefont
  {Witzel}},\ }\bibfield  {title} {\bibinfo {title} {{Dislocations under
  gradient flow and their effect on the renormalized coupling}},\ }\href
  {https://doi.org/10.1103/PhysRevD.103.034505} {\bibfield  {journal} {\bibinfo
   {journal} {Phys. Rev. D}\ }\textbf {\bibinfo {volume} {103}},\ \bibinfo
  {pages} {034505} (\bibinfo {year} {2021})},\ \Eprint
  {https://arxiv.org/abs/2004.00758} {arXiv:2004.00758 [hep-lat]} \BibitemShut
  {NoStop}%
\bibitem [{\citenamefont {Golterman}\ and\ \citenamefont
  {Shamir}(2015)}]{Golterman:2015zwa}%
  \BibitemOpen
  \bibfield  {author} {\bibinfo {author} {\bibfnamefont {M.}~\bibnamefont
  {Golterman}}\ and\ \bibinfo {author} {\bibfnamefont {Y.}~\bibnamefont
  {Shamir}},\ }\bibfield  {title} {\bibinfo {title} {{Top quark induced
  effective potential in a composite Higgs model}},\ }\href
  {https://doi.org/10.1103/PhysRevD.91.094506} {\bibfield  {journal} {\bibinfo
  {journal} {Phys. Rev.}\ }\textbf {\bibinfo {volume} {D91}},\ \bibinfo {pages}
  {094506} (\bibinfo {year} {2015})},\ \Eprint
  {https://arxiv.org/abs/1502.00390} {arXiv:1502.00390 [hep-ph]} \BibitemShut
  {NoStop}%
\bibitem [{\citenamefont {Golterman}\ and\ \citenamefont
  {Shamir}(2018)}]{Golterman:2017vdj}%
  \BibitemOpen
  \bibfield  {author} {\bibinfo {author} {\bibfnamefont {M.}~\bibnamefont
  {Golterman}}\ and\ \bibinfo {author} {\bibfnamefont {Y.}~\bibnamefont
  {Shamir}},\ }\bibfield  {title} {\bibinfo {title} {{Effective potential in
  ultraviolet completions for composite Higgs models}},\ }\href
  {https://doi.org/10.1103/PhysRevD.97.095005} {\bibfield  {journal} {\bibinfo
  {journal} {Phys. Rev.}\ }\textbf {\bibinfo {volume} {D97}},\ \bibinfo {pages}
  {095005} (\bibinfo {year} {2018})},\ \Eprint
  {https://arxiv.org/abs/1707.06033} {arXiv:1707.06033 [hep-ph]} \BibitemShut
  {NoStop}%
\bibitem [{\citenamefont {DeGrand}\ and\ \citenamefont
  {Shamir}(2015)}]{DeGrand:2015yna}%
  \BibitemOpen
  \bibfield  {author} {\bibinfo {author} {\bibfnamefont {T.}~\bibnamefont
  {DeGrand}}\ and\ \bibinfo {author} {\bibfnamefont {Y.}~\bibnamefont
  {Shamir}},\ }\bibfield  {title} {\bibinfo {title} {{One-loop anomalous
  dimension of top-partner hyperbaryons in a family of composite Higgs
  models}},\ }\href {https://doi.org/10.1103/PhysRevD.92.075039} {\bibfield
  {journal} {\bibinfo  {journal} {Phys. Rev.}\ }\textbf {\bibinfo {volume}
  {D92}},\ \bibinfo {pages} {075039} (\bibinfo {year} {2015})},\ \Eprint
  {https://arxiv.org/abs/1508.02581} {arXiv:1508.02581 [hep-ph]} \BibitemShut
  {NoStop}%
\bibitem [{\citenamefont {Buarque~Franzosi}\ and\ \citenamefont
  {Ferretti}(2019)}]{BuarqueFranzosi:2019eee}%
  \BibitemOpen
  \bibfield  {author} {\bibinfo {author} {\bibfnamefont {D.}~\bibnamefont
  {Buarque~Franzosi}}\ and\ \bibinfo {author} {\bibfnamefont {G.}~\bibnamefont
  {Ferretti}},\ }\bibfield  {title} {\bibinfo {title} {{Anomalous dimensions of
  potential top-partners}},\ }\href
  {https://doi.org/10.21468/SciPostPhys.7.3.027} {\bibfield  {journal}
  {\bibinfo  {journal} {SciPost Phys.}\ }\textbf {\bibinfo {volume} {7}},\
  \bibinfo {pages} {027} (\bibinfo {year} {2019})},\ \Eprint
  {https://arxiv.org/abs/1905.08273} {arXiv:1905.08273 [hep-ph]} \BibitemShut
  {NoStop}%
\bibitem [{\citenamefont {Hasenfratz}\ and\ \citenamefont
  {Knechtli}(2001)}]{Hasenfratz:2001hp}%
  \BibitemOpen
  \bibfield  {author} {\bibinfo {author} {\bibfnamefont {A.}~\bibnamefont
  {Hasenfratz}}\ and\ \bibinfo {author} {\bibfnamefont {F.}~\bibnamefont
  {Knechtli}},\ }\bibfield  {title} {\bibinfo {title} {{Flavor symmetry and the
  static potential with hypercubic blocking}},\ }\href
  {https://doi.org/10.1103/PhysRevD.64.034504} {\bibfield  {journal} {\bibinfo
  {journal} {Phys. Rev.}\ }\textbf {\bibinfo {volume} {D64}},\ \bibinfo {pages}
  {034504} (\bibinfo {year} {2001})},\ \Eprint
  {https://arxiv.org/abs/hep-lat/0103029} {arXiv:hep-lat/0103029 [hep-lat]}
  \BibitemShut {NoStop}%
\bibitem [{\citenamefont {Hasenfratz}\ \emph {et~al.}(2007)\citenamefont
  {Hasenfratz}, \citenamefont {Hoffmann},\ and\ \citenamefont
  {Schaefer}}]{Hasenfratz:2007rf}%
  \BibitemOpen
  \bibfield  {author} {\bibinfo {author} {\bibfnamefont {A.}~\bibnamefont
  {Hasenfratz}}, \bibinfo {author} {\bibfnamefont {R.}~\bibnamefont
  {Hoffmann}},\ and\ \bibinfo {author} {\bibfnamefont {S.}~\bibnamefont
  {Schaefer}},\ }\bibfield  {title} {\bibinfo {title} {{Hypercubic smeared
  links for dynamical fermions}},\ }\href
  {https://doi.org/10.1088/1126-6708/2007/05/029} {\bibfield  {journal}
  {\bibinfo  {journal} {JHEP}\ }\textbf {\bibinfo {volume} {05}},\ \bibinfo
  {pages} {029}},\ \Eprint {https://arxiv.org/abs/hep-lat/0702028}
  {arXiv:hep-lat/0702028 [hep-lat]} \BibitemShut {NoStop}%
\bibitem [{\citenamefont {Bernard}\ and\ \citenamefont
  {DeGrand}(2000)}]{Bernard:1999kc}%
  \BibitemOpen
  \bibfield  {author} {\bibinfo {author} {\bibfnamefont {C.~W.}\ \bibnamefont
  {Bernard}}\ and\ \bibinfo {author} {\bibfnamefont {T.~A.}\ \bibnamefont
  {DeGrand}},\ }\bibfield  {title} {\bibinfo {title} {{Perturbation theory for
  fat link fermion actions}},\ }\bibfield  {booktitle} {\emph {\bibinfo
  {booktitle} {{Lattice field theory. Proceedings, 17th International
  Symposium, Lattice'99, Pisa, Italy, June 29-July 3, 1999}}},\ }\href
  {https://doi.org/10.1016/S0920-5632(00)91822-X} {\bibfield  {journal}
  {\bibinfo  {journal} {Nucl. Phys. Proc. Suppl.}\ }\textbf {\bibinfo {volume}
  {83}},\ \bibinfo {pages} {845} (\bibinfo {year} {2000})},\ \Eprint
  {https://arxiv.org/abs/hep-lat/9909083} {arXiv:hep-lat/9909083 [hep-lat]}
  \BibitemShut {NoStop}%
\bibitem [{\citenamefont {Shamir}\ \emph {et~al.}(2011)\citenamefont {Shamir},
  \citenamefont {Svetitsky},\ and\ \citenamefont {Yurkovsky}}]{Shamir:2010cq}%
  \BibitemOpen
  \bibfield  {author} {\bibinfo {author} {\bibfnamefont {Y.}~\bibnamefont
  {Shamir}}, \bibinfo {author} {\bibfnamefont {B.}~\bibnamefont {Svetitsky}},\
  and\ \bibinfo {author} {\bibfnamefont {E.}~\bibnamefont {Yurkovsky}},\
  }\bibfield  {title} {\bibinfo {title} {{Improvement via hypercubic smearing
  in triplet and sextet QCD}},\ }\href
  {https://doi.org/10.1103/PhysRevD.83.097502} {\bibfield  {journal} {\bibinfo
  {journal} {Phys. Rev.}\ }\textbf {\bibinfo {volume} {D83}},\ \bibinfo {pages}
  {097502} (\bibinfo {year} {2011})},\ \Eprint
  {https://arxiv.org/abs/1012.2819} {arXiv:1012.2819 [hep-lat]} \BibitemShut
  {NoStop}%
\bibitem [{\citenamefont {DeGrand}\ \emph {et~al.}(2014)\citenamefont
  {DeGrand}, \citenamefont {Shamir},\ and\ \citenamefont
  {Svetitsky}}]{DeGrand:2014rwa}%
  \BibitemOpen
  \bibfield  {author} {\bibinfo {author} {\bibfnamefont {T.}~\bibnamefont
  {DeGrand}}, \bibinfo {author} {\bibfnamefont {Y.}~\bibnamefont {Shamir}},\
  and\ \bibinfo {author} {\bibfnamefont {B.}~\bibnamefont {Svetitsky}},\
  }\bibfield  {title} {\bibinfo {title} {{Suppressing dislocations in
  normalized hypercubic smearing}},\ }\href
  {https://doi.org/10.1103/PhysRevD.90.054501} {\bibfield  {journal} {\bibinfo
  {journal} {Phys. Rev.}\ }\textbf {\bibinfo {volume} {D90}},\ \bibinfo {pages}
  {054501} (\bibinfo {year} {2014})},\ \Eprint
  {https://arxiv.org/abs/1407.4201} {arXiv:1407.4201 [hep-lat]} \BibitemShut
  {NoStop}%
\bibitem [{\citenamefont {{MILC Collaboration}}()}]{MILC}%
  \BibitemOpen
  \bibfield  {author} {\bibinfo {author} {\bibnamefont {{MILC
  Collaboration}}},\ }\href@noop {} {}\bibinfo {howpublished}
  {\url{http://www.physics.utah.edu/~detar/milc/}}\BibitemShut {NoStop}%
\end{thebibliography}%
\end{document}